\documentclass[aps,twocolumn,superscriptaddress,amsmath,amsfonts]{revtex4-2}

\usepackage{graphicx,float}
\usepackage{braket}
\usepackage{siunitx}
\usepackage{appendix}
\usepackage{amssymb}
\usepackage{mathrsfs}
\usepackage{appendix}
\usepackage{slashed}
\usepackage{physics}
\usepackage{mathtools}
\usepackage{amsmath}
\usepackage{bbold}
\usepackage{xcolor}
\usepackage[mathscr]{euscript}

\begin{document}
\title{Onset of non-Gaussian quantum physics in pulsed squeezing with mesoscopic fields}

\author{Ryotatsu~Yanagimoto}
\thanks{These authors contributed equally to this work.\\Email: ryotatsu@stanford.edu, edwin.ng@ntt-research.com}
\affiliation{E.\,L.\,Ginzton Laboratory, Stanford University, Stanford, California 94305, USA}

\author{Edwin~Ng}
\thanks{These authors contributed equally to this work.\\Email: ryotatsu@stanford.edu, edwin.ng@ntt-research.com}
\affiliation{E.\,L.\,Ginzton Laboratory, Stanford University, Stanford, California 94305, USA}
\affiliation{Physics \& Informatics Laboratories, NTT Research, Inc., Sunnyvale, California 94085, USA}

\author{Atsushi~Yamamura}
\affiliation{E.\,L.\,Ginzton Laboratory, Stanford University, Stanford, California 94305, USA}

\author{Tatsuhiro~Onodera}
\affiliation{Physics \& Informatics Laboratories, NTT Research, Inc., Sunnyvale, California 94085, USA}
\affiliation{School of Applied and Engineering Physics, Cornell University, Ithaca, New York 14853, USA}

\author{Logan~G.~Wright}
\affiliation{Physics \& Informatics Laboratories, NTT Research, Inc., Sunnyvale, California 94085, USA}
\affiliation{School of Applied and Engineering Physics, Cornell University, Ithaca, New York 14853, USA}

\author{Marc~Jankowski}
\affiliation{E.\,L.\,Ginzton Laboratory, Stanford University, Stanford, California 94305, USA}
\affiliation{Physics \& Informatics Laboratories, NTT Research, Inc., Sunnyvale, California 94085, USA}

\author{M.~M.~Fejer}
\affiliation{E.\,L.\,Ginzton Laboratory, Stanford University, Stanford, California 94305, USA}

\author{Peter L. McMahon}
\affiliation{School of Applied and Engineering Physics, Cornell University, Ithaca, New York 14853, USA}

\author{Hideo~Mabuchi}
\affiliation{E.\,L.\,Ginzton Laboratory, Stanford University, Stanford, California 94305, USA}
\date{\today}
\begin{abstract}
    We study the emergence of non-Gaussian quantum features in pulsed squeezed light generation with a mesoscopic number (i.e., dozens to hundreds) of pump photons. Due to the strong optical nonlinearities necessarily involved in this regime, squeezing occurs alongside significant pump depletion, compromising the predictions made by conventional semiclassical models for squeezing. Furthermore, nonlinear interactions among multiple frequency modes render the system dynamics exponentially intractable in na\"ive quantum models, requiring a more sophisticated modeling framework. To this end, we construct a nonlinear Gaussian approximation to the squeezing dynamics, defining a ``Gaussian interaction frame'' (GIF) in which non-Gaussian quantum dynamics can be isolated and concisely described using a few dominant (i.e., principal) supermodes. Numerical simulations of our model reveal non-Gaussian distortions of squeezing in the mesoscopic regime, largely associated with signal-pump entanglement. We argue that the state of the art in nonlinear nanophotonics is quickly approaching this regime, providing an all-optical platform for experimental studies of the semiclassical-to-quantum transition in a rich paradigm of coherent, multimode nonlinear dynamics. Mesoscopic pulsed squeezing thus provides an intriguing case study of the rapid rise in dynamic complexity associated with semiclassical-to-quantum crossover, which we view as a correlate of the emergence of new information-processing capacities in the quantum regime.
\end{abstract}

\maketitle
\section{Introduction}
Harnessing the quantum-mechanical nature of optical fields promises a route towards achieving quantum-enhanced performance in a wide variety of applications in quantum information and engineering~\cite{Nielsen2000,Gisin2007,Giovannetti2011}. The generation and control of squeezed light particularly stand out as an important gateway to the field of quantum optics~\cite{Walls1983,Andersen2016}: In addition to its foundational role in establishing concepts like quantum entanglement and teleportation in optical systems~\cite{Furusawa1998,Ou1992}, the generation of high-quality squeezed light is an essential building block for quantum technologies such as continuous-variable quantum computation~\cite{Menicucci2006,Asavarant2019}, Gaussian boson sampling~\cite{Zhong2020,Arrazola2021}, and metrology~\cite{Pezze2008,LIGO2013}, among many others~\cite{Vahlbruch2007,Yuen1987,Hillery2000}.

In this context, the use of ultra-short pulses is an attractive technological option~\cite{Slusher1987,Anderson1995,Triginer2020,Eckstein2011}. High peak intensities significantly reduce the average pump powers needed to generate non-classical light~\cite{Florez2020,Yanagimoto2020}, mitigating the need for high-Q, low-bandwidth resonators ubiquitous in continuous-wave devices. Furthermore, the multimode nature of pulses enables manipulation of the quantum state of light over a high-dimensional set of spectral-temporal channels, e.g., via temporal pulse-shaping~\cite{Brecht2015,Grice1997,Ansari2018}. These advantages are further amplified by recent advances in the fabrication of photonic devices with nanoscale transverse field confinement, enabling highly nonlinear dispersion-engineered interactions~\cite{Jankowski2020,Jankowski2021-review}.

In fact, classical scaling arguments suggest that this emerging class of nonlinear-optical devices, with strong field confinement in both space and time, is poised to reach an unprecedented regime of attojoule-per-pulse operation~\cite{Jankowski2021-review}. It is natural to ask, therefore, when and how we might expect the \emph{quantum} nature of the propagating fields to play an important role in engineering photonics at this scale. Unfortunately, the conventional tools of quantum optics developed to study few-photon interactions~\cite{Yanagimoto2021_mps,Leung2009,Hafezi2012,Drummond1997} cannot readily be applied to this \emph{mesoscopic} regime, in which a complete description for the propagation dynamics critically requires understanding the interplay between both semiclassical features such as strong squeezing and non-classical features such as Wigner-function negativity~\cite{Kenfack2004,Walschaers2021,Lvovsky2020}. Physically, in the case of pulsed squeezed light generation, this mesoscopic regime is characterized by strong squeezing and significant pump depletion induced by dozens to hundreds of pump photons.

Further complicating matters, the multimode nature of quantum pulse propagation in a strongly nonlinear medium is highly nontrivial. As indicated even by linear analysis of pulsed squeezing, photons in a broadband pulse experience complicated spectral-temporal entanglement~\cite{Wasilewski2006,Lvovsky2007}, for which we only have analytic treatments in the regime of weak nonlinearities~\cite{Christ2013}. Multimode effects are also well known to cause difficulties for the realization of pulsed quantum logic gates~\cite{Shapiro2006}, and careful dispersion control and pulse shaping techniques have been developed in various contexts to realize (quasi-)single-mode gate operations~\cite{Eckstein2011,Yanagimoto2020,Brecht2015}. But in the mesoscopic regime, any similar attempts to harness ultrafast quantum coherence face a significant modeling problem: The combination of non-Gaussian features with multimode entanglement produce complex nonlinear dynamics that often require exponentially large Hilbert-space dimension to describe in conventional models~\cite{Lloyd1999,Braunstein2005}. Therefore, we require more sophisticated model-reduction techniques that make use of physical intuition to realize more concise representations of the dynamics in this regime~\cite{Yanagimoto2021-spie}.

Mature frameworks for describing multimode entanglement in pulsed squeezing are readily available through the use of Gaussian-state models~\cite{Olivares2012,Weedbrook2021}, in which quantum fluctuations and correlations are assumed to follow Gaussian statistics under weak nonlinearities. Gaussian-state models are especially efficient when combined with supermode techniques that can be used to identify reduced sets of spectral-temporal waveforms that dominate the multimode entanglement structure~\cite{Wasilewski2006,Lvovsky2007,Christ2013,Gouzien2020}. As we move towards stronger nonlinearities, however, quantum fluctuations begin to develop \emph{non-Gaussian} features~\cite{Kenfack2004,Walschaers2021,Lvovsky2020}, making a Gaussian-state assumption invalid. However, even in the presence of significant nonlinear dynamics, a Gaussian model can still serve as a useful leading-order approximation to the true quantum state, providing essential information, for example, on the supermodes in which non-Gaussian features can be expected to first appear~\cite{Onodera2018}.

In this paper, we leverage these opportunities to study full-quantum dynamics of pulsed squeezing in the mesoscopic regime by formulating a model-reduction approach that begins with a dynamical Gaussian approximation for the system dynamics going beyond conventional undepleted-pump approximations. This allows us to define a Gaussian interaction frame (GIF) in which the non-Gaussian quantum dynamics can be efficiently projected onto a low-dimensional set of principal supermodes dynamically inferred from the GIF. Numerical simulations of the reduced models provide new insights into the nature of quantum non-Gaussian features (e.g., Wigner-function negativity~\cite{Kenfack2004}) in squeezed light generation. We find that ``non-Gaussian corruption'' in the form of increased quadrature noise and loss of purity due to signal-pump entanglement can become an obstacle for quantum applications reliant on high-quality squeezing~\cite{Menicucci2006,Asavarant2019,Zhong2020,Arrazola2021}.

At the same time, the exotic quantum states produced in this mesoscopic regime also point towards novel quantum applications that make use of high-bandwidth, coherent non-Gaussianity~\cite{Lvovsky2020,Walschaers2021}, e.g., as resources for quantum information processing~\cite{Zhuang2018,Albarelli2018,Killoran2019}, quantum illumination~\cite{Fan2018}, and quantum metrology~\cite{Gessner2019}. Notably, in contrast to few-photon nonlinear quantum dynamics, the mesoscopic pulses studied here can retain large peak intensities that effectively enhance the nonlinear interactions, mitigating hardware requirements needed to realize these nonclassical features. In fact, state-of-the-art experimental numbers for nanophotonic nonlinear waveguides indicate this mesoscopic regime can be relevant in the near future.

\section{Single-mode squeezing beyond undepleted pump approximation}
\label{sec:singlemode}
In this section, we introduce a model-reduction framework based on the notion of a Gaussian interaction frame (GIF), in which nonlinear quantum dynamics can be concisely captured. Our approach is similar in spirit to Ref.~\cite{Tezak2017}, where the aim is to ``factor out'' semiclassical features of the system evolution using a Gaussian unitary $\hat{U}(t)$, leaving only non-Gaussian quantum state evolution in the interaction frame induced by $\hat{U}(t)$. In this approach, $\hat U(t)$ is defined by a set of time-dependent parameters, whose evolution can follow \emph{nonlinear} equations of motion; in vacuum squeezing, for instance, this nonlinearity can faithfully capture pump depletion (which is neglected in most Gaussian frameworks). While we focus on the case of vacuum squeezing in this paper, the GIF framework can, in principle, be extended to a broad class of Hamiltonians and states.

Before we show how our approach can be applied to the main problem of pulsed squeezing, which requires a multimode treatment, we first illustrate the central ideas by revisiting the quantum dynamics of \emph{single-mode} squeezing. We consider the canonical $\chi^{(2)}$ Hamiltonian
\begin{align}
    \label{eq:single-mode-chi2}
    \hat{H}=\frac{1}{2}\left(\hat{a}^2\hat{b}^\dagger+\hat{a}^{\dagger 2}\hat{b}\right)+\delta\hat{b}^\dagger\hat{b},
\end{align}
where $\delta$ is the normalized phase-mismatch, and $\hat{a}$ and $\hat{b}$ are annihilation operators for the signal, or fundamental-harmonic (FH), and pump, or second-harmonic (SH), modes, respectively. The model \eqref{eq:single-mode-chi2} is well-established for describing $\chi^{(2)}$-nonlinear interactions in single-mode resonators~\cite{Lu2020,Heuck2020b,Kinsler1991}, while we emphasize that for modeling pulsed squeezing, it is, at best, only a toy model, as the broadband nature of waveguided parametric interactions produces physics that are intrinsically multimode~\cite{Wasilewski2006,Lvovsky2007,Christ2013}. As we will see in this section, however, even the quantum treatment of single-mode squeezing can be significantly improved upon, which, as we will see, directly translates into strategies for mitigating the generic intractability of quantum multimode dynamics~\cite{Braunstein2005,Lloyd1999} in the pulsed setting. 

For vacuum squeezing, we begin with a coherent state in the pump mode and vacuum in the signal, i.e., the initial state $\ket{\varphi(0)}$ obeys $\expectationvalue{\hat{b}}{\varphi(0)}=\beta(0)$, where $\beta(0)$ is the initial pump amplitude. The state then evolves canonically according to the Schr\"odinger equation
\begin{align}
    \label{eq:schroedinger}
    \mathrm{i}\partial_t\ket{\varphi(t)}=\hat{H}\ket{\varphi(t)}.
\end{align}
In the semiclassical regime with $|\beta(0)|\gg 1$, the resulting dynamics are well-approximated as single-mode Gaussian squeezing of the signal while negligible depletion of the pump occurs. In principle, however, the nonlinear dynamics produced by \eqref{eq:single-mode-chi2} and \eqref{eq:schroedinger} can take $\ket{\varphi(t)}$ out of this Gaussian-state manifold, e.g., when there is significant pump depletion~\cite{Kinsler1991}.

In the approach that follows, we aim to reconcile these two pictures by employing a hybrid parameterization of $\ket{\varphi(t)}$ that has both a significant Gaussian (i.e., semiclassical) component and a relatively small but non-negligible non-Gaussian (i.e., fully quantum) component. Specifically, we ``factor out'' the semiclassical part of dynamics as a parametrized Gaussian unitary $\hat{U}(t)$, i.e., $\ket{\varphi(t)}=\hat{U}(t)\ket{\varphi_\text{I}(t)}$, where $\ket{\varphi_\text{I}(t)}$ represents the residual non-Gaussian features~\cite{Tezak2017}. Thus the dynamics of $\ket{\varphi(t)}$ are represented instead as the joint dynamics of $\hat U(t)$ and $\ket{\varphi_\text{I}(t)}$. Formally, the time evolution of $\ket{\varphi_\text{I}(t)}$ is given in the interaction frame of the unitary $\hat{U}(t)$ as
\begin{subequations}
    \label{eq:schroedinger-if}
\begin{align}
    \mathrm{i}\partial_t\ket{\varphi_\text{I}(t)}=\hat{H}_\text{I}(t)\ket{\varphi_\text{I}(t)},
\end{align}
with an interaction-frame Hamiltonian
\begin{align}
    \label{eq:single-mode-hi}
    \hat{H}_\text{I}(t)&=\hat{U}^\dagger(t)\hat{H}\hat{U}(t)-\mathrm{i}\hat{U}^\dagger(t)\partial_t \hat{U}(t).
\end{align}
\end{subequations}
If $\hat{U}(t)$ is appropriately chosen, only non-trivial quantum features remain in the interaction frame, making \eqref{eq:schroedinger-if} a ``compressed'' version of the lab-frame dynamics \eqref{eq:schroedinger}. In addition to this compression directly facilitating more efficient numerical simulations, the identification of a proper interaction frame can also be physically insightful: in regimes where $\ket{\varphi(t)}$ is only negligibly non-Gaussian, we can reasonably expect $\ket{\varphi_\text{I}(t)} \approx \ket{0}$, which then implies $\ket{\varphi_\text{G}(t)}=\hat{U}(t)\ket{0}$ is a Gaussian state which faithfully approximates the true state $\ket{\varphi(t)}=\hat{U}(t)\ket{\varphi_\text{I}(t)}$.

The rest of the section is structured as follows. In Sec.~\ref{subsec:compressed}, we specify a generic representation of $\hat{U}(t)$ as a squeezing unitary effectively pumped by a time-dependent function $\beta(t)$, whose value at $t=0$ coincides with the initial pump amplitude $\beta(0)$. In Sec.~\ref{subsec:undepleted}, we show that the conventional undepleted-pump approximation is equivalent to choosing $\beta(t) = e^{-\mathrm{i}\delta t}\beta(0)$. In Sec.~\ref{subsec:gif}, we introduce the GIF given by a more refined choice of $\beta(t)$ taking finite pump depletion into account. We show that, compared to the undepleted-pump approximation, the GIF provides a better Gaussian approximation to the true quantum dynamics while also prescribing a more efficient compressed representation for the non-Gaussian part of the state.

\subsection{Quantum dynamics in an interaction frame}
\label{subsec:compressed}
As alluded to above, we choose a compressed representation of the quantum state by specifying a time-evolving Gaussian unitary with the generic form
\begin{align}
    \label{eq:gassian-decomposition}
    \hat{U}(t)=\hat{D}(t)\hat{Q}(t),
\end{align}
where $\hat{D}(t)$ is a displacement operation and $\hat{Q}(t)$ is a symplectic transformation~\cite{Olivares2012} (i.e., phase shift and squeezing). If the Hamiltonian $\hat{H}$ were Gaussian, i.e., containing terms only up to second order in operators, there exist appropriate choices of $\hat{D}(t)$ and $\hat{Q}(t)$ that would make $\hat{H}_\text{I}(t)=0$, completely eliminating dynamics in the interaction frame and indicating that we have picked an optimal interaction frame. Conversely, residual dynamics in the interaction frame can be attributed to nontrivial non-Gaussian quantum evolution that could not be factored out by $\hat{U}(t)$. 

\begin{figure}[bt]
    \centering
    \includegraphics[width=0.38\textwidth]{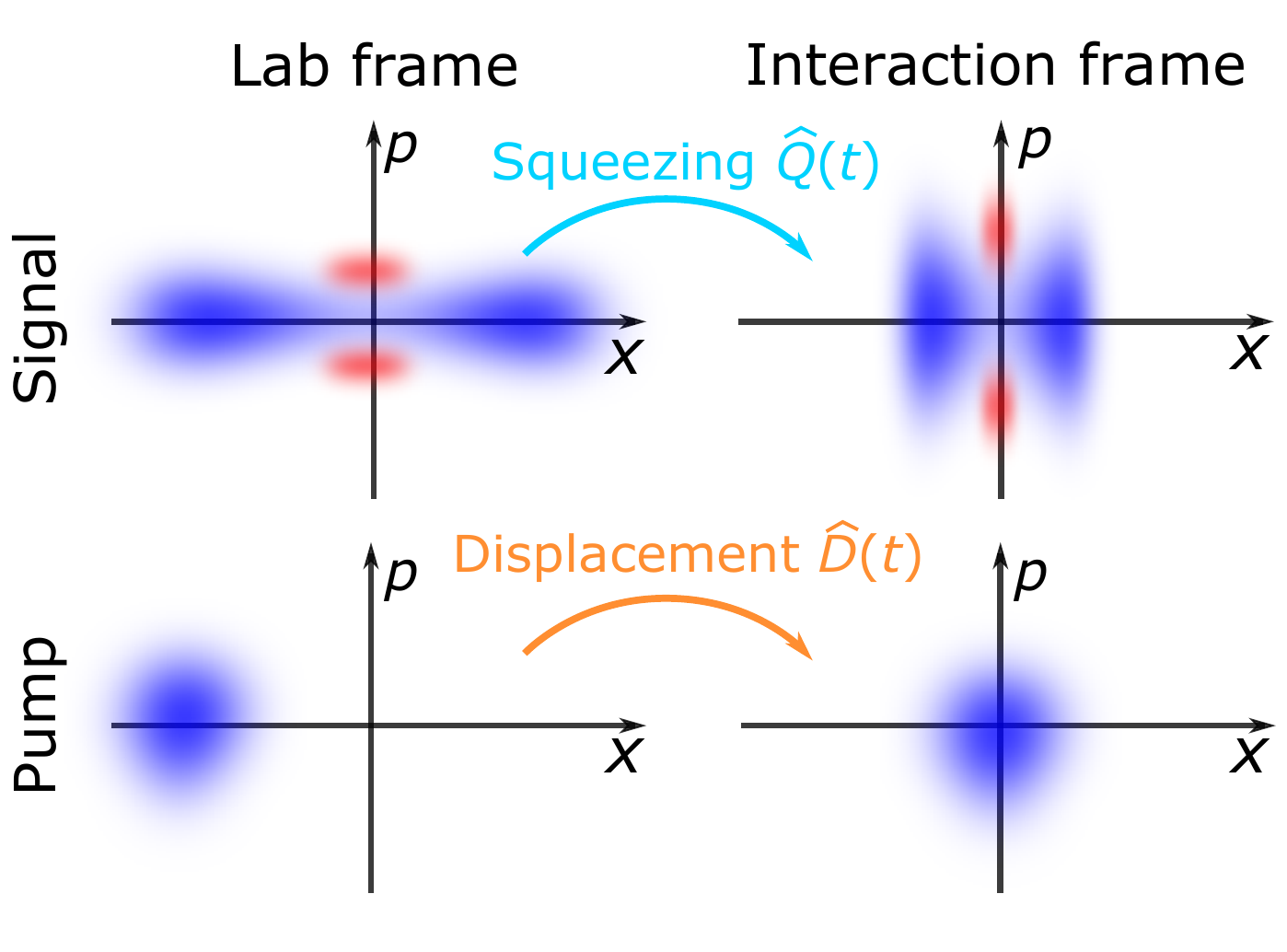}
    \caption{Illustration of a Gaussian interaction frame for single-mode squeezing, depicted in terms of phase-space portraits (i.e., Wigner functions). In the lab frame, the signal experiences squeezing, while the pump experiences displacement, both of which are generally time-dependent. Excitations from the origin can be reduced by moving to an interaction frame via an appropriate Gaussian transformation, here implemented by a squeezing transformation $\hat{Q}(t)$ on the signal and a displacement transformation $\hat{D}(t)$ on the pump~\cite{Tezak2017}.}
    \label{fig:qmanifold}
\end{figure}

We specialize $\hat D(t)$ and $\hat Q(t)$ to the case of vacuum squeezing, although as we discuss later, suitable modifications to the expressions below can also be made to accommodate more general conditions. For vacuum squeezing, we expect no displacement to occur in signal, since $\hat{a} \mapsto -\hat{a}$ is a symmetry of $\hat H$ and $\ket{\varphi(0)}$; thus $\hat D(t)$ should only involve pump operators. Additionally, in its role as the pump, the pump is expected to remain close to a coherent state, and thus, we exclude squeezing of the pump from $\hat{Q}(t)$. In Fig.~\ref{fig:qmanifold}, we illustrate how the use of pump displacement and signal squeezing can be used to compress excitations in the interaction frame. Concretely, we parametrize $\hat U(t)$ using a time-dependent complex-valued function $\beta(t)$ as
\begin{subequations}
    \label{eq:single-mode-if}
\begin{align}
    \label{eq:single-mode-displacement}
    \hat{D}(t)&=\exp\left(\beta(t)\hat{b}^\dagger-\beta^*(t)\hat{b}\right) \\
    \label{eq:single-mode-unitary-undepleted}
    \hat{Q}(t)&=\mathcal{T}\exp(-\mathrm{i}\int^t_0\mathrm{d}{t'}\hat{V}_\text{Q}(t')), \quad\text{where} \\
    \label{eq:single-mode-quadratic}
    \hat{V}_\text{Q}(t)&=\frac{1}{2}\left(\beta(t)\hat{a}^{\dagger2}+\beta^*(t)\hat{a}^2\right)+\delta\hat{b}^\dagger\hat{b},
\end{align}
\end{subequations}
where $\mathcal{T}$ denotes time-ordering~\cite{Quesada2014}. In this parameterization for $\hat Q(t)$, $\beta(t)$ effectively drives the signal squeezing, suggesting that a good choice for $\beta(t)$ would be some estimation of the true pump amplitude $\expectationvalue{\hat{b}}{\varphi(t)}$. The goal in the following is to identify a reasonable choice for $\beta(t)$ by attempting to minimize the dynamics in the interaction frame, or, equivalently, to maximally capture the Gaussian part of the system evolution. 

To evaluate $\hat{H}_\text{I}(t)$ using \eqref{eq:single-mode-hi} we use
\begin{subequations}
 \label{eq:single-mode-trans}
\begin{align}
 \label{eq:single-mode-fh-trans}
    \hat{U}^\dagger(t)\hat{a}\hat{U}(t)&=C(t)\hat{a}+S(t)\hat{a}^\dagger\\
    \label{eq:single-mode-sh-trans}
    \hat{U}^\dagger(t)\hat{b}\hat{U}(t)&=e^{-\mathrm{i}\delta t}\hat{b}+\beta(t),
\end{align}
\end{subequations}
where $C(t)$ and $S(t)$ are Green's functions with initial conditions $C(0)=1$ and $S(0)=0$; the general form \eqref{eq:single-mode-trans} is referred to as a Bologiubov transformation~\cite{Braunstein2005}. It can be shown that $C(t)$ and $S(t)$ are determined by $\beta(t)$ according to
\begin{subequations}
    \label{eq:single-mode-greens-eom}
    \begin{align}
        \partial_t C(t)&=-\mathrm{i}\beta(t)S^*(t)\\
        \partial_t S(t)&=-\mathrm{i}\beta(t)C^*(t).
    \end{align}
\end{subequations}
As a result, $\hat{H}_\text{I}(t)$ takes the simple form
\begin{subequations}
\begin{align}
    \label{eq:single-mode-gif-hamiltonian}
    \hat{H}_\text{I}(t)=\hat{H}_\text{NL}(t)+\hat{H}_\text{L}(t),
\end{align}
consisting of a cubic nonlinear term
\begin{align}
    \label{eq:single-mode-hc}
    \hat{H}_\text{NL}(t)=\frac{e^{\mathrm{i}\delta t}}{2}\hat{b}^\dagger \left(C(t)\hat{a}+S(t)\hat{a}^\dagger\right)^2+\mathrm{H.c.}
\end{align}
and a linear term
\begin{align}
    \hat{H}_\text{L}(t)=e^{\mathrm{i}\delta t}\left(\delta \beta(t)-\mathrm{i}\partial_t \beta(t)\right)\hat{b}^\dagger+\mathrm{H.c.}.
\end{align}
\end{subequations}
It is worth noting that due to our choice of $\hat Q(t)$, terms of the form $\hat{Q}^\dagger(t)\hat{V}_\text{Q}(t)\hat{Q}(t)$ internally cancel in the evaluation of $\hat{H}_\text{I}(t)$, leaving no quadratic contributions in \eqref{eq:single-mode-gif-hamiltonian}.
\subsection{The undepleted-pump approximation}
\label{subsec:undepleted}
One simple way of fixing $\beta(t)$ is to first ignore $\hat H_\text{NL}(t)$ altogether and choose $\beta(t)$ so that $\hat H_\text{L}(t) = 0$, assuming that higher-order nonlinear terms $\hat H_\text{NL}(t)$ have negligible contributions compared to $H_\text{L}(t)$. This choice corresponds to taking $\partial_t \beta(t)=-\mathrm{i}\delta \beta(t)$, with the solution
\begin{align}
    \label{eq:single-mode-fs-undepleted}
    \beta(t)=e^{-\mathrm{i}\delta t}\beta(0).
\end{align}
This functional form of $\beta(t)$ is particularly notable because the assumptions that (i) $\expectationvalue{\hat{b}}{\varphi(t)}=\beta(t)$ as given by \eqref{eq:single-mode-fs-undepleted} and (ii) $\hat{H}_\text{NL}(t)=0$ are precisely equivalent to the undepleted-pump approximation conventionally used to treat parametric downconversion and low-efficiency squeezing. (More formally, the undepleted-pump approximation is often performed via the substitution $\hat{b}\mapsto \beta(t)$ as given by \eqref{eq:single-mode-fs-undepleted} in \eqref{eq:single-mode-chi2}.)

\begin{figure}[bt]
    \centering
    \includegraphics[width=0.5\textwidth]{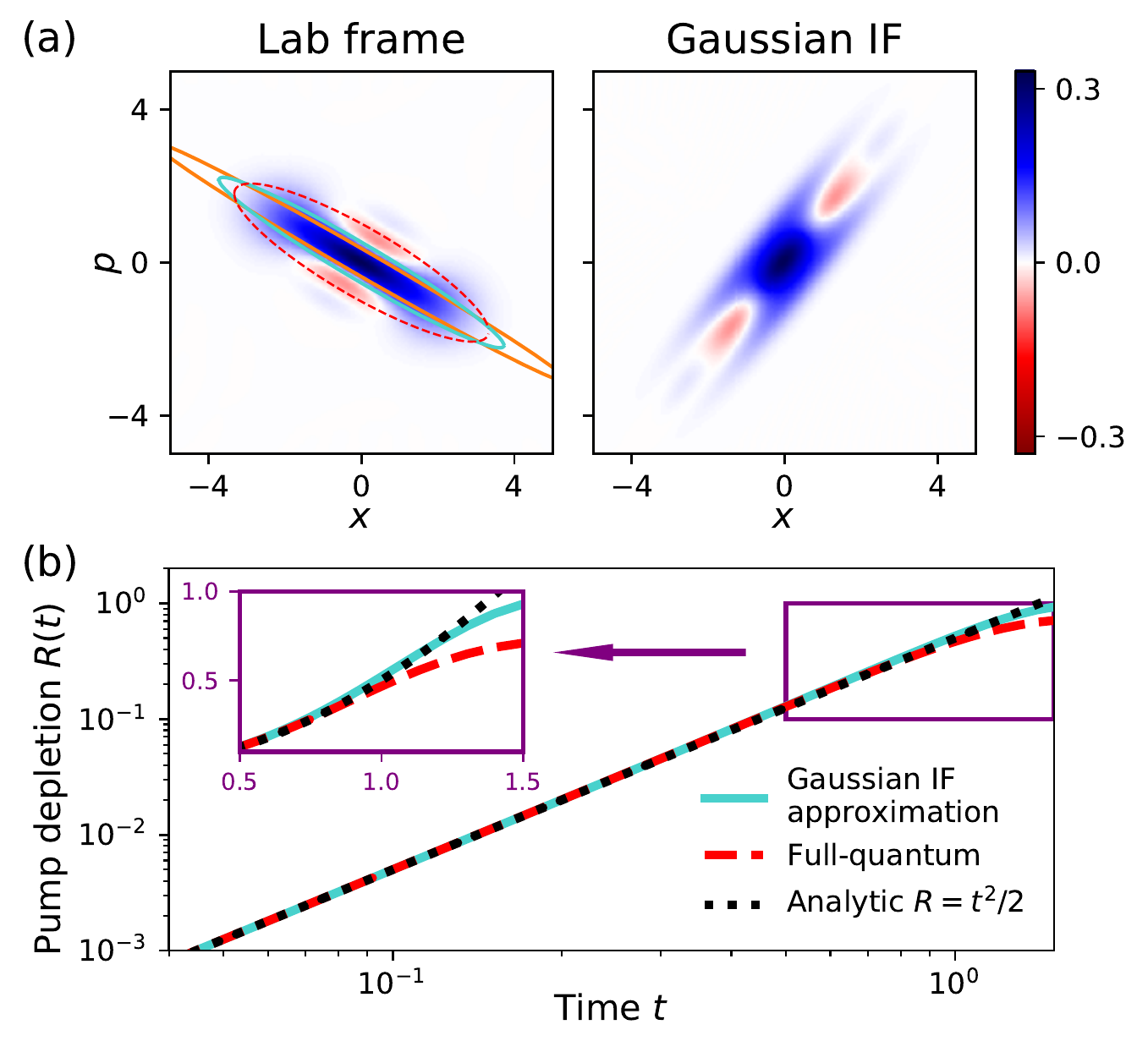}
    \caption{(a) Snapshot of the signal Wigner function at time $t=1.5$ in the lab frame (left) and the Gaussian interaction frame (IF) (right) under single-mode squeezing. The orange, turquoise, and dashed red ellipses superimposed on the left plot represent the $1/e^2$-width of the variances as predicted by the undepleted-pump approximation, GIF approximation, and the full-quantum simulation, respectively. (b) The pump depletion ratio $R(t)$ as a function of time as predicted by the Gaussian IF approximation and the full-quantum simulation. At small $t$, the GIF approximation yields an analytic scaling $R(t)\approx t^2/2$, shown as black dots. We use a phase-mismatch of $\delta=-0.5$ and an initial pump excitation of $\beta(0)=1.0$. Full-quantum simulations are performed using QuantumOptics.jl~\cite{Kraemer2018}, a general-purpose package for numerical simulation of quantum optics.}
    \label{fig:singlemode}
\end{figure}

In Fig.~\ref{fig:singlemode}, we compare the squeezing dynamics calculated with the undepleted-pump approximation against that of a full-quantum simulation in the few-pump-photon regime. From the Wigner functions for the signal mode in Fig.~\ref{fig:singlemode}(a), we see the undepleted-pump approximation greatly overestimates the amount of squeezing. Furthermore, since the approximation takes $\hat H_\text{NL}(t) = 0$ by construction, it also cannot capture the Wigner-function negativity exhibited in the full-quantum simulation. True to the name, the breakdown of the undepleted-pump approximation is most clearly seen in the pump depletion ratio
\begin{align}
\label{eq:depletionratio}
    R(t)=1-\frac{N_\text{SH}(t)}{N_\text{SH}(0)},
\end{align}
where $N_\text{SH}(t)=\langle\hat{b}^\dagger\hat{b}\rangle$ is the total pump photon number at time $t$. As shown in Fig.~\ref{fig:singlemode}(b), a full-quantum treatment exhibits pump depletion with a characteristic asymptotic scaling $R(t)\sim t^2$ for $t \rightarrow 0$, whereas $R(t)=0$ under the undepleted-pump approximation. In fact, perhaps unsurprisingly, the absence of pump depletion leads to an unphysical growth in the generalized particle number $N(t)=N_\text{FH}(t)+2N_\text{SH}(t)$, a conserved quantity (sometimes called the Manly-Rowe invariant~\cite{Drummond1997}) in $\chi^{(2)}$ nonlinear interactions. These considerations show how the undepleted-pump approximation may be insufficient to model squeezed light generation in the highly nonlinear regime.

\subsection{The Gaussian interaction frame}
\label{subsec:gif}
It turns out that by choosing $\beta(t)$ differently, we can further reduce the dynamics in the interaction frame beyond $\hat H_\text{I}(t) = \hat H_\text{NL}(t)$. Since the operators in \eqref{eq:single-mode-hc} are not normally ordered, $\hat{H}_\text{NL}(t)$ induces trivial linear displacements when acting on vacuum, which could have been absorbed into $\beta(t)$. Physically, this refinement in the choice of $\beta(t)$ can be seen as an improved estimate for $\expectationvalue{\hat{b}}{\varphi(t)}$, incorporating some of the effects of pump depletion.

In this latter approach, we instead write $\hat{H}_\text{I}(t)$ as
\begin{subequations}
\begin{align}
    \hat{H}_\text{I}(t)=\hat{H}_\text{GIF}(t)+\hat{H}'_\text{L}(t),
\end{align}
with a normally ordered nonlinear term
\begin{align}
    \label{eq:single-mode-vc}
    \hat{H}_\text{GIF}(t)&=\frac{e^{\mathrm{i}\delta t}}{2}\hat{b}^\dagger\left(S^2(t)\hat{a}^{\dagger 2}+2(CS)(t)\hat{a}^\dagger\hat{a}+C^2(t)\hat{a}^2\right)\nonumber\\
    &{}+\mathrm{H.c.}
\end{align}
and a different residual linear term
\begin{align}
    \hat{H}_\text{L}'(t)=\hat{H}_\text{L}(t)+\left(\frac{e^{\mathrm{i}\delta t}}{2}C(t)S(t)\hat{b}^\dagger+\mathrm{H.c.}\right).
\end{align}
\end{subequations}
Now, to completely eliminate linear dynamics in the interaction frame, we can set $\beta(t)$ so that $\hat{H}_\text{L}'(t)=0$ (as opposed to just $\hat{H}_\text{L}(t)=0$ in the undepleted-pump approximation), which leads to
\begin{align}
    \label{eq:single-mode-feom}
    \partial_t \beta(t)=-\mathrm{i}\delta \beta(t)-\frac{\mathrm{i}}{2}C(t)S(t),
\end{align}
upon which we have $\hat{H}_\text{I}(t)=\hat{H}_\text{GIF}(t)$. As promised, the second term in \eqref{eq:single-mode-feom} accounts for depletion of pump photons as excitations are transferred to signal mode. 

In the following, we refer to the interaction frame induced by choosing $\beta(t)$ according to \eqref{eq:single-mode-feom} as the \emph{Gaussian interaction frame (GIF)}. Additionally, in similar spirit to the undepleted-pump approximation, we can make a \emph{GIF approximation} by assuming (i) $\expectationvalue{\hat{b}}{\varphi(t)}=\beta(t)$ as given by \eqref{eq:single-mode-feom} and (ii) $\hat{H}_\text{GIF}(t)=0$. Physically, the GIF can be defined as an interaction frame in which the linear and quadratic operator terms are eliminated from the interaction-frame Hamiltonian under the normal operator ordering.

At this point, it should be clear that the notion of a GIF as defined above for the case of pulsed squeezing can be quite broadly generalized. By modifying the forms of $\hat D(t)$ and $\hat Q(t)$ and making appropriate choices for the dynamics of various parameters analogous to $\beta(t)$, we can in principle define more general GIFs that can handle, e.g., non-zero displacement in the signal, pump-signal entanglement, $\chi^{(3)}$ nonlinearities, and so on.

In Fig.~\ref{fig:singlemode}, we also consider how the GIF approximation fares in estimating the squeezing dynamics. The signal Wigner functions in Fig.~\ref{fig:singlemode}(a) show that the overestimation of squeezing is alleviated compared to the undepleted-pump approximation. Moreover, as shown in Fig.~\ref{fig:singlemode}(b), the GIF approximation reproduces early-time dynamics of the pump depletion ratio $R(t)$. In fact, one can analytically show that $R(t)\sim t^2/2$ asymptotically for the GIF approximation as $t\rightarrow 0$ (see Appendix~\ref{sec:asymptotic} for a complete derivation), which shows excellent agreement with the full-quantum results. Another notable property of the GIF approximation is that it explicitly preserves the generalized particle number $N(t)$, which can be easily verified by showing $\partial_t N(t)=0$. The conservation of $N(t)$ and the absence of residual linear terms in $\hat H_\text{I}(t)$ make the GIF a particularly attractive frame in which to study the quantum dynamics of squeezing.

Finally, we can also perform quantum simulation in the GIF as well, by evolving $\ket{\varphi_\text{I}(t)}$ under $\hat{H}_\text{GIF}(t)$ (which is equivalent to full-quantum simulation for the single-mode case considered in this section). As seen to the right in Fig.~\ref{fig:singlemode}(a), the Wigner function of the signal \emph{in the GIF} has a characteristic shape in which the squeezing present in the lab frame (to the left) has been compensated. For the rest of the paper, we show phase-space portraits in the GIF rather than the lab frame unless otherwise specified.

\section{Broadband squeezing in the Gaussian interaction frame} 
\label{sec:gif}
In this section, we apply the GIF framework developed in the previous section to the main target of this research: pulsed squeezed light generation in broadband $\chi^{(2)}$ nonlinear waveguides. For the most part, we reuse the intuitions and notations developed for the single-mode case while focusing this section primarily on the unique features of the multimode system, which, as we show, exhibits qualitatively different dynamics. 

Formally, the single-mode model of squeezing considered in Sec.~\ref{sec:singlemode} is ``zero dimensional'' in a sense that it involves a few discrete modes, represented by annihilation operators $\hat{a}$ and $\hat{b}$. On the other hand, pulse propagation in a nonlinear waveguide is a ``one-dimensional'' problem, where we need to consider continuous quantum fields $\hat{\Phi}_z$ and $\hat{\Psi}_z$ at every position along a spatial coordinate $z$ for the signal and pump, respectively. Intuitively, $\hat{\Phi}_z$ and $\hat{\Psi}_z$ are continuum analogues of $\hat{a}$ and $\hat{b}$, respectively, and their commutation relations $[\hat{\Phi}_z,\hat{\Phi}_{z'}^\dagger]=[\hat{\Psi}_z,\hat{\Psi}_{z'}^\dagger]=\delta(z-z')$ reflect the continuous nature of the photon-polariton fields they annihilate. For a $\chi^{(2)}$ nonlinear waveguide, the dynamics of these quantum fields (in a frame comoving with the signal) are generically governed by a Hamiltonian~\cite{Drummond2014,Drummond1997,Werner1995,Helt2020,Quesada2020-pdc} 
\begin{align}
    \label{eq:chi2-position}
    &\hat{H}=\frac{1}{2}\int\mathrm{d}z\left(\hat{\Phi}_z^{\dagger 2}\hat{\Psi}_z+\hat{\Phi}_z^2\hat{\Psi}_z^\dagger\right)-\frac{1}{2}\int\mathrm{d}z\left(\hat{\Phi}_z^\dagger\partial_z^2\hat{\Phi}_z\right)\nonumber\\
    &+\int\mathrm{d}z\left(d_0\hat{\Psi}_z^\dagger\hat{\Psi}_z-\mathrm{i}d_1\hat{\Psi}_z^\dagger\partial_z\hat{\Psi}_z-\frac{d_2}{2}\hat{\Psi}_z^\dagger\partial_z^2\hat{\Psi}_z\right),
\end{align}
where we have performed appropriate normalization of time and space coordinates. In this model, we approximate the energy dispersions of signal and pump up to second order around their carriers, and $d_0$ and $d_1$ respectively describe the normalized phase mismatch and group-velocity mismatch, while $d_2$ represents the ratio of the group velocity dispersion of pump to that of signal. Operator dynamics under \eqref{eq:chi2-position} follow
\begin{subequations}
    \begin{align}
        \mathrm{i}\partial_t\hat{\Phi}_z&=-\frac{1}{2}\partial_z^2\hat{\Phi}_z+\hat{\Phi}_z^\dagger\hat{\Psi}_z\\
        \mathrm{i}\partial_t\hat{\Psi}_z&=d_0\hat{\Psi}_z-\mathrm{i}d_1\partial_z\hat{\Psi}_z-\frac{d_2}{2}\partial_z^2\hat{\Psi}_z+\frac{1}{2}\hat{\Phi}_z^2,
    \end{align}
\end{subequations}
where we see that c-number substitutions $\hat{\Phi}_z\mapsto\Phi_z$ and $\hat{\Psi}_z\mapsto\Psi_z$ produce the usual normalized coupled-wave equations for classical $\chi^{(2)}$ nonlinear waveguide propagation~\cite{Agrawal2019}.

While the Hamiltonian \eqref{eq:chi2-position} in the spatial domain is convenient for capturing temporal information about the pulse dynamics, a more concise description of pulse squeezing is attained in the wavespace domain. To this end, we consider Fourier-transformed fields $\hat{\phi}_s=\int\mathrm{d}z\, e^{-2\pi \mathrm{i}sz}\hat{\Phi}_z$ and $\hat{\psi}_s=\int\mathrm{d}z\, e^{-2\pi \mathrm{i}sz}\hat{\Psi}_z$, so that
\begin{align}\begin{split}
    \label{eq:chi2-full}
    \hat{H}&=\frac{1}{2}\int\mathrm{d}s\,\mathrm{d}s'\left(\hat{\psi}_{s+s'}\hat{\phi}_s^\dagger\hat{\phi}_{s'}^\dagger+\hat{\psi}_{s+s'}^\dagger\hat{\phi}_s\hat{\phi}_{s'}\right)\\
    &+\int\mathrm{d}s\,\left(\gamma_s\hat{\phi}_s^\dagger\hat{\phi}_s+\delta_s\hat{\psi}_s^\dagger\hat{\psi}_s\right),
\end{split}\end{align}
where $\hat{\phi}_s$ and $\hat{\psi}_s$ are respectively annihilation operators for continuum monochromatic signal and pump modes, obeying commutation relations $[\hat{\phi}_s,\hat{\phi}_{s'}^\dagger]=[\hat{\psi}_s,\hat{\psi}_{s'}^\dagger]=\delta(s-s')$. We label these wavespace modes by the wavenumber $s$, and the energy dispersion is now characterized by $\gamma_s=\frac12(2\pi s)^2$ and $\delta_s=d_0+d_1(2\pi s)+\frac12d_2 (2\pi s)^2$, respectively.

For the pump, we consider an initial coherent-state pump pulse with 
\begin{align}
    \expectationvalue{\hat{\psi}_s}{\varphi(0)}=\beta_s(0),
\end{align}
while the signal begins in vacuum.

To parametrize $\hat{U}(t)$ in \eqref{eq:gassian-decomposition}, we again introduce a time-dependent function $\beta_s(t)$, under which we define
\begin{subequations}
    \label{eq:if}
    \begin{align}
        \label{eq:displacement}
    \hat{D}(t)&=\exp\left(\int\mathrm{d}s\,\left(\beta_s(t)\hat{\psi}_s^\dagger-\beta_s^*(t)\hat{\psi}_s\right)\right), \\
    \label{eq:unitary-undepleted}
    \hat{Q}(t)&=\mathcal{T}\exp(-\mathrm{i}\int^t_0\mathrm{d}{t'}\,\hat{V}_\text{Q}(t')), \quad\text{where} \\
    \label{eq:chi2-undepleted}
    \hat{V}_\text{Q}(t)&=\frac{1}{2}\int\mathrm{d}s\,\mathrm{d}s'\left(\beta_{s+s'}(t)\hat{\phi}_s^\dagger\hat{\phi}_{s'}^\dagger+\beta_{s+s'}^*(t)\hat{\phi}_s\hat{\phi}_{s'}\right)\nonumber\\
    &+\int\mathrm{d}s\,\left(\gamma_s\hat{\phi}_s^\dagger\hat{\phi}_s+\delta_s\hat{\psi}_s^\dagger\hat{\psi}_s\right).
    \end{align}
\end{subequations}
To calculate the interaction-frame Hamiltonian $\hat{H}_\text{I}(t)$ according to \eqref{eq:single-mode-hi}, we note the operators transform as~\cite{Wasilewski2006,Lvovsky2007,Christ2013}
\begin{subequations}
\begin{align}
    \label{eq:transformation}
    \hat{U}^\dagger(t)\hat{\psi}_s\hat{U}(t)&=e^{-\mathrm{i}\delta_st}\hat{\psi}_s+\beta_s(t)\\
    \hat{U}^\dagger(t)\hat{\phi}_s\hat{U}(t)&=\int\mathrm{d}p\,\left(C_{s,p}(t)\hat{\phi}_p+S_{s,p}(t)\hat{\phi}^\dagger_p\right),
\end{align}
\end{subequations}
where the multimode Green's functions $C_{s,p}(t)$ and $S_{s,p}(t)$ are determined by $\beta_s(t)$ according to
\begin{subequations}
    \label{eq:greens-eom}
\begin{align}
    \partial_t C_{s,p}(t)&=-\mathrm{i}\int \mathrm{d}q\, \beta_{s+q}(t)S_{q,p}^*(t)-\mathrm{i}\gamma_sC_{s,p}(t)\\
    \partial_t S_{s,p}(t)&=-\mathrm{i}\int \mathrm{d}q\, \beta_{s+q}(t)C_{q,p}^*(t)-\mathrm{i}\gamma_sS_{s,p}(t).
\end{align}
\end{subequations}

As in the single-mode case, $\hat{H}_\text{I}(t)$ can be decomposed in two different ways:
\begin{subequations}
\begin{align}
    \label{eq:gif-hamiltonian-hc}
    \hat{H}_\text{I}(t)&=\hat{H}_\text{NL}(t)+\hat{H}_\text{L}(t)\\
    \label{eq:gif-hamiltonian-hgif}
    &=\hat{H}_\text{GIF}(t)+\hat{H}_\text{L}'(t)
\end{align}
\end{subequations}
depending on operator ordering. The first decomposition \eqref{eq:gif-hamiltonian-hc} contains a linear term
\begin{align}
    \hat{H}_\text{L}(t)=\int\mathrm{d}s\,e^{\mathrm{i}\delta_s t}\left(\delta_s\beta_s(t)-\mathrm{i}\partial_t \beta_s(t)\right)\hat{\psi}_s^\dagger+\mathrm{H.c.},
\end{align}
and by choosing $\beta_s(t)$ so that $\hat{H}_\text{L}(t)=0$, we obtain
\begin{align}
    \label{eq:fs-undepleted}
    \beta_s(t)=e^{-\mathrm{i}\delta_st}\beta_s(0).
\end{align}
Then, as in the single-mode case, we obtain dynamics equivalent to a conventional undepleted-pump approximation for pulsed squeezing upon neglecting $\hat{H}_\text{NL}(t)$.

Again, because $\hat{H}_\text{NL}(t)$ is \emph{not} normally ordered, it can induce linear dynamics upon acting on a vacuum, which are eliminated in \eqref{eq:gif-hamiltonian-hgif} with a normally-ordered cubic term
\begin{subequations}
    \begin{align}
        \label{eq:vc}
        \hat{H}_\text{GIF}&(t)=\frac{1}{2} \int\mathrm{d}s\,\mathrm{d}s'\mathrm{d}p\,\mathrm{d}p'\, e^{\mathrm{i}\delta_{s+s'}t}\hat{\psi}_{s+s'}^\dagger \nonumber\\
        &{}\times\left(S_{s,p}(t)S_{s',p'}(t)\hat{\phi}_p^\dagger\hat{\phi}_{p'}^\dagger+2S_{s,p}(t)C_{s',p'}(t)\hat{\phi}_p^\dagger\hat{\phi}_{p'}\right.\nonumber\\
        &\qquad\left.{}+C_{s,p}(t)C_{s',p'}(t)\hat{\phi}_p\hat{\phi}_{p'}\right)+\mathrm{H.c.}
    \end{align}
    and a residual linear term
    \begin{align}
        \hat{H}_\text{L}'(t)&=\frac{1}{2}\int\mathrm{d}s\,\mathrm{d}p\,\mathrm{d}q\, e^{\mathrm{i}\delta_st}\left(C_{p,q}(t)S_{s-p,q}(t)\hat{\psi}_s^\dagger+\mathrm{H.c.}\right)\nonumber\\
        &+\hat{H}_\text{L}(t).
    \end{align}
\end{subequations}
By enforcing $\hat{H}_\text{L}'(t)=0$ to minimize trivial linear dynamics in the interaction frame, we obtain a modified choice of $\beta_s(t)$ as
\begin{align}
    \label{eq:feom-depleted}
    \partial_t \beta_s(t)=-\mathrm{i}\delta_s\beta_s(t)-\frac{\mathrm{i}}{2}\int\mathrm{d}p\,\mathrm{d}q\, C_{p,q}(t)S_{s-p,q}(t),
\end{align}
under which $\hat{H}_\text{I}(t)=\hat{H}_\text{GIF}(t)$. The interaction frame parametrized by \eqref{eq:feom-depleted} defines the multimode Gaussian interaction frame (GIF) for pulsed squeezing. As in the single-mode case, we use the term GIF approximation to refer to the Gaussian dynamics obtained by neglecting $\hat{H}_\text{GIF}(t)$.

\begin{figure}[bt]
    \centering
    \includegraphics[width=0.48\textwidth]{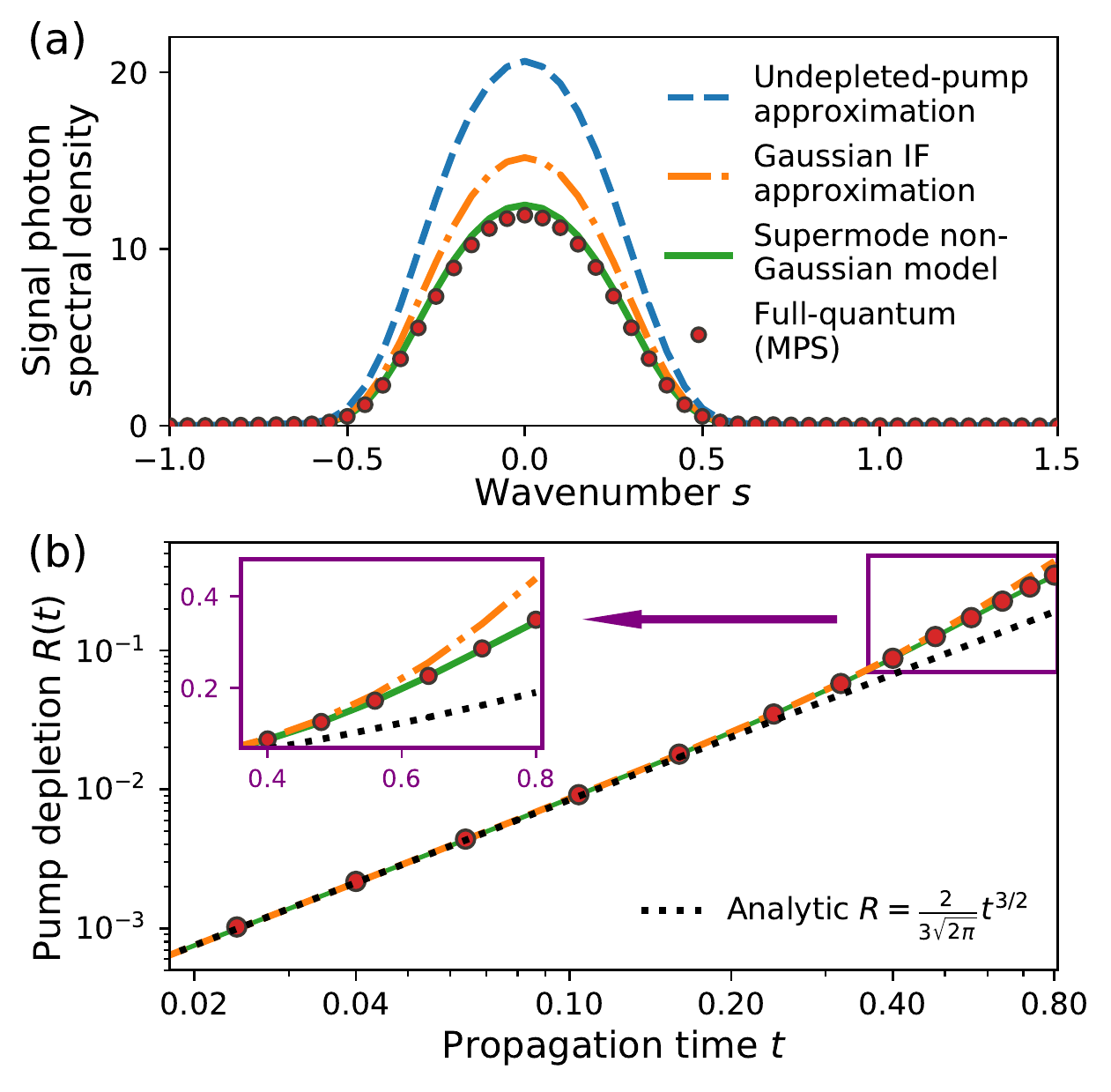}
    \caption{(a) Signal photon spectral density $\langle\hat{\phi}_s^\dagger\hat{\phi}_s\rangle$ at time $t=0.8$ as predicted by various models, including a full-quantum model using MPS. (b) Pump depletion ratio $R(t)$ as predicted by those same models (not including undepleted-pump). We initialize the system with a Gaussian-shaped pump pulse described by \eqref{eq:sh-init} with $N_\text{SH}(0)=10$. The dispersion parameters are chosen to be $d_0=d_1=0$ and $d_2=0.3$. Of the models considered here, the undepleted-pump approximation and the GIF approximation are Gaussian models, while the supermode non-Gaussian model (with $M_\text{FH}=2$) and the full-quantum MPS-based simulations can account for non-Gaussian features. Black dots in (b) show the $t \rightarrow 0$ asymptotic scaling of $R(t)\approx 2t^{3/2}/3\sqrt{2\pi}$ analytically derived with the GIF approximation.}
    \label{fig:gif}
\end{figure}

In Fig.~\ref{fig:gif}, we study pulsed squeezing in the few-pump photon regime where quantum non-Gaussianity can have a qualitative effect on the dynamics. Here, we compare the undepleted-pump approximation and the GIF approximation against a full-quantum matrix product state (MPS) simulation~\cite{Vidal2003,Garcia-Ripoll2006,Yanagimoto2021_mps}. For concreteness, we consider a Gaussian-shaped pump 
\begin{align}
    \label{eq:sh-init}
    \beta_s(0)=N_\text{SH}^{1/2}(0)\pi^{1/4}e^{-(\pi s)^2/2}
\end{align} 
with initial pump photon number $N_\text{SH}(0)=10$. Figure~\ref{fig:gif}(a) shows the signal photon spectral density, $\langle\hat{\phi}_s^\dagger\hat{\phi}_s\rangle$, where we observe the undepleted-pump approximation significantly overestimates signal power, while inclusion of finite pump depletion via the GIF approximation mitigates the discrepancy.

As in Fig.~\ref{fig:singlemode}(b), Fig.~\ref{fig:gif}(b) shows the pump depletion ratio $R(t)$. Interestingly, the multimode case exhibits a fractional asymptotic scaling of $R(t)\sim t^{3/2}$ at $t\rightarrow 0$, which is qualitatively different from the polynomial scaling $R(t)\sim t^2$ of the single-mode model. Such scaling differences highlight the effects of multimode physics intrinsic to pulse propagation~\cite{Yanagimoto2020_fano}. In fact, using the GIF approximation, we can analytically derive that $R(t)\approx 2t^{3/2}/3\sqrt{2\pi}$ regardless of the pulse shape, under appropriate conditions (see Appendix~\ref{sec:asymptotic} for details), which shows excellent agreement with the MPS simulation. As for the single-mode GIF approximation, the generalized particle number $N(t)$ is also explicitly preserved in the multimode case. Thus, the dynamics of the GIF unitary faithfully capture the Gaussian features of pulsed squeezing, including leading-order corrections associated with pump depletion.

At later times, however, we see that the GIF approximation eventually also overestimates both the signal spectrum and the pump depletion ratio, due to the contribution from $\hat H_\text{GIF}(t)$ producing non-Gaussian features (e.g., Wigner-function negativity) in $\ket{\varphi_\text{I}(t)}$ at later times. Thus, to accurately describe squeezed light generation in a highly non-classical regime, we also need to take into account the non-Gaussian dynamics in the GIF.
\section{Modeling non-Gaussianity within GIF-principal supermodes}
\label{sec:nongaussian-supermode}
In contrast to the single-mode case, where the Gaussian interaction frame has resulted in a much simpler parametrization of the quantum dynamics, we still do not yet have a tractable numerical model for pulsed squeezing. This is because a generic representation of $\ket{\varphi_\text{I}(t)}$ still requires a Hilbert-space dimension exponential in the number of modes, making na\"ive numerical simulations intractable.

For one-dimensional waveguides with few-photon pulses, it has been shown MPS is a natural and efficient way to reduce multimode complexity by exploiting the locality of 1D interactions~\cite{Yanagimoto2021_mps}, which predominantly produces short-range entanglement. However, scaling of MPS methods to larger photon numbers remains difficult, because strong spectral-temporal multimode squeezing (e.g., due to the Gaussian part of the dynamics) can lead to significant growth of long-range entanglement and require larger MPS bond dimension~\cite{Garcia-Ripoll2006}.

An alternative method, which has been particularly well-developed for treating pulsed squeezing in the undepleted-pump limit, is to account for spectral-temporal multimode interactions using the concept of pulse \emph{supermodes}~\cite{Wasilewski2006,Lvovsky2007,Christ2013,Patera2010,Gouzien2020,Raymer2020-review}. In this approach, we express the dynamics in a truncated basis of low-order, principal waveforms comprising the pulse envelope. An efficient choice of such supermodes can concisely extract the essential features of the pulse dynamics. Such an approach has been leveraged, for instance, in Ref.~\cite{Onodera2018} to unravel the nonlinear quantum dynamics of ultrafast optical parametric oscillators.

Following this approach, we introduce in this section a prescription for choosing \emph{GIF-principal supermodes}, which provide a time-dependent supermode decomposition for $\hat H_\text{GIF}(t)$ and $\ket{\varphi_\text{I}(t)}$ in the GIF to capture the predominant non-Gaussian dynamics. The use of such a ``morphing supermode''~\cite{Gouzien2020} subspace can facilitate efficient numerical simulation of quantum pulse propagation even in regimes of large photon number, where MPS-based techniques are not suitable.

Let us denote general signal and pump supermodes in the GIF as
\begin{align}
    &\hat{a}_m(t)=\int\mathrm{d}sA_{m,s}(t)\hat{\phi}_s, &\hat{b}_m(t)=\int\mathrm{d}sB_{m,s}(t)\hat{\psi}_s
\end{align}
with integer indices $m\in\{0,1,2,\cdots\}$ and normalization conditions $\int\mathrm{d}s\,A^*_{m,s}(t)A_{m',s}(t)=\delta_{m,m'}$ and $\sum_{m=0}^\infty A^*_{m,s}(t)A^*_{m,s'}(t)=\delta(s-s')$, and similarly for pump. Equivalently, $\hat{\phi}_s=\sum_{m=0}^\infty A_{m,s}^*(t)\hat{a}_m(t)$ (and similarly for $\hat b_m(t)$ and $\hat\psi_s$). The Hamiltonian can be rewritten in this supermode basis as
\begin{align}
\label{eq:supermode-gif}
    &\hat{H}_\text{GIF}(t)=\frac{1}{2}\sum_{\ell,m,n=0}^\infty\left\{\hat{b}_\ell^\dagger(t)\left(\mu_{\ell,m,n}(t)\hat{a}_m^\dagger(t)\hat{a}_n^\dagger(t)\right.\right.\\
    &\left.\left.{}+2\nu_{\ell,m,n}(t)\hat{a}_m^\dagger(t)\hat{a}_n(t)+\xi_{\ell,m,n}(t)\hat{a}_m(t)\hat{a}_n(t)\right)+\mathrm{H.c.}\right\}\nonumber
\end{align}
with tensors
\begin{align}
&\mu_{\ell,m,n}=\int\mathrm{d}s\,\mathrm{d}p\,\mathrm{d}q\,\mathrm{d}r\, e^{\mathrm{i}\delta_st}S_{p,q}S_{s-p,r}B_{\ell,s}A_{m,q}A_{n,r}\nonumber\\
&\nu_{\ell,m,n}=\int\mathrm{d}s\,\mathrm{d}p\,\mathrm{d}q\,\mathrm{d}r\, e^{\mathrm{i}\delta_st}S_{p,q}C_{s-p,r}B_{\ell,s}A_{m,q}A^*_{n,r}\nonumber\\
&\xi_{\ell,m,n}=\int\mathrm{d}s\,\mathrm{d}p\,\mathrm{d}q\,\mathrm{d}r\, e^{\mathrm{i}\delta_st}C_{p,q}C_{s-p,r}B_{\ell,s}A^*_{m,q}A^*_{n,r}.
\end{align}

To attain a reduced system description, we choose $M_\text{FH}$ and $M_\text{SH}$ supermodes for signal and pump, respectively, and neglect the dynamics of the other supermodes. At any given time $t$, these selected $(M_\text{FH}+M_\text{SH})$ supermodes span a non-Gaussian supermode subspace $\mathcal{S}(t)$, and we denote the rest of the Hilbert space as $\mathcal{E}(t)$. Formally, we approximate $\ket{\varphi(t)}$ with the reduced state
\begin{align}
    \ket{\widetilde{\varphi}(t)}=\hat{U}(t)\ket{\widetilde{\varphi}_\mathrm{I}(t)}=\hat{U}(t)\ket{\varphi_\mathrm{\mathcal{S}}(t)}\otimes \ket{0_\mathrm{\mathcal{E}}}.
\end{align}
Intuitively, $\ket{\widetilde{\varphi}_\text{I}(t)}$ is a supermode-truncated approximation of $\ket{\varphi_\text{I}(t)}$ \emph{in the GIF}, which involves only non-trivial dynamics in the subspace $\mathcal S(t)$, while dynamics in $\mathcal E(t)$ remain trivial (i.e., that component of the state remains in vacuum). Thus, aside from the Gaussian frame dynamics $\hat U(t)$, we only need to specify the dynamics for $\ket{\varphi_\mathcal{S}(t)}$, which is given by
\begin{align}
\label{eq:dynamical-ng}
    \mathrm{i}\partial_t\ket{\varphi_\mathcal{S}(t)}=\left(\widetilde{H}_\text{GIF}(t)+\widetilde{H}_\text{inertial}(t)\right)\ket{\varphi_\mathcal{S}(t)},
\end{align}
where the tilde on the Hamiltonian terms denotes that they have been projected onto $\mathcal{S}(t)$. Since $\mathcal S(t)$ is spanned by a time-dependent basis, we must include the inertial terms
\begin{align}
    &\widetilde{H}_\text{inertial}(t)=\sum_{m,n=0}^{M_\text{FH}-1}\int\mathrm{d}s\,(\mathrm{i}A_{n,s}^*(t)\partial_tA_{m,s}(t))\hat{a}_{m}^\dagger(t)\hat{a}_{n}(t)\nonumber\\
    &+\sum_{m,n=0}^{M_\text{SH}-1}\int\mathrm{d}s\,(\mathrm{i}B_{n,s}^*(t)\partial_tB_{m,s}(t))\hat{b}_{m}^\dagger(t)\hat{b}_{n}(t).
\end{align}

It is essential to choose $A_{m,s}(t)$ and $B_{m,s}(t)$, such that $\mathcal{S}(t)$ maximally contains the non-Gaussian dynamics. Upon a close inspection of $\hat{H}_\text{GIF}(t)$ given in \eqref{eq:vc}, only terms composed solely of creation operators can act non-trivially on a vacuum. This observation motivates us to choose $\mathcal{S}(t)$ by looking at the dynamics induced by
\begin{align}
\begin{split}
    \label{eq:major}
    \hat{G}=\frac{1}{2}\int\mathrm{d}s\,\mathrm{d}s'\,\mathrm{d}p\,\mathrm{d}p'\, e^{\mathrm{i}\delta_{s+s'}t}\hat{\psi}_{s+s'}^\dagger\left(S_{s,p}(t)\hat{\phi}_p^\dagger S_{s',p'}(t)\hat{\phi}_{p'}^\dagger\right),
    \end{split}
\end{align}
where the Green's functions $S_{s,p}(t)$ play the role of effective coupling matrices between signal fields and pump fields. Mathematically, the effects of these couplings can be concisely unraveled with a singular value decomposition (SVD). Specifically, we can write~\cite{Wasilewski2006,Lvovsky2007,Christ2013}
\begin{subequations}
    \label{eq:svd}
\begin{align}
    &C_{s,p}(t)=\sum_{m=0}^\infty W^*_{m,s}\cosh{\lambda_m}\,V_{m,p}\\
    \label{eq:s-svd}
    &S_{s,p}(t)=\sum_{m=0}^\infty W^*_{m,s}\sinh{\lambda_m}\,V^*_{m,p}
\end{align}
\end{subequations}
where $\sinh{\lambda_m}(t)\geq 0$ are the singular values of $S_{s,q}(t)$ sorted in descending order. Intuitively, the $m$th right singular vector $V_{m,p}(t)$ represents a signal spectral waveform that is coupled to pump fields via $S_{s,p}(t)$ with corresponding strength $\sinh \lambda_m(t)$. Therefore, a reasonable way to choose $M_\text{FH}$ signal supermodes is $A_{m,s}(t)=V_{m,s}(t)$ for $m\in\{0,1,\dots,M_\text{FH}-1\}$.

\begin{figure}[bt]
    \centering
    \includegraphics[width=0.5\textwidth]{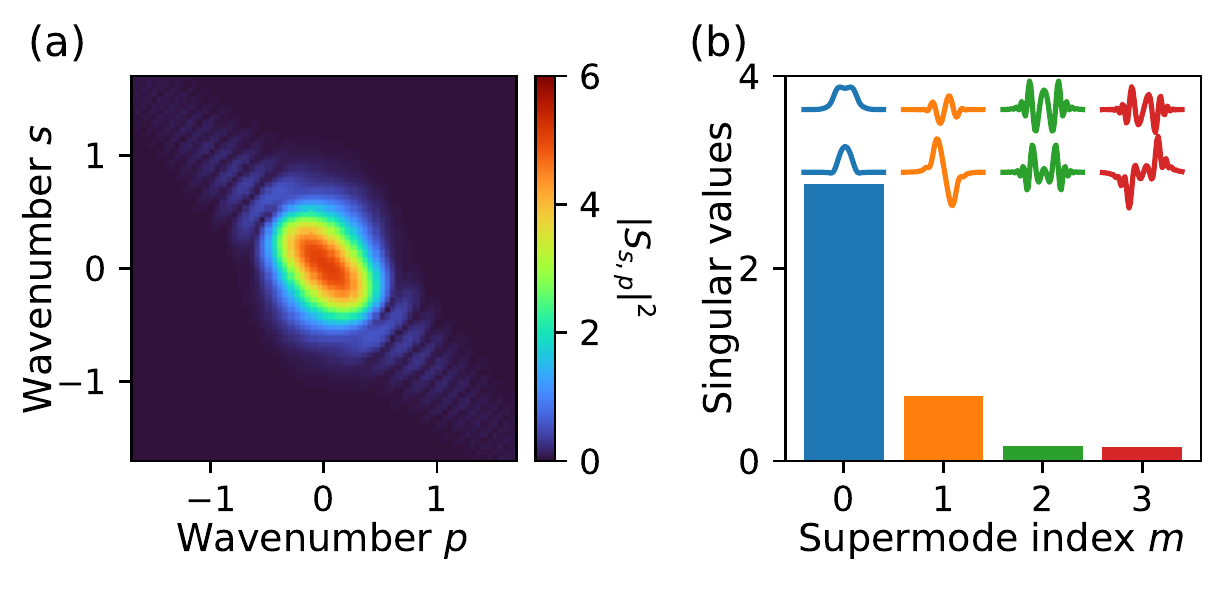}
    \caption{(a) Structure of the magnitude of Green's function $|S_{s,p}(t)|^2$ and (b) the major singular values $\sinh\lambda_m$ of $S_{s,p}(t)$, both shown at time $t=0.8$ for the waveguide parameters and initial condition from Fig.~\ref{fig:gif}. In (b), waveforms of the GIF-principal signal supermodes $A_{m,s}(t)=V_{m,s}(t)$ are shown above the bars, where the upper and lower rows depict the real and imaginary parts of $A_{m,s}(t)$, respectively.}
    \label{fig:supermode}
\end{figure}

In Fig.~\ref{fig:supermode}(a), we show the structure of the Green's function $S_{s,p}(t)$ given an initial Gaussian-shaped pump \eqref{eq:sh-init}. In this example, the structure of the singular vectors is primarily determined by the initial shape of the pump pulse with corrections from the dispersion and the depletion of the pump~\cite{Ansari2018}. As shown in Fig.~\ref{fig:supermode}(b), rapid decay of the singular value spectrum indicates that only a few signal supermodes are needed to capture dynamics in the GIF.

Finally, to complete the prescription, we also need to identify pump supermodes that are most strongly coupled to the signal supermodes chosen above. After truncating to the first $M_\text{FH}$ signal supermodes, \eqref{eq:major} reduces to
\begin{subequations}
\label{eq:truncated}
\begin{align}
    \hat{G}\approx\sum_{m,m'=0}^{M_\text{FH}-1} \int \mathrm{d}s\, K^*_{(m,m'),s}(t)\hat{\psi}_s^\dagger\hat{a}_m^\dagger(t)\hat{a}_{m'}^\dagger(t),
\end{align}
where 
\begin{align}
\begin{split}
K^*_{(m,m'),s}(t)&=\frac{1}{2}\sinh\lambda_m\sinh\lambda_{m'}\int\mathrm{d}p\,e^{\mathrm{i}\delta_{s}t}W^*_{m,p}W^*_{m',s-p}.
\end{split}
\end{align} 
\end{subequations}
In general, there exist $(2M_\text{FH}-1)$ independent vectors $K_{(m,m'),s}(t)$ involved in \eqref{eq:truncated}, indicating $\hat G$ directly couples $M_\text{SH}=(2M_\text{FH}-1)$ supermodes in the pump to the signal supermodes. Thus, $K_{(m,m'),s}(t)$ defines a non-orthogonal set of vectors (in $s$) indexed by $(m,m')$ that we can orthonormalize into a new set $K_{k,s}(t)$ indexed by $k\in\{0,1,\dots,M_\text{SH}-1\}$. We finally take this orthonormalized set to be the pump supermodes, i.e., $B_{k,s}(t)=K_{k,s}(t)$. We refer to the set of supermodes $A_{m,s}(t)$ and $B_{m,s}(t)$ constructed via this prescription as the \emph{GIF-principal supermodes}.

\begin{figure}[bt]
    \centering
    \includegraphics[width=0.48\textwidth]{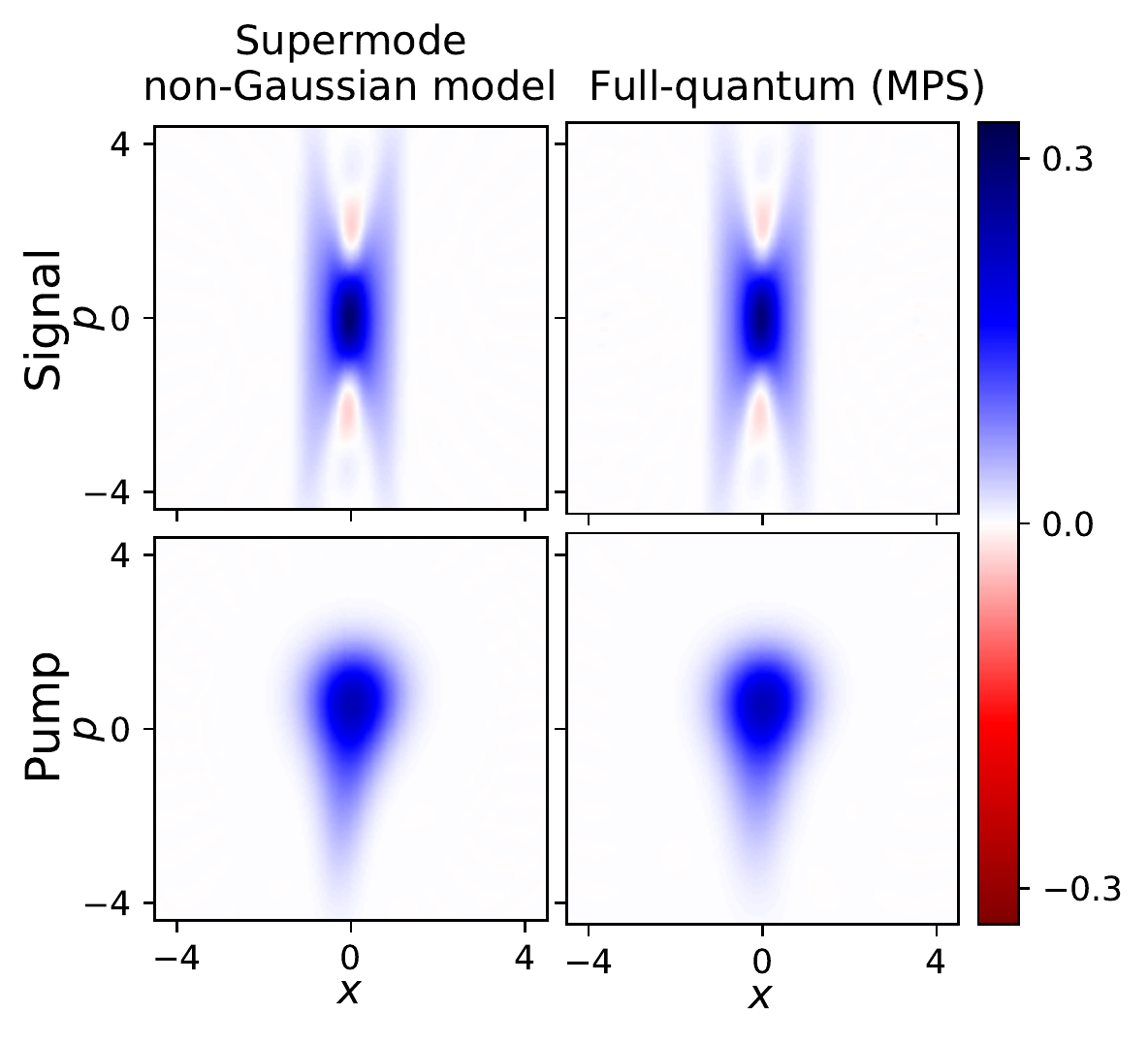}
    \caption{Wigner functions of the GIF-principal supermodes in the interaction frame at time $t=0.8$ calculated using the supermode non-Gaussian model with $M_\text{FH}=2$ (left column) and a full-quantum MPS simulation (right column). Top and bottom rows show the most major signal and pump supermodes, respectively, where the most major pump supermode is $B_{0,s}(t)\propto K_{(0,0),s}(t)$. The waveguide paramters and initial condition is the same as that considered in Fig.~\ref{fig:gif}. See Appendix~\ref{sec:supermodes} for methods used to calculate the phase-space portraits of quantum states encoded in the MPS representation.}
    \label{fig:wigner}
\end{figure}

Using the GIF-principal supermodes, we can perform efficient numerical simulation of non-Gaussian dynamics in the GIF-principal supermode subspace $\mathcal{S}(t)$. As shown in Fig.~\ref{fig:gif}, this non-Gaussian (GIF-principal) supermode model exhibits better agreement with the full-quantum (MPS) simulation than any Gaussian approximation, including the GIF approximation, indicating that physics beyond Gaussian quantum optics need to be taken into account. The emergence of non-Gaussian features can be easily seen in the Wigner function in the GIF: As shown in Fig.~\ref{fig:wigner}, the most major signal GIF-principal supermode exhibits a visible amount of Wigner-function negativity in the GIF, a genuine quantum non-Gaussian feature that cannot be captured by any semiclassical model~\cite{Kenfack2004,Walschaers2021}. Furthermore, the Wigner function of the pump takes a droplet-like non-Gaussian shape, highlighting the breakdown of the conventional assumption that the pump remains in a coherent state. The good agreement between $\ket{\widetilde\varphi_\text{I}(t)}$ given by a GIF-principal supermode simulation with $\ket{\varphi_\text{I}(t)}$ given by a full-quantum (MPS) simulation shows the validity of the supermode reduction to describe non-Gaussian effects in pulsed squeezing deep in the few-pump photon regime.

\section{Non-Gaussian quantum physics in mesoscopic pulsed squeezing}
\label{sec:mesoscopic}
Having formulated a tractable quantum model using GIF-principal supermodes, we now proceed to study how non-Gaussian quantum dynamics can affect the generation of pulsed squeezed light in the mesoscopic regime~\cite{Stanojevic2009,Weber2015,Imry1997,Harder2016}, where the number of photons involved is neither very large (which falls under an undepleted-pump approximation) nor very small (which results in highly quantum behavior qualitatively different from squeezing~\cite{Yanagimoto2020_fano,Leung2009,Antonosyan2014}). Experimentally, this would correspond to situations in which considerable (anti-)squeezing (e.g., ${}\sim\SI{15}{dB}$) can be generated with only dozens to hundreds of pump photons. Such an operating regime is of primary interest from a technological perspective because it is squarely on the near-term horizon of ongoing efforts to increase power efficiency in nonlinear nanophotonics~\cite{Ferrera2008,Liu2018,McKenna2021,Jankowski2021-pdc}. In addition, a mesoscopic system where non-Gaussian quantum states exist alongside significant mean-field/Gaussian excitations can provide an excellent platform for quantum photonics as multi-photon interactions can effectively enhance material nonlinearities~\cite{Yanagimoto2020,Leroux2018,Qin2018,Ramelow2019-ring}.

The analysis of pulsed squeezed light generation conventionally proceeds by considering the \emph{squeezing supermodes} of the system. In the lab frame, they are defined by supermode annihilation operators $\hat{f}_m(t)=\int\mathrm{d}s\, W_{m,s}(t)\hat{\phi}_s$~\cite{Wasilewski2006,Lvovsky2007,Christ2013}. These squeezing supermodes are directly connected to the GIF-principal supermodes (defined in the GIF) via
\begin{align}
\label{eq:squeezing-supermode}
    \hat{U}^\dagger\hat{f}_m\hat{U}=\cosh{\lambda_m}\hat{a}_m+\sinh{\lambda_m}\hat{a}^\dagger_m.
\end{align}
Using \eqref{eq:squeezing-supermode}, we can evaluate lab-frame squeezing by calculating the quadrature variances
\begin{subequations}
\label{eq:quadratures}
\begin{align}
    \expectationvalue{{X}^2(\hat{f}_m)}{\varphi}&=e^{+2\lambda_m}\expectationvalue{{X}^2(\hat{a}_m)}{\varphi_\mathrm{I}}\\
    \expectationvalue{{P}^2(\hat{f}_m)}{\varphi}&=e^{-2\lambda_m}\expectationvalue{{P}^2(\hat{a}_m)}{\varphi_\mathrm{I}},
\end{align}
\end{subequations}
where $X(\hat{c})=(\hat{c}+\hat{c}^\dagger)/\sqrt{2}$ and $P(\hat{c})=(\hat{c}-\hat{c}^\dagger)/\sqrt{2}\mathrm{i}$. Notice that if non-Gaussian contributions can be ignored, i.e., $\ket{\varphi_\text{I}}=\ket{0}$, each squeezing supermode is populated with a pure squeezed-vacuum state. On the other hand, finite excitation in the GIF, i.e., $\ket{\varphi_\text{I}}\ne\ket{0}$, can result in nontrivial, \emph{non-Gaussian} corrections to the observed lab-frame supermode squeezing. Notice that the relationship \eqref{eq:squeezing-supermode} indicates that, physically, the signal waveforms that predominantly accumulate non-Gaussian quantum features actually coincide with the conventional squeezing supermodes.

\begin{figure}[bth]
    \centering
    \includegraphics[width=0.48\textwidth]{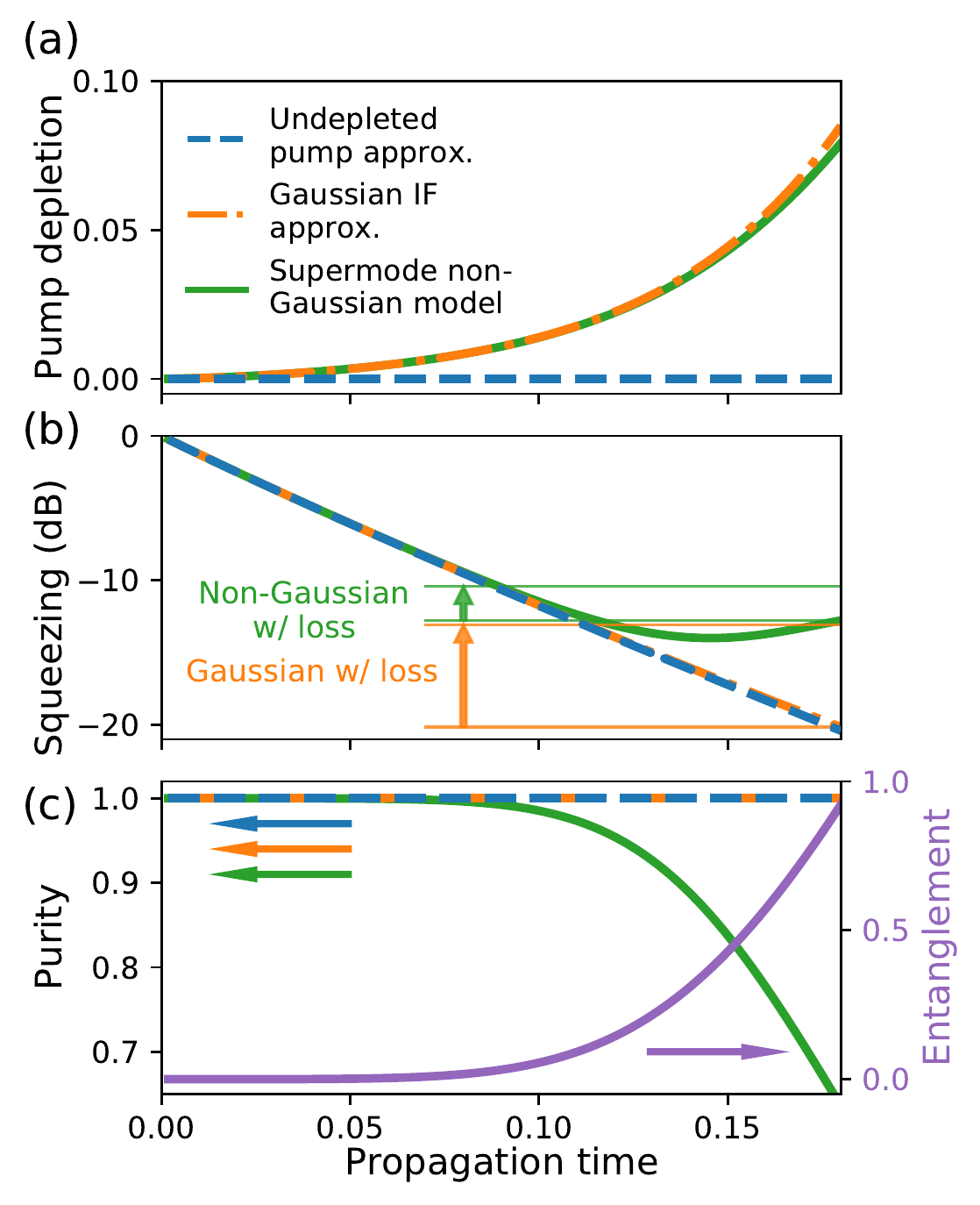}
    \caption{Squeezing dynamics under various models as a function of propagation time $t$. (a) Pump depletion ratio $R(t)$; (b) Squeezing of the quadrature variance $\expectationvalue{{P}^2(\hat{f}_0)}{\varphi}$ for the most major squeezing supermode $\hat f_0$; (c) purity in $\hat f_0$ and its correlation with the entanglement entropy between signal and pump (calculated with all modes). Of the models considered here, the undepleted-pump approximation and the GIF approximation are Gaussian models, while the supermode non-Gaussian (with $M_\text{FH}=2$) can account for non-Gaussian features. In this simulation, we use an initial Gaussian-shaped pump \eqref{eq:sh-init} with initial photon number $N_\text{SH}(0)=200$ propagating in a waveguide with dispersion parameters $d_0=d_1=0$ and $d_2=1.0$. In (b), we also show how squeezing levels degrade under assuming $\SI{4}{\percent}$ discrete loss occurring at the end of the waveguide, for the Gaussian IF approximation (orange) and the supermode non-Gaussian model (green).}
    \label{fig:mesoscopic}
\end{figure}

\begin{figure}[bth]
    \centering
    \includegraphics[width=0.5\textwidth]{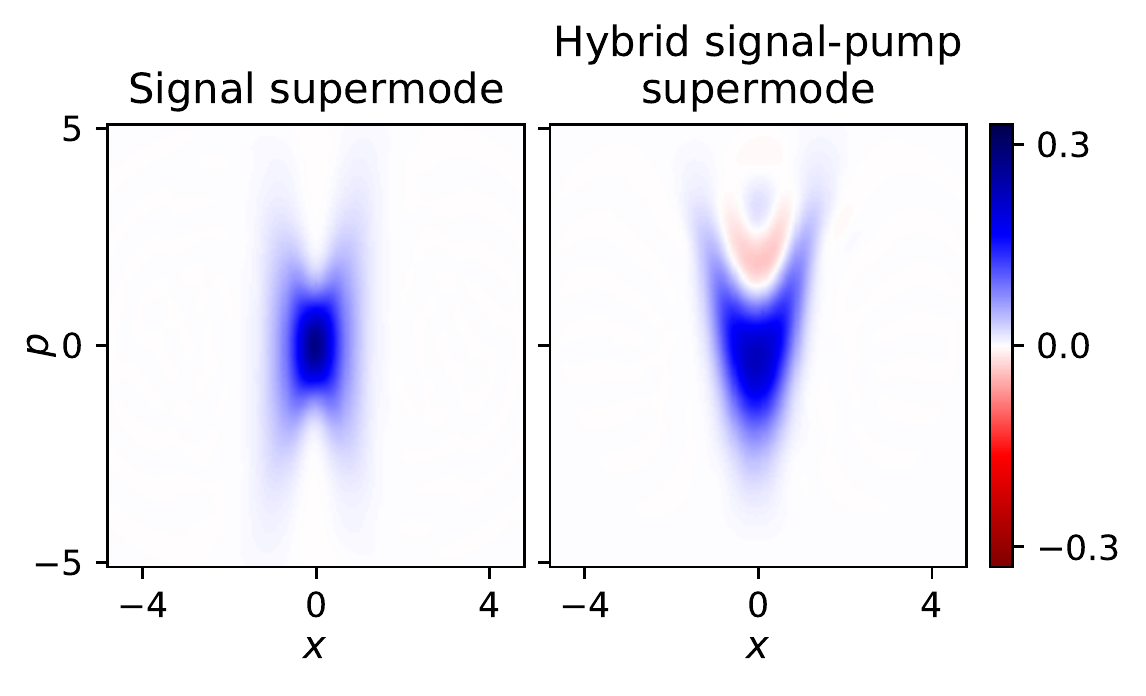}
    \caption{Wigner function of the most major GIF-principal signal supermode $\hat a_0$ (left) and that of a hybrid signal-pump GIF supermode (right), both at $t=0.18$. The hybrid supermode chosen corresponds to $\cos\phi\,\hat{a}_0+e^{\mathrm{i}\theta}\sin\phi \,\hat{b}_0$ with $\phi=0.19\pi$ and $\theta=0.02\pi$, where $\hat{b}_0$ is the most major pump supermode $B_{0,s}(t)\propto K_{(0,0),s}(t)$. We use the same waveguide parameters and initial conditions as in Fig.~\ref{fig:mesoscopic}.}
    \label{fig:mesoscopic_wigner}
\end{figure}

In Fig.~\ref{fig:mesoscopic}, we numerically show how, for squeezed light generation in the mesoscopic regime, pump depletion can generate quantum non-Gaussianity and ultimately lead to significant corrections in both squeezing levels and state purity. We consider a Gaussian-shaped pump \eqref{eq:sh-init} with a mesoscopic excitation of $N_\text{SH}(0)=200$ and perform numerical quantum simulations using GIF-principal supermodes according to \eqref{eq:dynamical-ng}, comparing the results against Gaussian predictions by both the undepleted-pump approximation and the GIF approximation. 

As shown in Fig.~\ref{fig:mesoscopic}(a), the system exhibits pump depletion as large as \SI{10}{\percent}, extracting \num{20} photons from the pump and potentially generating \num{40} photons worth of signal vacuum squeezing. Under a multimode Gaussian model, we would naively expect $> \SI{20}{dB}$ of squeezing in the most major squeezing supermode, as shown in Fig.~\ref{fig:mesoscopic}(b). However, when we include non-Gaussian corrections, the squeezing saturates at a much lower level as a result of excess quadrature noise contributed by non-Gaussian excitations in the GIF according to \eqref{eq:quadratures}. It is particularly notable that the non-Gaussian corruption has comparable or even larger impact than linear loss, which conventionally sets the primary limit on squeezed-light generation in most experiments: As shown in Fig.~\ref{fig:mesoscopic}(b), the addition of $\SI{4}{\percent}$ of linear loss to the GIF (i.e., Gaussian) approximation results in a degraded squeezing level similar to that of the \emph{lossless} non-Gaussian model, indicating that non-Gaussian degradation dominates for system losses below $\approx\SI{4}{\percent}$. Furthermore, when linear loss is added to the non-Gaussian model, we note they simply add up.

As shown in Fig.~\ref{fig:mesoscopic}(c), we also see the purity of the most major squeezing supermode (after partial-tracing out the pump) decays rapidly with the onset of pump depletion and non-Gaussianity. It is worth emphasizing that the loss in purity primarily stems from entanglement between signal and pump in the GIF-principal supermodes rather than trivial correlations between different signal supermodes. To see this more clearly, we superimpose the entanglement entropy~\cite{Nielsen2000} between signal and pump in Fig.~\ref{fig:mesoscopic}(c), where we observe a strong correlation between the decrease in the purity and the increase of the signal-pump entanglement.

Compared to the fewer-photon operation studied in Figs.~\ref{fig:gif} and \ref{fig:wigner}, significant pump depletion and its accompanying non-Gaussian features appear earlier, which reflects greater enhancement of the nonlinear rate by the larger peak photon density.

It is worth noting that the degradations in squeezing and purity are inherently non-Gaussian in origin and cannot be mitigated by additional Gaussian operations. Specifically, we have observed that no other supermode exhibits squeezing significantly better than the most major squeezing supermode used here. These observations underscore the importance of properly accounting for non-Gaussian quantum physics as we design and engineer nonlinear photonic devices in the mesoscopic regime.

At the same time, however, the appearance of physics beyond the Gaussian limit suggests significant potential of pulsed mesoscopic optics as a novel platform for quantum engineering and information processing, e.g., as an all-optical, high-bandwidth source of quantum non-Gaussian light~\cite{Lvovsky2020,Walschaers2021}. To highlight this emergence of genuine non-classical features in the quantum state, we can examine in more detail phase-space portraits of the dominant supermodes in the GIF. In Fig.~\ref{fig:mesoscopic_wigner}, we show the Wigner function of the most major GIF-principal signal supermode and of a hybridized signal-pump GIF-principal supermode. The latter inclusion of pump via hybridization reveals the presence of significant Wigner-function negativity, indicating that the non-Gaussian features of the state consist of strong non-classical correlations between signal and pump, which is a prevailing feature in strongly nonlinear $\chi^{(2)}$ interactions such as the quantum propagation of simultons~\cite{Yanagimoto2021_mps} (i.e., quadratic solitons).

\section{Experimental prospects}
In this section, we discuss experimental parameter regimes of $\chi^{(2)}$ nonlinear waveguides where non-Gaussian quantum physics are expected to play critical roles. For this purpose, we relate the normalized propagation time to experimentally relevant parameters; A unit normalized time of $t=1$ corresponds to a physical propagation distance on a waveguide of~\cite{Yanagimoto2021_mps}
\begin{align}
    L_{\chi^{(2)}}=\sqrt[\leftroot{-1}\uproot{2}\scriptstyle 3]{\frac{|k_\text{FH}''|}{(\hbar\omega_\text{SH}\eta)^2}},
\end{align}
where the pump frequency $\omega_\text{SH}/2\pi=c/\lambda_\text{SH}$ is related to the pump carrier wavelength $\lambda_\text{SH}$, and $k_\text{FH}''$ is the group velocity dispersion of the signal. Experimental measure of the magnitude of $\chi^{(2)}$ nonlinearity is given via the normalized small-signal conversion efficiency of the second harmonic generation $\eta$ with the units of $[\mathrm{power}^{-1}\cdot\mathrm{length}^{-2}]$. 

We note that $L_{\chi^{(2)}}$ is to be interpreted as a single-photon nonlinear length over which a single pump photon downconverts to a pair of signal photons. However, beyond the few-photon regime, non-Gaussian features could emerge over a much shorter propagation length $L_\text{eff}$, since multi-photon (i.e., peak-intensity) effects can enhance the effective nonlinear dynamical rate. Here, we phenomenologically denote $L_\text{eff}=rL_{\chi^{(2)}}$, where $r$ is a dynamics-dependent factor representing this enhancement. For the mesoscopic pump pulse studied in Sec.~\ref{sec:mesoscopic}, we observe non-classical features at a distance roughly an order of magnitude shorter than $L_{\chi^{(2)}}$, or with $r\approx 0.18$. When the characteristic (i.e., $\SI{3}{dB}$) power attenuation length $L_\text{loss}$ of a waveguide is much longer than $L_\text{eff}$, the coherence of the quantum nonlinear dynamics of interest are expected to survive.

For a given material, scaling the waveguide geometries with respect to the operating wavelength results in $\eta\sim\lambda_{\text{SH}}^{-4}$ and $L_{\chi^{(2)}}\sim\lambda_{\text{SH}}^{10/3}$, whose strong scaling with $\lambda_{\text{SH}}$ motivates operations in the regime of shorter wavelength to enhance nonlinearity~\cite{Jankowski2021-review,Yanagimoto2020_fano}. Towards this end, a recent experiment at $\lambda_\text{SH}=\SI{456.5}{nm}$ on PPLN nanophotonics has demonstrated conversion efficiency as high as $\eta=\SI{330}{W^{-1}cm^{-2}}~(\SI{33000}{\%W^{-1}cm^{-2}})$~\cite{Park2021}, which, assuming an advanced dispersion engineering $|k_\text{FH}''|=\SI{1}{fs^2mm^{-1}}$, leads to $L_\text{eff}=\SI{1.4}{cm}$. Notably, $L_\text{loss}=\SI{1}{m}$, which corresponds to propagation loss of $\sim\SI{1}{\percent}$ over a distance $L_\text{eff}$ with $r=0.18$, has been achieved on the same material platform~\cite{Zhang2017}. These numbers suggest that coherent non-Gaussian quantum physics studied in this work could become relevant for engineering and designing highly-nonlinear nanophotonics in the near future.

\section{Conclusion}
\label{sec:conclusion}
In this research, we study how non-Gaussian quantum effects can arise in the generation of pulsed squeezed light. To handle the complicated multimode quantum dynamics involved in this analysis, we present a novel modeling approach where we first identify a semiclassical approximation to the squeezing dynamics and then utilize that information to isolate the non-Gaussian quantum dynamics through the construction of the GIF. Compared to a conventional undepleted-pump approximation, the semiclassical dynamics captured by the GIF include nonlinear effects such as pump depletion caused by strong optical nonlinearity.

By predicting the nominal structure of the multimode squeezing, the GIF crucially allows us to identify a small set of principal supermodes that predominantly participate in the non-Gaussian part of the dynamics. After truncating the quantum model in the GIF to a few-mode subspace spanned by these GIF-principal supermodes, we can perform efficient quantum simulations of the non-Gaussian dynamics and hence obtain corrections to the nominal Gaussian description.

Using this framework, we study pulsed squeezing in the mesoscopic regime involving dozens to hundreds of pump photons. In this regime of operation, the pump has a mean field sufficiently large to drive strong squeezing, while at the same time it also experiences enough depletion for the quantum-mechanical (i.e., discrete) nature of photons to emerge as non-Gaussian features such as Wigner-function negativity. An interesting consequence---with implications for applications like CV quantum computation and Gaussian boson sampling---is that non-Gaussian effects can ``degrade'' the quality of the squeezed light, e.g., by inducing excess quadrature noise and state impurity due to signal-pump entanglement, which cannot be mitigated by additional Gaussian operations. 

On the other hand, these observations also herald a unique opportunity to harness coherent non-Gaussianity for a wide range of all-optical, high-bandwidth quantum applications. Notably, the recent state of the art in nonlinear nanophotonics suggests experimental access to such non-Gaussian physics is on the horizon. We expect our results to guide experimental efforts towards this important milestone, while simultaneously advancing our theoretical capabilities for understanding and manipulating the rich, multimode non-Gaussian physics inherent to broadband quantum optics.

\section*{Acknowledgments}
R.\,Y., E.\,N., and H.\,M.\ developed the project into its final form starting from its initial conception by all authors. R.\,Y., E.\,N., A.\,Y., and T.\,O.\ developed the numerical techniques. R.\,Y.\ performed the simulations and generated the figures. E.\,N.\ and H.\,M.\ advised and directed the project. R.\,Y.\ and E.\,N.\ wrote the paper with detailed input and feedback from all authors.

This work has been supported by the Army Research Office under Grant No. W911NF-16-1-0086, and the National Science Foundation under awards CCF-1918549 and PHY-2011363. The authors wish to thank NTT Research for their financial and technical support. R.\,Y.\ is supported by a Stanford Q-FARM Ph.D.\ Fellowship and the Masason Foundation. A.\,Y.\ is supported by the Masason Foundation. The authors thank Ryan Hamerly and Jatadhari Mishra for helpful discussions.

\begin{appendix}
    \section{ASYMPTOTIC SCALING OF PUMP DEPLETION}
\label{sec:asymptotic}
In this section, we use the equations of motion provided by the GIF approximation to derive analytic expressions for the asymptotic scaling as $t\rightarrow0$ of the pump depletion ratio $R(t)$, defined in \eqref{eq:depletionratio}. We start with the single-mode example and then generalize to the full multimode model of squeezing.

According to \eqref{eq:single-mode-greens-eom} and \eqref{eq:single-mode-feom}, the single-mode GIF is governed by equations
\begin{align}
    \partial_t \beta(t)&=-\mathrm{i}\delta \beta(t)-\frac{\mathrm{i}}{2}C(t)S(t)\nonumber\\
    \partial_t C(t)&=-\mathrm{i}\beta(t)S^*(t)\\
    \partial_t S(t)&=-\mathrm{i}\beta(t)C^*(t)\nonumber.
\end{align}
By applying Picard iteration~\cite{Ricardo2020} twice for the initial conditions $C(0)=1$ and $S(0)=0$, we obtain 
\begin{align}
    \beta^{(2)}(t)= \beta(0)-\mathrm{i}\delta \beta(0) t-\left(\frac{\delta^2}{2}+\frac{1}{4}\right)\beta(0)t^2
\end{align}
leading to 
\begin{align}
    R(t)\approx 1-\frac{|\beta^{(2)}(t)|^2}{|\beta(0)|^2}= \frac{t^2}{2}
\end{align}
as $t\rightarrow 0$. Here, parenthetical superscripts denote the order of the Picard iterate.

We next consider multimode pulsed squeezing, focusing in particular on quasi-degenerate operation in which $d_0\approx 0$. According to the equations of motion for the GIF approximation \eqref{eq:greens-eom} and \eqref{eq:feom-depleted}, we have
\begin{align}
\partial_t \bar{\beta}_s(t)&=-\frac{\mathrm{i}}{2}\int\mathrm{d}p\,\mathrm{d}q\, e^{\mathrm{i}(\delta_s-\gamma_p-\gamma_{s-p})t}\bar{C}_{p,q}(t)\bar{S}_{s-p,q}(t)\nonumber\\
\partial_t \bar{C}_{s,p}(t)&=-\mathrm{i}\int \mathrm{d}q\, e^{\mathrm{i}(\gamma_s+\gamma_q-\delta_{s+q})t}\bar{\beta}_{s+q}(t)\bar{S}_{q,p}^*(t)\\
\partial_t \bar{S}_{s,p}(t)&=-\mathrm{i}\int \mathrm{d}q\, e^{\mathrm{i}(\gamma_s+\gamma_q-\delta_{s+q})t} \bar{\beta}_{s+q}(t)\bar{C}_{q,p}^*(t)\nonumber,
\end{align}
where we have introduced $\beta_s(t)=e^{-\mathrm{i}\delta_st}\bar{\beta}_s(t)$, $C_{s,p}(t)=e^{-\mathrm{i}\gamma_st}\bar{C}_{s,p}(t)$, and $S_{s,p}(t)=e^{-\mathrm{i}\gamma_st}\bar{S}_{s,p}(t)$. Starting from initial conditions $\beta_s(0)=\beta$, $C_{s,p}(0)=\delta(s-p)$, and $S_{s,p}(0)=0$, the first-order Picard iterates are
\begin{align}
\bar{\beta}^{(1)}_s(t)&=\bar{\beta}_s(0)\nonumber\\
\bar{C}^{(1)}_{s,p}(t)&=\delta(s-p)\\
\bar{S}^{(1)}_{s,p}(t)&=-\mathrm{i}t\exp(\frac{-\mathrm{i}\Delta_{s,p}t}{2})\,\mathrm{sinc}\left(\frac{\Delta_{s,p}t}{2}\right)\bar{\beta}_{s+p}(0)\nonumber,
\end{align}
with $\Delta_{s,p}=\delta_{s+p}-\gamma_s-\gamma_p$. The second-order Picard iterate for $\bar{\beta}_s(t)$ is
\begin{align}
    \label{eq:app-f2}
    &\bar{\beta}^{(2)}_s(t)-\bar{\beta}_s(0)\\
    &=-\frac{\mathrm{i}}{2}\int^t_0\mathrm{d}t'\int\mathrm{d}p\,\mathrm{d}q\, e^{\mathrm{i}\Delta_{s-p,p}t'}\bar{C}^{(1)}_{p,q}(t')\bar{S}^{(1)}_{s-p,q}(t')\nonumber\\
    &=-\frac{\bar{\beta}_{s}(0)}{2}\int^t_0\mathrm{d}t'\int\mathrm{d}p\, t'\exp(\frac{\mathrm{i}\Delta_{s-p,p}t'}{2})\mathrm{sinc}\left(\frac{\Delta_{s-p,p}t'}{2}\right)\nonumber
\end{align}
While \eqref{eq:app-f2} requires numerical evaluation in general, we can obtain an analytic solution when the pump bandwidth is relatively narrow compared to the signal bandwidth; in this case, only excitations near the carrier (i.e., $s\approx 0$) contribute to the final result. If we therefore ignore the $s$ dependence in the integral by writing $\Delta_{s-p,p}\approx \Delta_{-p,p}$,
\begin{align}
    \bar{\beta}_s^{(2)}(t)&\approx\beta_s(0)\left(1-\frac{1-\mathrm{i}}{3\sqrt{2\pi}}t^{3/2}\right),
\end{align}
where we have also employed $d_0 \approx 0$. As a result, the asymptotic scaling as $t\rightarrow 0$ for the multimode pump depletion ratio is
\begin{align}
    R(t)\approx 1-\frac{1}{N_\text{SH}(0)}\int\mathrm{d}s\bigl|\bar{\beta}_s^{(2)}(t)\bigr|^2\approx \frac{2}{3\sqrt{2\pi}}t^{3/2}.
\end{align}

\section{PHASE-SPACE PORTRAITS OF SUPERMODES}
\label{sec:supermodes}
In this section, we derive the relationship between the phase-space distributions of the GIF-principal supermodes (in the GIF) and the squeezing supermodes (in the lab frame) for the signal. This is useful for comparing phase-space portraits produced by simulations in the non-Gaussian supermode model with those produced by MPS simulations. 

By virtue of \eqref{eq:squeezing-supermode}, we have
\begin{align}
\label{eq:trans}
    X(\hat{a}_m)=r_m^2\hat{U}^\dagger X(\hat{f}_m)\hat{U},
\end{align}
where $r_m=e^{-\lambda_m}$, and $X(\hat{c})=(\hat{c}+\hat{c}^\dagger)/\sqrt{2}$ is a quadrature operator for a given mode annihilation operator $\hat{c}$. Denoting an eigenstate of $X(\hat{c})$ with eigenvalue $x$ by $\ket{X(\hat{c})=x}$, we have
\begin{align}
\label{eq:xeigen}
\bigl\vert X(\hat{a}_m)=x\bigr\rangle=r_m\hat{U}^\dagger\bigl\vert X(\hat{f}_m)=r_m^2x\bigr\rangle,
\end{align}
which is what allows us to relate the phase-space representation of the GIF-principal supermode $\hat{a}_m$ to that of the squeezing supermode $\hat{f}_m$.

The Wigner quasi-probability distribution of the GIF-principal supermode $\hat{a}_m$ is~\cite{Case2008}
\begin{align}
\label{eq:wig-gif}
    &W_{\text{GIF},m}(x,p)=\frac{1}{2\pi}\int\mathrm{d}y\,e^{-\mathrm{i}py}\\
    &\times\mathrm{Tr}\Bigl\{\bigl\langle X(\hat{a}_m)=x+y/2\big\vert\varphi_\text{I}\bigr\rangle\bigl\langle\varphi_\text{I}\big\vert X(\hat{a}_m)=x-y/2\bigr\rangle\Bigr\},\nonumber
\end{align}
where the trace naturally excludes contributions by all other supermodes. By substituting \eqref{eq:xeigen} in \eqref{eq:wig-gif}, we have
\begin{widetext}
\begin{align}
\label{eq:wigner}
    W_{\text{GIF},m}(x,p)&=\frac{1}{2\pi}\int\mathrm{d}y\,r_m^2e^{-\mathrm{i}py}\mathrm{Tr}\left\{\bigl\langle X(\hat{f}_m)=r_m^2(x+y/2)\big\vert\varphi\bigr\rangle\bigl\langle\varphi\big\vert X(\hat{f}_m)=r_m^2(x-y/2)\bigr\rangle\right\}\nonumber\\
    &=\frac{1}{2\pi}\int\mathrm{d}y'\,e^{-\mathrm{i}p'y'}\mathrm{Tr}\left\{\bigl\langle X(\hat{f}_m)=x'+y'/2\big\vert\varphi\bigr\rangle\bigl\langle\varphi\big\vert X(\hat{f}_m)=x'-y'/2\bigr\rangle\right\}\nonumber\\
    &=W_{\text{lab},m}(x',p'),
\end{align}
\end{widetext}
where $x'=r_m^2x$ and $p'=r_m^{-2}p$. As a result, the Wigner function of the squeezing supermode $W_{\text{lab},m}(x',p')$ is directly related to $W_{\text{GIF},m}(x,p)$ by a simple scaling of phase-space coordinates.

In our MPS simulations, we obtain an MPS representation of the lab-frame state $\ket{\varphi}$. To calculate phase-space portraits in the GIF from an MPS, we can first calculate $W_{\text{lab},m}$ using the demultiplexing algorithm in Ref.~\cite{Yanagimoto2021_mps}. Then, simple scaling of coordinates via \eqref{eq:wigner} produces the corresponding phase-space portrait in the GIF.

\end{appendix}

\bibliography{myfile}

\begin{thebibliography}{82}%
\makeatletter
\providecommand \@ifxundefined [1]{%
 \@ifx{#1\undefined}
}%
\providecommand \@ifnum [1]{%
 \ifnum #1\expandafter \@firstoftwo
 \else \expandafter \@secondoftwo
 \fi
}%
\providecommand \@ifx [1]{%
 \ifx #1\expandafter \@firstoftwo
 \else \expandafter \@secondoftwo
 \fi
}%
\providecommand \natexlab [1]{#1}%
\providecommand \enquote  [1]{``#1''}%
\providecommand \bibnamefont  [1]{#1}%
\providecommand \bibfnamefont [1]{#1}%
\providecommand \citenamefont [1]{#1}%
\providecommand \href@noop [0]{\@secondoftwo}%
\providecommand \href [0]{\begingroup \@sanitize@url \@href}%
\providecommand \@href[1]{\@@startlink{#1}\@@href}%
\providecommand \@@href[1]{\endgroup#1\@@endlink}%
\providecommand \@sanitize@url [0]{\catcode `\\12\catcode `\$12\catcode
  `\&12\catcode `\#12\catcode `\^12\catcode `\_12\catcode `\%12\relax}%
\providecommand \@@startlink[1]{}%
\providecommand \@@endlink[0]{}%
\providecommand \url  [0]{\begingroup\@sanitize@url \@url }%
\providecommand \@url [1]{\endgroup\@href {#1}{\urlprefix }}%
\providecommand \urlprefix  [0]{URL }%
\providecommand \Eprint [0]{\href }%
\providecommand \doibase [0]{https://doi.org/}%
\providecommand \selectlanguage [0]{\@gobble}%
\providecommand \bibinfo  [0]{\@secondoftwo}%
\providecommand \bibfield  [0]{\@secondoftwo}%
\providecommand \translation [1]{[#1]}%
\providecommand \BibitemOpen [0]{}%
\providecommand \bibitemStop [0]{}%
\providecommand \bibitemNoStop [0]{.\EOS\space}%
\providecommand \EOS [0]{\spacefactor3000\relax}%
\providecommand \BibitemShut  [1]{\csname bibitem#1\endcsname}%
\let\auto@bib@innerbib\@empty
\bibitem [{\citenamefont {Nielsen}\ and\ \citenamefont
  {Chuang}(2000)}]{Nielsen2000}%
  \BibitemOpen
  \bibfield  {author} {\bibinfo {author} {\bibfnamefont {M.~A.}\ \bibnamefont
  {Nielsen}}\ and\ \bibinfo {author} {\bibfnamefont {I.~L.}\ \bibnamefont
  {Chuang}},\ }\href@noop {} {\emph {\bibinfo {title} {Quantum Computation and
  Quantum Information}}}\ (\bibinfo  {publisher} {Cambridge University Press},\
  \bibinfo {year} {2000})\BibitemShut {NoStop}%
\bibitem [{\citenamefont {Gisin}\ and\ \citenamefont {Thew}(2007)}]{Gisin2007}%
  \BibitemOpen
  \bibfield  {author} {\bibinfo {author} {\bibfnamefont {N.}~\bibnamefont
  {Gisin}}\ and\ \bibinfo {author} {\bibfnamefont {R.}~\bibnamefont {Thew}},\
  }\bibfield  {title} {\bibinfo {title} {{Quantum communication}},\ }\href@noop
  {} {\bibfield  {journal} {\bibinfo  {journal} {Nat. Photon.}\ }\textbf
  {\bibinfo {volume} {1}},\ \bibinfo {pages} {165} (\bibinfo {year}
  {2007})}\BibitemShut {NoStop}%
\bibitem [{\citenamefont {Giovannetti}\ \emph {et~al.}(2011)\citenamefont
  {Giovannetti}, \citenamefont {Lloyd},\ and\ \citenamefont
  {Maccone}}]{Giovannetti2011}%
  \BibitemOpen
  \bibfield  {author} {\bibinfo {author} {\bibfnamefont {V.}~\bibnamefont
  {Giovannetti}}, \bibinfo {author} {\bibfnamefont {S.}~\bibnamefont {Lloyd}},\
  and\ \bibinfo {author} {\bibfnamefont {L.}~\bibnamefont {Maccone}},\
  }\bibfield  {title} {\bibinfo {title} {{Advances in quantum metrology}},\
  }\href@noop {} {\bibfield  {journal} {\bibinfo  {journal} {Nat. Photon.}\
  }\textbf {\bibinfo {volume} {5}},\ \bibinfo {pages} {222} (\bibinfo {year}
  {2011})}\BibitemShut {NoStop}%
\bibitem [{\citenamefont {Walls}(1983)}]{Walls1983}%
  \BibitemOpen
  \bibfield  {author} {\bibinfo {author} {\bibfnamefont {D.~F.}\ \bibnamefont
  {Walls}},\ }\bibfield  {title} {\bibinfo {title} {{Squeezed states of
  light}},\ }\href@noop {} {\bibfield  {journal} {\bibinfo  {journal} {Nature}\
  }\textbf {\bibinfo {volume} {306}},\ \bibinfo {pages} {141} (\bibinfo {year}
  {1983})}\BibitemShut {NoStop}%
\bibitem [{\citenamefont {Andersen}\ \emph {et~al.}(2016)\citenamefont
  {Andersen}, \citenamefont {Gehring}, \citenamefont {Marquardt},\ and\
  \citenamefont {Leuchs}}]{Andersen2016}%
  \BibitemOpen
  \bibfield  {author} {\bibinfo {author} {\bibfnamefont {U.~L.}\ \bibnamefont
  {Andersen}}, \bibinfo {author} {\bibfnamefont {T.}~\bibnamefont {Gehring}},
  \bibinfo {author} {\bibfnamefont {C.}~\bibnamefont {Marquardt}},\ and\
  \bibinfo {author} {\bibfnamefont {G.}~\bibnamefont {Leuchs}},\ }\bibfield
  {title} {\bibinfo {title} {{30 years of squeezed light generation}},\
  }\href@noop {} {\bibfield  {journal} {\bibinfo  {journal} {Phys. Scr.}\
  }\textbf {\bibinfo {volume} {91}},\ \bibinfo {pages} {053001} (\bibinfo
  {year} {2016})}\BibitemShut {NoStop}%
\bibitem [{\citenamefont {Furusawa}\ \emph {et~al.}(1998)\citenamefont
  {Furusawa}, \citenamefont {S\o{}rensen}, \citenamefont {Braunstein},
  \citenamefont {Fuchs}, \citenamefont {Kimble},\ and\ \citenamefont
  {Polzik}}]{Furusawa1998}%
  \BibitemOpen
  \bibfield  {author} {\bibinfo {author} {\bibfnamefont {A.}~\bibnamefont
  {Furusawa}}, \bibinfo {author} {\bibfnamefont {J.~L.}\ \bibnamefont
  {S\o{}rensen}}, \bibinfo {author} {\bibfnamefont {S.~L.}\ \bibnamefont
  {Braunstein}}, \bibinfo {author} {\bibfnamefont {C.~A.}\ \bibnamefont
  {Fuchs}}, \bibinfo {author} {\bibfnamefont {H.~J.}\ \bibnamefont {Kimble}},\
  and\ \bibinfo {author} {\bibfnamefont {E.~S.}\ \bibnamefont {Polzik}},\
  }\bibfield  {title} {\bibinfo {title} {{Unconditional Quantum
  Teleportation}},\ }\href@noop {} {\bibfield  {journal} {\bibinfo  {journal}
  {Science}\ }\textbf {\bibinfo {volume} {282}},\ \bibinfo {pages} {706}
  (\bibinfo {year} {1998})}\BibitemShut {NoStop}%
\bibitem [{\citenamefont {Ou}\ \emph {et~al.}(1992)\citenamefont {Ou},
  \citenamefont {Pereira}, \citenamefont {Kimble},\ and\ \citenamefont
  {Peng}}]{Ou1992}%
  \BibitemOpen
  \bibfield  {author} {\bibinfo {author} {\bibfnamefont {Z.~Y.}\ \bibnamefont
  {Ou}}, \bibinfo {author} {\bibfnamefont {S.~F.}\ \bibnamefont {Pereira}},
  \bibinfo {author} {\bibfnamefont {H.~J.}\ \bibnamefont {Kimble}},\ and\
  \bibinfo {author} {\bibfnamefont {K.~C.}\ \bibnamefont {Peng}},\ }\bibfield
  {title} {\bibinfo {title} {{Realization of the Einstein-Podolsky-Rosen
  paradox for continuous variables}},\ }\href@noop {} {\bibfield  {journal}
  {\bibinfo  {journal} {Phys. Rev. Lett.}\ }\textbf {\bibinfo {volume} {68}},\
  \bibinfo {pages} {3663} (\bibinfo {year} {1992})}\BibitemShut {NoStop}%
\bibitem [{\citenamefont {Menicucci}\ \emph {et~al.}(2006)\citenamefont
  {Menicucci}, \citenamefont {{van Loock}}, \citenamefont {Gu}, \citenamefont
  {Weedbrook}, \citenamefont {Ralph},\ and\ \citenamefont
  {Nielsen}}]{Menicucci2006}%
  \BibitemOpen
  \bibfield  {author} {\bibinfo {author} {\bibfnamefont {N.~C.}\ \bibnamefont
  {Menicucci}}, \bibinfo {author} {\bibfnamefont {P.}~\bibnamefont {{van
  Loock}}}, \bibinfo {author} {\bibfnamefont {M.}~\bibnamefont {Gu}}, \bibinfo
  {author} {\bibfnamefont {C.}~\bibnamefont {Weedbrook}}, \bibinfo {author}
  {\bibfnamefont {T.~C.}\ \bibnamefont {Ralph}},\ and\ \bibinfo {author}
  {\bibfnamefont {M.~A.}\ \bibnamefont {Nielsen}},\ }\bibfield  {title}
  {\bibinfo {title} {{Universal Quantum Computation with Continuous-Variable
  Cluster States}},\ }\href@noop {} {\bibfield  {journal} {\bibinfo  {journal}
  {Phys. Rev. Lett.}\ }\textbf {\bibinfo {volume} {97}},\ \bibinfo {pages}
  {110501} (\bibinfo {year} {2006})}\BibitemShut {NoStop}%
\bibitem [{\citenamefont {Asavanant}\ \emph {et~al.}(2019)\citenamefont
  {Asavanant}, \citenamefont {Shiozawa}, \citenamefont {Yokoyama},
  \citenamefont {Charoensombutamon}, \citenamefont {Emura}, \citenamefont
  {Alexander}, \citenamefont {Takeda}, \citenamefont {Yoshikawa}, \citenamefont
  {Menicucci}, \citenamefont {Yonezawa},\ and\ \citenamefont
  {Furusawa}}]{Asavarant2019}%
  \BibitemOpen
  \bibfield  {author} {\bibinfo {author} {\bibfnamefont {W.}~\bibnamefont
  {Asavanant}}, \bibinfo {author} {\bibfnamefont {Y.}~\bibnamefont {Shiozawa}},
  \bibinfo {author} {\bibfnamefont {S.}~\bibnamefont {Yokoyama}}, \bibinfo
  {author} {\bibfnamefont {B.}~\bibnamefont {Charoensombutamon}}, \bibinfo
  {author} {\bibfnamefont {H.}~\bibnamefont {Emura}}, \bibinfo {author}
  {\bibfnamefont {R.~N.}\ \bibnamefont {Alexander}}, \bibinfo {author}
  {\bibfnamefont {S.}~\bibnamefont {Takeda}}, \bibinfo {author} {\bibfnamefont
  {J.}~\bibnamefont {Yoshikawa}}, \bibinfo {author} {\bibfnamefont {N.~C.}\
  \bibnamefont {Menicucci}}, \bibinfo {author} {\bibfnamefont {H.}~\bibnamefont
  {Yonezawa}},\ and\ \bibinfo {author} {\bibfnamefont {A.}~\bibnamefont
  {Furusawa}},\ }\bibfield  {title} {\bibinfo {title} {{Generation of
  time-domain-multiplexed two-dimensional cluster state}},\ }\href@noop {}
  {\bibfield  {journal} {\bibinfo  {journal} {Science}\ }\textbf {\bibinfo
  {volume} {366}},\ \bibinfo {pages} {373} (\bibinfo {year}
  {2019})}\BibitemShut {NoStop}%
\bibitem [{\citenamefont {Zhong}\ \emph {et~al.}(2020)\citenamefont {Zhong},
  \citenamefont {Wang}, \citenamefont {Deng}, \citenamefont {Chen},
  \citenamefont {Peng}, \citenamefont {Luo}, \citenamefont {Qin}, \citenamefont
  {Wu}, \citenamefont {Ding}, \citenamefont {Hu}, \citenamefont {Hu},
  \citenamefont {Yang}, \citenamefont {Zhang}, \citenamefont {Li},
  \citenamefont {Li}, \citenamefont {Jiang}, \citenamefont {Gan}, \citenamefont
  {Yang}, \citenamefont {You}, \citenamefont {Wang}, \citenamefont {Li},
  \citenamefont {Liu}, \citenamefont {Lu},\ and\ \citenamefont
  {Pan}}]{Zhong2020}%
  \BibitemOpen
  \bibfield  {author} {\bibinfo {author} {\bibfnamefont {H.-S.}\ \bibnamefont
  {Zhong}}, \bibinfo {author} {\bibfnamefont {H.}~\bibnamefont {Wang}},
  \bibinfo {author} {\bibfnamefont {Y.-H.}\ \bibnamefont {Deng}}, \bibinfo
  {author} {\bibfnamefont {M.-C.}\ \bibnamefont {Chen}}, \bibinfo {author}
  {\bibfnamefont {L.-C.}\ \bibnamefont {Peng}}, \bibinfo {author}
  {\bibfnamefont {Y.-H.}\ \bibnamefont {Luo}}, \bibinfo {author} {\bibfnamefont
  {J.}~\bibnamefont {Qin}}, \bibinfo {author} {\bibfnamefont {D.}~\bibnamefont
  {Wu}}, \bibinfo {author} {\bibfnamefont {X.}~\bibnamefont {Ding}}, \bibinfo
  {author} {\bibfnamefont {Y.}~\bibnamefont {Hu}}, \bibinfo {author}
  {\bibfnamefont {P.}~\bibnamefont {Hu}}, \bibinfo {author} {\bibfnamefont
  {X.-Y.}\ \bibnamefont {Yang}}, \bibinfo {author} {\bibfnamefont {W.-J.}\
  \bibnamefont {Zhang}}, \bibinfo {author} {\bibfnamefont {H.}~\bibnamefont
  {Li}}, \bibinfo {author} {\bibfnamefont {Y.}~\bibnamefont {Li}}, \bibinfo
  {author} {\bibfnamefont {X.}~\bibnamefont {Jiang}}, \bibinfo {author}
  {\bibfnamefont {L.}~\bibnamefont {Gan}}, \bibinfo {author} {\bibfnamefont
  {G.}~\bibnamefont {Yang}}, \bibinfo {author} {\bibfnamefont {L.}~\bibnamefont
  {You}}, \bibinfo {author} {\bibfnamefont {Z.}~\bibnamefont {Wang}}, \bibinfo
  {author} {\bibfnamefont {L.}~\bibnamefont {Li}}, \bibinfo {author}
  {\bibfnamefont {N.-L.}\ \bibnamefont {Liu}}, \bibinfo {author} {\bibfnamefont
  {C.-Y.}\ \bibnamefont {Lu}},\ and\ \bibinfo {author} {\bibfnamefont {J.-W.}\
  \bibnamefont {Pan}},\ }\bibfield  {title} {\bibinfo {title} {{Quantum
  computational advantage using photons }},\ }\href@noop {} {\bibfield
  {journal} {\bibinfo  {journal} {Science}\ }\textbf {\bibinfo {volume}
  {370}},\ \bibinfo {pages} {1460} (\bibinfo {year} {2020})}\BibitemShut
  {NoStop}%
\bibitem [{\citenamefont {Arrazola}\ \emph {et~al.}(2021)\citenamefont
  {Arrazola}, \citenamefont {Bergholm}, \citenamefont {Br\'adler},
  \citenamefont {Bromley}, \citenamefont {Collins}, \citenamefont {Dhand},
  \citenamefont {Fumagalli}, \citenamefont {Gerrits}, \citenamefont {Goussev},
  \citenamefont {Helt}, \citenamefont {Hundal}, \citenamefont {Isacsson},
  \citenamefont {Israel}, \citenamefont {Izaac}, \citenamefont {Jahangiri},
  \citenamefont {Janik}, \citenamefont {Killoran}, \citenamefont {Kumar},
  \citenamefont {Lavoie}, \citenamefont {Lita}, \citenamefont {Mahler},
  \citenamefont {Menotti}, \citenamefont {Morrison}, \citenamefont {Nam},
  \citenamefont {Neuhaus}, \citenamefont {Qi}, \citenamefont {Quesada},
  \citenamefont {Repingon}, \citenamefont {Sabapathy}, \citenamefont {Schuld},
  \citenamefont {Su}, \citenamefont {Swinarton}, \citenamefont {Sz\'ava},
  \citenamefont {Tan}, \citenamefont {Tan}, \citenamefont {Vaidya},
  \citenamefont {Vernon}, \citenamefont {Zabaneh},\ and\ \citenamefont
  {Zhang}}]{Arrazola2021}%
  \BibitemOpen
  \bibfield  {author} {\bibinfo {author} {\bibfnamefont {J.~M.}\ \bibnamefont
  {Arrazola}}, \bibinfo {author} {\bibfnamefont {V.}~\bibnamefont {Bergholm}},
  \bibinfo {author} {\bibfnamefont {K.}~\bibnamefont {Br\'adler}}, \bibinfo
  {author} {\bibfnamefont {T.~R.}\ \bibnamefont {Bromley}}, \bibinfo {author}
  {\bibfnamefont {M.~J.}\ \bibnamefont {Collins}}, \bibinfo {author}
  {\bibfnamefont {I.}~\bibnamefont {Dhand}}, \bibinfo {author} {\bibfnamefont
  {A.}~\bibnamefont {Fumagalli}}, \bibinfo {author} {\bibfnamefont
  {T.}~\bibnamefont {Gerrits}}, \bibinfo {author} {\bibfnamefont
  {A.}~\bibnamefont {Goussev}}, \bibinfo {author} {\bibfnamefont {L.~G.}\
  \bibnamefont {Helt}}, \bibinfo {author} {\bibfnamefont {J.}~\bibnamefont
  {Hundal}}, \bibinfo {author} {\bibfnamefont {T.}~\bibnamefont {Isacsson}},
  \bibinfo {author} {\bibfnamefont {R.~B.}\ \bibnamefont {Israel}}, \bibinfo
  {author} {\bibfnamefont {J.}~\bibnamefont {Izaac}}, \bibinfo {author}
  {\bibfnamefont {S.}~\bibnamefont {Jahangiri}}, \bibinfo {author}
  {\bibfnamefont {R.}~\bibnamefont {Janik}}, \bibinfo {author} {\bibfnamefont
  {N.}~\bibnamefont {Killoran}}, \bibinfo {author} {\bibfnamefont {S.~P.}\
  \bibnamefont {Kumar}}, \bibinfo {author} {\bibfnamefont {J.}~\bibnamefont
  {Lavoie}}, \bibinfo {author} {\bibfnamefont {A.~E.}\ \bibnamefont {Lita}},
  \bibinfo {author} {\bibfnamefont {D.~H.}\ \bibnamefont {Mahler}}, \bibinfo
  {author} {\bibfnamefont {M.}~\bibnamefont {Menotti}}, \bibinfo {author}
  {\bibfnamefont {B.}~\bibnamefont {Morrison}}, \bibinfo {author}
  {\bibfnamefont {S.~W.}\ \bibnamefont {Nam}}, \bibinfo {author} {\bibfnamefont
  {L.}~\bibnamefont {Neuhaus}}, \bibinfo {author} {\bibfnamefont {H.~Y.}\
  \bibnamefont {Qi}}, \bibinfo {author} {\bibfnamefont {N.}~\bibnamefont
  {Quesada}}, \bibinfo {author} {\bibfnamefont {A.}~\bibnamefont {Repingon}},
  \bibinfo {author} {\bibfnamefont {K.~K.}\ \bibnamefont {Sabapathy}}, \bibinfo
  {author} {\bibfnamefont {M.}~\bibnamefont {Schuld}}, \bibinfo {author}
  {\bibfnamefont {D.}~\bibnamefont {Su}}, \bibinfo {author} {\bibfnamefont
  {J.}~\bibnamefont {Swinarton}}, \bibinfo {author} {\bibfnamefont
  {A.}~\bibnamefont {Sz\'ava}}, \bibinfo {author} {\bibfnamefont
  {K.}~\bibnamefont {Tan}}, \bibinfo {author} {\bibfnamefont {P.}~\bibnamefont
  {Tan}}, \bibinfo {author} {\bibfnamefont {V.~D.}\ \bibnamefont {Vaidya}},
  \bibinfo {author} {\bibfnamefont {Z.}~\bibnamefont {Vernon}}, \bibinfo
  {author} {\bibfnamefont {Z.}~\bibnamefont {Zabaneh}},\ and\ \bibinfo {author}
  {\bibfnamefont {Y.}~\bibnamefont {Zhang}},\ }\bibfield  {title} {\bibinfo
  {title} {{Quantum circuits with many photons on a programmable nanophotonic
  chip}},\ }\href@noop {} {\bibfield  {journal} {\bibinfo  {journal} {Nature}\
  }\textbf {\bibinfo {volume} {591}},\ \bibinfo {pages} {54} (\bibinfo {year}
  {2021})}\BibitemShut {NoStop}%
\bibitem [{\citenamefont {Pezz\'e}\ and\ \citenamefont
  {Smerzi}(2008)}]{Pezze2008}%
  \BibitemOpen
  \bibfield  {author} {\bibinfo {author} {\bibfnamefont {L.}~\bibnamefont
  {Pezz\'e}}\ and\ \bibinfo {author} {\bibfnamefont {A.}~\bibnamefont
  {Smerzi}},\ }\bibfield  {title} {\bibinfo {title} {{Mach-Zehnder
  Interferometry at the Heisenberg Limit with Coherent and Squeezed-Vacuum
  Light}},\ }\href@noop {} {\bibfield  {journal} {\bibinfo  {journal} {Phys.
  Rev. Lett.}\ }\textbf {\bibinfo {volume} {100}},\ \bibinfo {pages} {073601}
  (\bibinfo {year} {2008})}\BibitemShut {NoStop}%
\bibitem [{\citenamefont {{The LIGO Scientific
  Collaboration}}(2013)}]{LIGO2013}%
  \BibitemOpen
  \bibfield  {author} {\bibinfo {author} {\bibnamefont {{The LIGO Scientific
  Collaboration}}},\ }\bibfield  {title} {\bibinfo {title} {{Enhanced
  sensitivity of the LIGO gravitational wave detector by using squeezed states
  of light}},\ }\href@noop {} {\bibfield  {journal} {\bibinfo  {journal} {Nat.
  Photon.}\ }\textbf {\bibinfo {volume} {7}},\ \bibinfo {pages} {613} (\bibinfo
  {year} {2013})}\BibitemShut {NoStop}%
\bibitem [{\citenamefont {Vahlbruch}\ \emph {et~al.}(2007)\citenamefont
  {Vahlbruch}, \citenamefont {Chelkowski}, \citenamefont {Danzmann},\ and\
  \citenamefont {Schnabel}}]{Vahlbruch2007}%
  \BibitemOpen
  \bibfield  {author} {\bibinfo {author} {\bibfnamefont {H.}~\bibnamefont
  {Vahlbruch}}, \bibinfo {author} {\bibfnamefont {S.}~\bibnamefont
  {Chelkowski}}, \bibinfo {author} {\bibfnamefont {K.}~\bibnamefont
  {Danzmann}},\ and\ \bibinfo {author} {\bibfnamefont {R.}~\bibnamefont
  {Schnabel}},\ }\bibfield  {title} {\bibinfo {title} {{Quantum engineering of
  squeezed states for quantum communication and metrology}},\ }\href@noop {}
  {\bibfield  {journal} {\bibinfo  {journal} {New J. Phys.}\ }\textbf {\bibinfo
  {volume} {9}},\ \bibinfo {pages} {371} (\bibinfo {year} {2007})}\BibitemShut
  {NoStop}%
\bibitem [{\citenamefont {Yuen}\ and\ \citenamefont
  {Shapiro}(1978)}]{Yuen1987}%
  \BibitemOpen
  \bibfield  {author} {\bibinfo {author} {\bibfnamefont {H.~P.}\ \bibnamefont
  {Yuen}}\ and\ \bibinfo {author} {\bibfnamefont {J.~H.}\ \bibnamefont
  {Shapiro}},\ }\bibfield  {title} {\bibinfo {title} {{Optical communication
  with two-photon coherent states--Part I: Quantum-state propagation and
  quantum-noise}},\ }\href@noop {} {\bibfield  {journal} {\bibinfo  {journal}
  {IEEE Trans. Inf. Theory}\ }\textbf {\bibinfo {volume} {24}},\ \bibinfo
  {pages} {657} (\bibinfo {year} {1978})}\BibitemShut {NoStop}%
\bibitem [{\citenamefont {Hillery}(2000)}]{Hillery2000}%
  \BibitemOpen
  \bibfield  {author} {\bibinfo {author} {\bibfnamefont {M.}~\bibnamefont
  {Hillery}},\ }\bibfield  {title} {\bibinfo {title} {{Quantum cryptography
  with squeezed states}},\ }\href@noop {} {\bibfield  {journal} {\bibinfo
  {journal} {Phys. Rev. A}\ }\textbf {\bibinfo {volume} {61}},\ \bibinfo
  {pages} {022309} (\bibinfo {year} {2000})}\BibitemShut {NoStop}%
\bibitem [{\citenamefont {Slusher}\ \emph {et~al.}(1987)\citenamefont
  {Slusher}, \citenamefont {Grangier}, \citenamefont {LaPorta}, \citenamefont
  {Yurke},\ and\ \citenamefont {Potasek}}]{Slusher1987}%
  \BibitemOpen
  \bibfield  {author} {\bibinfo {author} {\bibfnamefont {R.~E.}\ \bibnamefont
  {Slusher}}, \bibinfo {author} {\bibfnamefont {P.}~\bibnamefont {Grangier}},
  \bibinfo {author} {\bibfnamefont {A.}~\bibnamefont {LaPorta}}, \bibinfo
  {author} {\bibfnamefont {B.}~\bibnamefont {Yurke}},\ and\ \bibinfo {author}
  {\bibfnamefont {M.~J.}\ \bibnamefont {Potasek}},\ }\bibfield  {title}
  {\bibinfo {title} {{Pulsed Squeezed Light}},\ }\href@noop {} {\bibfield
  {journal} {\bibinfo  {journal} {Phys. Rev. Lett.}\ }\textbf {\bibinfo
  {volume} {59}},\ \bibinfo {pages} {2566} (\bibinfo {year}
  {1987})}\BibitemShut {NoStop}%
\bibitem [{\citenamefont {Anderson}\ \emph {et~al.}(1995)\citenamefont
  {Anderson}, \citenamefont {Beck}, \citenamefont {Raymer},\ and\ \citenamefont
  {Bierlein}}]{Anderson1995}%
  \BibitemOpen
  \bibfield  {author} {\bibinfo {author} {\bibfnamefont {M.~E.}\ \bibnamefont
  {Anderson}}, \bibinfo {author} {\bibfnamefont {M.}~\bibnamefont {Beck}},
  \bibinfo {author} {\bibfnamefont {M.~G.}\ \bibnamefont {Raymer}},\ and\
  \bibinfo {author} {\bibfnamefont {J.~D.}\ \bibnamefont {Bierlein}},\
  }\bibfield  {title} {\bibinfo {title} {{Quadrature squeezing with ultrashort
  pulses in nonlinear-optical waveguides}},\ }\href@noop {} {\bibfield
  {journal} {\bibinfo  {journal} {Opt. Lett.}\ }\textbf {\bibinfo {volume}
  {20}},\ \bibinfo {pages} {620} (\bibinfo {year} {1995})}\BibitemShut
  {NoStop}%
\bibitem [{\citenamefont {Triginer}\ \emph {et~al.}(2020)\citenamefont
  {Triginer}, \citenamefont {Vidrighin}, \citenamefont {Quesada}, \citenamefont
  {Eckstein}, \citenamefont {Moore}, \citenamefont {{Steven Kolthammer}},
  \citenamefont {Sipe},\ and\ \citenamefont {Walmsley}}]{Triginer2020}%
  \BibitemOpen
  \bibfield  {author} {\bibinfo {author} {\bibfnamefont {G.}~\bibnamefont
  {Triginer}}, \bibinfo {author} {\bibfnamefont {M.~D.}\ \bibnamefont
  {Vidrighin}}, \bibinfo {author} {\bibfnamefont {N.}~\bibnamefont {Quesada}},
  \bibinfo {author} {\bibfnamefont {A.}~\bibnamefont {Eckstein}}, \bibinfo
  {author} {\bibfnamefont {M.}~\bibnamefont {Moore}}, \bibinfo {author}
  {\bibfnamefont {W.}~\bibnamefont {{Steven Kolthammer}}}, \bibinfo {author}
  {\bibfnamefont {J.~E.}\ \bibnamefont {Sipe}},\ and\ \bibinfo {author}
  {\bibfnamefont {I.~A.}\ \bibnamefont {Walmsley}},\ }\bibfield  {title}
  {\bibinfo {title} {{Understanding High-Gain Twin-Beam Sources Using Cascaded
  Stimulated Emission}},\ }\href@noop {} {\bibfield  {journal} {\bibinfo
  {journal} {Phys. Rev. X}\ }\textbf {\bibinfo {volume} {10}},\ \bibinfo
  {pages} {031063} (\bibinfo {year} {2020})}\BibitemShut {NoStop}%
\bibitem [{\citenamefont {Eckstein}\ \emph {et~al.}(2011)\citenamefont
  {Eckstein}, \citenamefont {Christ}, \citenamefont {Mosley},\ and\
  \citenamefont {Silberhorn}}]{Eckstein2011}%
  \BibitemOpen
  \bibfield  {author} {\bibinfo {author} {\bibfnamefont {A.}~\bibnamefont
  {Eckstein}}, \bibinfo {author} {\bibfnamefont {A.}~\bibnamefont {Christ}},
  \bibinfo {author} {\bibfnamefont {P.~J.}\ \bibnamefont {Mosley}},\ and\
  \bibinfo {author} {\bibfnamefont {C.}~\bibnamefont {Silberhorn}},\ }\bibfield
   {title} {\bibinfo {title} {{Highly Efficient Single-Pass Source of Pulsed
  Single-Mode Twin Beams of Light}},\ }\href@noop {} {\bibfield  {journal}
  {\bibinfo  {journal} {Phys. Rev. Lett.}\ }\textbf {\bibinfo {volume} {106}},\
  \bibinfo {pages} {013603} (\bibinfo {year} {2011})}\BibitemShut {NoStop}%
\bibitem [{\citenamefont {Fl\'orez}\ \emph {et~al.}(2020)\citenamefont
  {Fl\'orez}, \citenamefont {Lundeen},\ and\ \citenamefont
  {Chekhova}}]{Florez2020}%
  \BibitemOpen
  \bibfield  {author} {\bibinfo {author} {\bibfnamefont {J.}~\bibnamefont
  {Fl\'orez}}, \bibinfo {author} {\bibfnamefont {J.~S.}\ \bibnamefont
  {Lundeen}},\ and\ \bibinfo {author} {\bibfnamefont {M.~V.}\ \bibnamefont
  {Chekhova}},\ }\bibfield  {title} {\bibinfo {title} {{Pump depletion in
  parametric down-conversion with low pump energies}},\ }\href@noop {}
  {\bibfield  {journal} {\bibinfo  {journal} {Opt. Lett.}\ }\textbf {\bibinfo
  {volume} {45}},\ \bibinfo {pages} {4264} (\bibinfo {year}
  {2020})}\BibitemShut {NoStop}%
\bibitem [{\citenamefont {Yanagimoto}\ \emph
  {et~al.}(2020{\natexlab{a}})\citenamefont {Yanagimoto}, \citenamefont
  {Onodera}, \citenamefont {Ng}, \citenamefont {Wright}, \citenamefont
  {McMahon},\ and\ \citenamefont {Mabuchi}}]{Yanagimoto2020}%
  \BibitemOpen
  \bibfield  {author} {\bibinfo {author} {\bibfnamefont {R.}~\bibnamefont
  {Yanagimoto}}, \bibinfo {author} {\bibfnamefont {T.}~\bibnamefont {Onodera}},
  \bibinfo {author} {\bibfnamefont {E.}~\bibnamefont {Ng}}, \bibinfo {author}
  {\bibfnamefont {L.~G.}\ \bibnamefont {Wright}}, \bibinfo {author}
  {\bibfnamefont {P.~L.}\ \bibnamefont {McMahon}},\ and\ \bibinfo {author}
  {\bibfnamefont {H.}~\bibnamefont {Mabuchi}},\ }\bibfield  {title} {\bibinfo
  {title} {{Engineering a Kerr-Based Deterministic Cubic Phase Gate via
  Gaussian Operations}},\ }\href@noop {} {\bibfield  {journal} {\bibinfo
  {journal} {Phys. Rev. Lett.}\ }\textbf {\bibinfo {volume} {124}},\ \bibinfo
  {pages} {240503} (\bibinfo {year} {2020}{\natexlab{a}})}\BibitemShut
  {NoStop}%
\bibitem [{\citenamefont {Brecht}\ \emph {et~al.}(2015)\citenamefont {Brecht},
  \citenamefont {Reddy}, \citenamefont {Silberhorn},\ and\ \citenamefont
  {Raymer}}]{Brecht2015}%
  \BibitemOpen
  \bibfield  {author} {\bibinfo {author} {\bibfnamefont {B.}~\bibnamefont
  {Brecht}}, \bibinfo {author} {\bibfnamefont {D.~V.}\ \bibnamefont {Reddy}},
  \bibinfo {author} {\bibfnamefont {C.}~\bibnamefont {Silberhorn}},\ and\
  \bibinfo {author} {\bibfnamefont {M.~G.}\ \bibnamefont {Raymer}},\ }\bibfield
   {title} {\bibinfo {title} {{Photon Temporal Modes: A Complete Framework for
  Quantum Information Science}},\ }\href@noop {} {\bibfield  {journal}
  {\bibinfo  {journal} {Phys. Rev. X}\ }\textbf {\bibinfo {volume} {5}},\
  \bibinfo {pages} {041017} (\bibinfo {year} {2015})}\BibitemShut {NoStop}%
\bibitem [{\citenamefont {Grice}\ and\ \citenamefont
  {Walmsley}(1997)}]{Grice1997}%
  \BibitemOpen
  \bibfield  {author} {\bibinfo {author} {\bibfnamefont {W.~P.}\ \bibnamefont
  {Grice}}\ and\ \bibinfo {author} {\bibfnamefont {I.~A.}\ \bibnamefont
  {Walmsley}},\ }\bibfield  {title} {\bibinfo {title} {{Spectral information
  and distinguishability in type-II down-conversion with a broadband pump}},\
  }\href@noop {} {\bibfield  {journal} {\bibinfo  {journal} {Phys. Rev. A}\
  }\textbf {\bibinfo {volume} {56}},\ \bibinfo {pages} {1627} (\bibinfo {year}
  {1997})}\BibitemShut {NoStop}%
\bibitem [{\citenamefont {Ansari}\ \emph {et~al.}(2018)\citenamefont {Ansari},
  \citenamefont {Donohue}, \citenamefont {Brecht},\ and\ \citenamefont
  {Silberhorn}}]{Ansari2018}%
  \BibitemOpen
  \bibfield  {author} {\bibinfo {author} {\bibfnamefont {V.}~\bibnamefont
  {Ansari}}, \bibinfo {author} {\bibfnamefont {J.~M.}\ \bibnamefont {Donohue}},
  \bibinfo {author} {\bibfnamefont {B.}~\bibnamefont {Brecht}},\ and\ \bibinfo
  {author} {\bibfnamefont {C.}~\bibnamefont {Silberhorn}},\ }\bibfield  {title}
  {\bibinfo {title} {{Tailoring nonlinear processes for quantum optics with
  pulsed temporal-mode encodings}},\ }\href@noop {} {\bibfield  {journal}
  {\bibinfo  {journal} {Optica}\ }\textbf {\bibinfo {volume} {5}},\ \bibinfo
  {pages} {534} (\bibinfo {year} {2018})}\BibitemShut {NoStop}%
\bibitem [{\citenamefont {Jankowski}\ \emph {et~al.}(2020)\citenamefont
  {Jankowski}, \citenamefont {Langrock}, \citenamefont {Desiatov},
  \citenamefont {Marandi}, \citenamefont {Wang}, \citenamefont {Zhang},
  \citenamefont {Phillips}, \citenamefont {Lon{\u c}ar},\ and\ \citenamefont
  {Fejer}}]{Jankowski2020}%
  \BibitemOpen
  \bibfield  {author} {\bibinfo {author} {\bibfnamefont {M.}~\bibnamefont
  {Jankowski}}, \bibinfo {author} {\bibfnamefont {C.}~\bibnamefont {Langrock}},
  \bibinfo {author} {\bibfnamefont {B.}~\bibnamefont {Desiatov}}, \bibinfo
  {author} {\bibfnamefont {A.}~\bibnamefont {Marandi}}, \bibinfo {author}
  {\bibfnamefont {C.}~\bibnamefont {Wang}}, \bibinfo {author} {\bibfnamefont
  {M.}~\bibnamefont {Zhang}}, \bibinfo {author} {\bibfnamefont {C.~R.}\
  \bibnamefont {Phillips}}, \bibinfo {author} {\bibfnamefont {M.}~\bibnamefont
  {Lon{\u c}ar}},\ and\ \bibinfo {author} {\bibfnamefont {M.~M.}\ \bibnamefont
  {Fejer}},\ }\bibfield  {title} {\bibinfo {title} {{Ultrabroadband nonlinear
  optics in nanophotonic periodically poled lithium niobate waveguides}},\
  }\href@noop {} {\bibfield  {journal} {\bibinfo  {journal} {Optica}\ }\textbf
  {\bibinfo {volume} {7}},\ \bibinfo {pages} {40} (\bibinfo {year}
  {2020})}\BibitemShut {NoStop}%
\bibitem [{\citenamefont {Jankowski}\ \emph
  {et~al.}(2021{\natexlab{a}})\citenamefont {Jankowski}, \citenamefont
  {Mishra},\ and\ \citenamefont {Fejer}}]{Jankowski2021-review}%
  \BibitemOpen
  \bibfield  {author} {\bibinfo {author} {\bibfnamefont {M.}~\bibnamefont
  {Jankowski}}, \bibinfo {author} {\bibfnamefont {J.}~\bibnamefont {Mishra}},\
  and\ \bibinfo {author} {\bibfnamefont {M.~M.}\ \bibnamefont {Fejer}},\
  }\bibfield  {title} {\bibinfo {title} {{Dispersion-engineered $\chi^{(2)}$
  nanophotonics: a flexible tool for nonclassical light}},\ }\href@noop {}
  {\bibfield  {journal} {\bibinfo  {journal} {J. Phys. Photon.}\ }\textbf
  {\bibinfo {volume} {3}},\ \bibinfo {pages} {042005} (\bibinfo {year}
  {2021}{\natexlab{a}})}\BibitemShut {NoStop}%
\bibitem [{\citenamefont {Yanagimoto}\ \emph
  {et~al.}(2021{\natexlab{a}})\citenamefont {Yanagimoto}, \citenamefont {Ng},
  \citenamefont {Wright}, \citenamefont {Onodera},\ and\ \citenamefont
  {Mabuchi}}]{Yanagimoto2021_mps}%
  \BibitemOpen
  \bibfield  {author} {\bibinfo {author} {\bibfnamefont {R.}~\bibnamefont
  {Yanagimoto}}, \bibinfo {author} {\bibfnamefont {E.}~\bibnamefont {Ng}},
  \bibinfo {author} {\bibfnamefont {L.~G.}\ \bibnamefont {Wright}}, \bibinfo
  {author} {\bibfnamefont {T.}~\bibnamefont {Onodera}},\ and\ \bibinfo {author}
  {\bibfnamefont {H.}~\bibnamefont {Mabuchi}},\ }\bibfield  {title} {\bibinfo
  {title} {{Efficient simulation of ultrafast quantum nonlinear optics with
  matrix product states}},\ }\href@noop {} {\bibfield  {journal} {\bibinfo
  {journal} {Optica}\ }\textbf {\bibinfo {volume} {8}},\ \bibinfo {pages}
  {1306} (\bibinfo {year} {2021}{\natexlab{a}})}\BibitemShut {NoStop}%
\bibitem [{\citenamefont {Leung}\ \emph {et~al.}(2009)\citenamefont {Leung},
  \citenamefont {Munro}, \citenamefont {Nemoto},\ and\ \citenamefont
  {Ralph}}]{Leung2009}%
  \BibitemOpen
  \bibfield  {author} {\bibinfo {author} {\bibfnamefont {P.~M.}\ \bibnamefont
  {Leung}}, \bibinfo {author} {\bibfnamefont {W.~J.}\ \bibnamefont {Munro}},
  \bibinfo {author} {\bibfnamefont {K.}~\bibnamefont {Nemoto}},\ and\ \bibinfo
  {author} {\bibfnamefont {T.~C.}\ \bibnamefont {Ralph}},\ }\bibfield  {title}
  {\bibinfo {title} {{Spectral effects of strong $\chi^{(2)}$ nonlinearity for
  quantum processing}},\ }\href@noop {} {\bibfield  {journal} {\bibinfo
  {journal} {Phys. Rev. A}\ }\textbf {\bibinfo {volume} {79}},\ \bibinfo
  {pages} {042307} (\bibinfo {year} {2009})}\BibitemShut {NoStop}%
\bibitem [{\citenamefont {Hafezi}\ \emph {et~al.}(2012)\citenamefont {Hafezi},
  \citenamefont {Chang}, \citenamefont {Gritsev}, \citenamefont {Demler},\ and\
  \citenamefont {Lukin}}]{Hafezi2012}%
  \BibitemOpen
  \bibfield  {author} {\bibinfo {author} {\bibfnamefont {M.}~\bibnamefont
  {Hafezi}}, \bibinfo {author} {\bibfnamefont {D.~E.}\ \bibnamefont {Chang}},
  \bibinfo {author} {\bibfnamefont {V.}~\bibnamefont {Gritsev}}, \bibinfo
  {author} {\bibfnamefont {E.}~\bibnamefont {Demler}},\ and\ \bibinfo {author}
  {\bibfnamefont {M.~D.}\ \bibnamefont {Lukin}},\ }\bibfield  {title} {\bibinfo
  {title} {{Quantum transport of strongly interacting photons in a
  one-dimensional nonlinear waveguide}},\ }\href@noop {} {\bibfield  {journal}
  {\bibinfo  {journal} {Phys. Rev. A}\ }\textbf {\bibinfo {volume} {85}},\
  \bibinfo {pages} {013822} (\bibinfo {year} {2012})}\BibitemShut {NoStop}%
\bibitem [{\citenamefont {Drummond}\ and\ \citenamefont
  {He}(1997)}]{Drummond1997}%
  \BibitemOpen
  \bibfield  {author} {\bibinfo {author} {\bibfnamefont {P.~D.}\ \bibnamefont
  {Drummond}}\ and\ \bibinfo {author} {\bibfnamefont {H.}~\bibnamefont {He}},\
  }\bibfield  {title} {\bibinfo {title} {{Optical mesons}},\ }\href@noop {}
  {\bibfield  {journal} {\bibinfo  {journal} {Phys. Rev. A}\ }\textbf {\bibinfo
  {volume} {56}},\ \bibinfo {pages} {R1107(R)} (\bibinfo {year}
  {1997})}\BibitemShut {NoStop}%
\bibitem [{\citenamefont {Kenfack}\ and\ \citenamefont
  {{\.Z}yczkowski}(2004)}]{Kenfack2004}%
  \BibitemOpen
  \bibfield  {author} {\bibinfo {author} {\bibfnamefont {A.}~\bibnamefont
  {Kenfack}}\ and\ \bibinfo {author} {\bibfnamefont {K.}~\bibnamefont
  {{\.Z}yczkowski}},\ }\bibfield  {title} {\bibinfo {title} {{Negativity of the
  Wigner function as an indicator of non-classicality}},\ }\href@noop {}
  {\bibfield  {journal} {\bibinfo  {journal} {J. Opt. B: Quantum Semiclass.
  Opt.}\ }\textbf {\bibinfo {volume} {6}},\ \bibinfo {pages} {396} (\bibinfo
  {year} {2004})}\BibitemShut {NoStop}%
\bibitem [{\citenamefont {Walschaers}(2021)}]{Walschaers2021}%
  \BibitemOpen
  \bibfield  {author} {\bibinfo {author} {\bibfnamefont {M.}~\bibnamefont
  {Walschaers}},\ }\bibfield  {title} {\bibinfo {title} {{Non-Gaussian Quantum
  States and Where to Find Them}},\ }\href@noop {} {\bibfield  {journal}
  {\bibinfo  {journal} {{PRX Quantum}}\ }\textbf {\bibinfo {volume} {2}},\
  \bibinfo {pages} {030204} (\bibinfo {year} {2021})}\BibitemShut {NoStop}%
\bibitem [{\citenamefont {Lvovsky}\ \emph {et~al.}(2020)\citenamefont
  {Lvovsky}, \citenamefont {Grangier}, \citenamefont {Ourjoumtsev},
  \citenamefont {Parigi}, \citenamefont {Sasaki},\ and\ \citenamefont
  {Tualle-Brouri}}]{Lvovsky2020}%
  \BibitemOpen
  \bibfield  {author} {\bibinfo {author} {\bibfnamefont {A.~I.}\ \bibnamefont
  {Lvovsky}}, \bibinfo {author} {\bibfnamefont {P.}~\bibnamefont {Grangier}},
  \bibinfo {author} {\bibfnamefont {A.}~\bibnamefont {Ourjoumtsev}}, \bibinfo
  {author} {\bibfnamefont {V.}~\bibnamefont {Parigi}}, \bibinfo {author}
  {\bibfnamefont {M.}~\bibnamefont {Sasaki}},\ and\ \bibinfo {author}
  {\bibfnamefont {R.}~\bibnamefont {Tualle-Brouri}},\ }\href@noop {} {\bibinfo
  {title} {{Production and applications of non-Gaussian quantum states of
  light}}} (\bibinfo {year} {2020}),\ \Eprint
  {https://arxiv.org/abs/2006.16985} {arXiv:2006.16985 [quant-ph]} \BibitemShut
  {NoStop}%
\bibitem [{\citenamefont {Wasilewski}\ \emph {et~al.}(2006)\citenamefont
  {Wasilewski}, \citenamefont {Lvovsky}, \citenamefont {Banaszek},\ and\
  \citenamefont {Radzewicz}}]{Wasilewski2006}%
  \BibitemOpen
  \bibfield  {author} {\bibinfo {author} {\bibfnamefont {W.}~\bibnamefont
  {Wasilewski}}, \bibinfo {author} {\bibfnamefont {A.~I.}\ \bibnamefont
  {Lvovsky}}, \bibinfo {author} {\bibfnamefont {K.}~\bibnamefont {Banaszek}},\
  and\ \bibinfo {author} {\bibfnamefont {C.}~\bibnamefont {Radzewicz}},\
  }\bibfield  {title} {\bibinfo {title} {{Pulsed squeezed light: Simultaneous
  squeezing of multiple modes}},\ }\href@noop {} {\bibfield  {journal}
  {\bibinfo  {journal} {Phys. Rev. A}\ }\textbf {\bibinfo {volume} {73}},\
  \bibinfo {pages} {063819} (\bibinfo {year} {2006})}\BibitemShut {NoStop}%
\bibitem [{\citenamefont {Lvovsky}\ \emph {et~al.}(2007)\citenamefont
  {Lvovsky}, \citenamefont {Wasilewski},\ and\ \citenamefont
  {Banaszek}}]{Lvovsky2007}%
  \BibitemOpen
  \bibfield  {author} {\bibinfo {author} {\bibfnamefont {A.~I.}\ \bibnamefont
  {Lvovsky}}, \bibinfo {author} {\bibfnamefont {W.}~\bibnamefont
  {Wasilewski}},\ and\ \bibinfo {author} {\bibfnamefont {K.}~\bibnamefont
  {Banaszek}},\ }\bibfield  {title} {\bibinfo {title} {{Decomposing a pulsed
  optical parametric amplifier into independent squeezers}},\ }\href@noop {}
  {\bibfield  {journal} {\bibinfo  {journal} {J. Mod. Opt.}\ }\textbf {\bibinfo
  {volume} {54}},\ \bibinfo {pages} {721} (\bibinfo {year} {2007})}\BibitemShut
  {NoStop}%
\bibitem [{\citenamefont {Christ}\ \emph {et~al.}(2013)\citenamefont {Christ},
  \citenamefont {Brecht}, \citenamefont {Mauerer},\ and\ \citenamefont
  {Silberhorn}}]{Christ2013}%
  \BibitemOpen
  \bibfield  {author} {\bibinfo {author} {\bibfnamefont {A.}~\bibnamefont
  {Christ}}, \bibinfo {author} {\bibfnamefont {B.}~\bibnamefont {Brecht}},
  \bibinfo {author} {\bibfnamefont {W.}~\bibnamefont {Mauerer}},\ and\ \bibinfo
  {author} {\bibfnamefont {C.}~\bibnamefont {Silberhorn}},\ }\bibfield  {title}
  {\bibinfo {title} {{Theory of quantum frequency conversion and type-II
  parametric down-conversion in the high-gain regime}},\ }\href@noop {}
  {\bibfield  {journal} {\bibinfo  {journal} {New J. Phys.}\ }\textbf {\bibinfo
  {volume} {15}},\ \bibinfo {pages} {053038} (\bibinfo {year}
  {2013})}\BibitemShut {NoStop}%
\bibitem [{\citenamefont {Shapiro}(2006)}]{Shapiro2006}%
  \BibitemOpen
  \bibfield  {author} {\bibinfo {author} {\bibfnamefont {J.~H.}\ \bibnamefont
  {Shapiro}},\ }\bibfield  {title} {\bibinfo {title} {{Single-photon Kerr
  nonlinearities do not help quantum computation}},\ }\href@noop {} {\bibfield
  {journal} {\bibinfo  {journal} {Phys. Rev. A}\ }\textbf {\bibinfo {volume}
  {73}},\ \bibinfo {pages} {062305} (\bibinfo {year} {2006})}\BibitemShut
  {NoStop}%
\bibitem [{\citenamefont {Lloyd}\ and\ \citenamefont
  {Braunstein}(1999)}]{Lloyd1999}%
  \BibitemOpen
  \bibfield  {author} {\bibinfo {author} {\bibfnamefont {S.}~\bibnamefont
  {Lloyd}}\ and\ \bibinfo {author} {\bibfnamefont {S.~L.}\ \bibnamefont
  {Braunstein}},\ }\bibfield  {title} {\bibinfo {title} {{Quantum Computation
  over Continuous Variables}},\ }\href@noop {} {\bibfield  {journal} {\bibinfo
  {journal} {Phys. Rev. Lett.}\ }\textbf {\bibinfo {volume} {82}},\ \bibinfo
  {pages} {1784} (\bibinfo {year} {1999})}\BibitemShut {NoStop}%
\bibitem [{\citenamefont {Braunstein}\ and\ \citenamefont {{van
  Loock}}(2005)}]{Braunstein2005}%
  \BibitemOpen
  \bibfield  {author} {\bibinfo {author} {\bibfnamefont {S.~L.}\ \bibnamefont
  {Braunstein}}\ and\ \bibinfo {author} {\bibfnamefont {P.}~\bibnamefont {{van
  Loock}}},\ }\bibfield  {title} {\bibinfo {title} {{Quantum information with
  continuous variables}},\ }\href@noop {} {\bibfield  {journal} {\bibinfo
  {journal} {Rev. Mod. Phys.}\ }\textbf {\bibinfo {volume} {77}},\ \bibinfo
  {pages} {513} (\bibinfo {year} {2005})}\BibitemShut {NoStop}%
\bibitem [{\citenamefont {Yanagimoto}\ \emph
  {et~al.}(2021{\natexlab{b}})\citenamefont {Yanagimoto}, \citenamefont {Ng},
  \citenamefont {Onodera},\ and\ \citenamefont
  {Mabuchi}}]{Yanagimoto2021-spie}%
  \BibitemOpen
  \bibfield  {author} {\bibinfo {author} {\bibfnamefont {R.}~\bibnamefont
  {Yanagimoto}}, \bibinfo {author} {\bibfnamefont {E.}~\bibnamefont {Ng}},
  \bibinfo {author} {\bibfnamefont {T.}~\bibnamefont {Onodera}},\ and\ \bibinfo
  {author} {\bibfnamefont {H.}~\bibnamefont {Mabuchi}},\ }\bibfield  {title}
  {\bibinfo {title} {{Towards an engineering framework for ultrafast quantum
  nonlinear optics}},\ }\href@noop {} {\bibfield  {journal} {\bibinfo
  {journal} {Proc. SPIE}\ }\textbf {\bibinfo {volume} {11684}},\ \bibinfo
  {pages} {11684D} (\bibinfo {year} {2021}{\natexlab{b}})}\BibitemShut
  {NoStop}%
\bibitem [{\citenamefont {Olivares}(2012)}]{Olivares2012}%
  \BibitemOpen
  \bibfield  {author} {\bibinfo {author} {\bibfnamefont {S.}~\bibnamefont
  {Olivares}},\ }\bibfield  {title} {\bibinfo {title} {{Quantum optics in the
  phase space: A tutorial on Gaussian states}},\ }\href@noop {} {\bibfield
  {journal} {\bibinfo  {journal} {Eur. Phys. J. Special Topics}\ }\textbf
  {\bibinfo {volume} {203}},\ \bibinfo {pages} {3} (\bibinfo {year}
  {2012})}\BibitemShut {NoStop}%
\bibitem [{\citenamefont {Weedbrook}\ \emph {et~al.}(2012)\citenamefont
  {Weedbrook}, \citenamefont {Pirandola}, \citenamefont {Garc\'ia-Patr\'on},
  \citenamefont {Cerf}, \citenamefont {Ralph}, \citenamefont {Shapiro},\ and\
  \citenamefont {Lloyd}}]{Weedbrook2021}%
  \BibitemOpen
  \bibfield  {author} {\bibinfo {author} {\bibfnamefont {C.}~\bibnamefont
  {Weedbrook}}, \bibinfo {author} {\bibfnamefont {S.}~\bibnamefont
  {Pirandola}}, \bibinfo {author} {\bibfnamefont {R.}~\bibnamefont
  {Garc\'ia-Patr\'on}}, \bibinfo {author} {\bibfnamefont {N.~J.}\ \bibnamefont
  {Cerf}}, \bibinfo {author} {\bibfnamefont {T.~C.}\ \bibnamefont {Ralph}},
  \bibinfo {author} {\bibfnamefont {J.~H.}\ \bibnamefont {Shapiro}},\ and\
  \bibinfo {author} {\bibfnamefont {S.}~\bibnamefont {Lloyd}},\ }\bibfield
  {title} {\bibinfo {title} {{Gaussian quantum information}},\ }\href@noop {}
  {\bibfield  {journal} {\bibinfo  {journal} {Rev. Mod. Phys.}\ }\textbf
  {\bibinfo {volume} {84}},\ \bibinfo {pages} {621} (\bibinfo {year}
  {2012})}\BibitemShut {NoStop}%
\bibitem [{\citenamefont {Gouzien}\ \emph {et~al.}(2020)\citenamefont
  {Gouzien}, \citenamefont {Tanzilli}, \citenamefont {D’Auria},\ and\
  \citenamefont {Patera}}]{Gouzien2020}%
  \BibitemOpen
  \bibfield  {author} {\bibinfo {author} {\bibfnamefont {E.}~\bibnamefont
  {Gouzien}}, \bibinfo {author} {\bibfnamefont {S.}~\bibnamefont {Tanzilli}},
  \bibinfo {author} {\bibfnamefont {V.}~\bibnamefont {D’Auria}},\ and\
  \bibinfo {author} {\bibfnamefont {G.}~\bibnamefont {Patera}},\ }\bibfield
  {title} {\bibinfo {title} {{Morphing Supermodes: A Full Characterization for
  Enabling Multimode Quantum Optics}},\ }\href@noop {} {\bibfield  {journal}
  {\bibinfo  {journal} {Phys. Rev. Lett.}\ }\textbf {\bibinfo {volume} {125}},\
  \bibinfo {pages} {103601} (\bibinfo {year} {2020})}\BibitemShut {NoStop}%
\bibitem [{\citenamefont {Onodera}\ \emph {et~al.}(2018)\citenamefont
  {Onodera}, \citenamefont {Ng}, \citenamefont {L\"orch}, \citenamefont
  {Yamamura}, \citenamefont {Hamerly}, \citenamefont {McMahon}, \citenamefont
  {Marandi},\ and\ \citenamefont {Mabuchi}}]{Onodera2018}%
  \BibitemOpen
  \bibfield  {author} {\bibinfo {author} {\bibfnamefont {T.}~\bibnamefont
  {Onodera}}, \bibinfo {author} {\bibfnamefont {E.}~\bibnamefont {Ng}},
  \bibinfo {author} {\bibfnamefont {N.}~\bibnamefont {L\"orch}}, \bibinfo
  {author} {\bibfnamefont {A.}~\bibnamefont {Yamamura}}, \bibinfo {author}
  {\bibfnamefont {R.}~\bibnamefont {Hamerly}}, \bibinfo {author} {\bibfnamefont
  {P.~L.}\ \bibnamefont {McMahon}}, \bibinfo {author} {\bibfnamefont
  {A.}~\bibnamefont {Marandi}},\ and\ \bibinfo {author} {\bibfnamefont
  {H.}~\bibnamefont {Mabuchi}},\ }\href@noop {} {\bibinfo {title} {{Nonlinear
  Quantum Behavior of Ultrashort-Pulse Optical Parametric Oscillators}}}
  (\bibinfo {year} {2018}),\ \Eprint {https://arxiv.org/abs/1811.10583}
  {arXiv:1811.10583 [quant-ph]} \BibitemShut {NoStop}%
\bibitem [{\citenamefont {Zhuang}\ \emph {et~al.}(2018)\citenamefont {Zhuang},
  \citenamefont {Shor},\ and\ \citenamefont {Shapiro}}]{Zhuang2018}%
  \BibitemOpen
  \bibfield  {author} {\bibinfo {author} {\bibfnamefont {Q.}~\bibnamefont
  {Zhuang}}, \bibinfo {author} {\bibfnamefont {P.~W.}\ \bibnamefont {Shor}},\
  and\ \bibinfo {author} {\bibfnamefont {J.~H.}\ \bibnamefont {Shapiro}},\
  }\bibfield  {title} {\bibinfo {title} {{Resource theory of non-Gaussian
  operations}},\ }\href@noop {} {\bibfield  {journal} {\bibinfo  {journal}
  {Phys. Rev. A}\ }\textbf {\bibinfo {volume} {97}},\ \bibinfo {pages} {052317}
  (\bibinfo {year} {2018})}\BibitemShut {NoStop}%
\bibitem [{\citenamefont {Albarelli}\ \emph {et~al.}(2018)\citenamefont
  {Albarelli}, \citenamefont {Genoni}, \citenamefont {Paris},\ and\
  \citenamefont {Ferraro}}]{Albarelli2018}%
  \BibitemOpen
  \bibfield  {author} {\bibinfo {author} {\bibfnamefont {F.}~\bibnamefont
  {Albarelli}}, \bibinfo {author} {\bibfnamefont {M.~G.}\ \bibnamefont
  {Genoni}}, \bibinfo {author} {\bibfnamefont {M.~G.~A.}\ \bibnamefont
  {Paris}},\ and\ \bibinfo {author} {\bibfnamefont {A.}~\bibnamefont
  {Ferraro}},\ }\bibfield  {title} {\bibinfo {title} {{Resource theory of
  quantum non-Gaussianity and Wigner negativity}},\ }\href@noop {} {\bibfield
  {journal} {\bibinfo  {journal} {Phys. Rev. A}\ }\textbf {\bibinfo {volume}
  {98}},\ \bibinfo {pages} {052350} (\bibinfo {year} {2018})}\BibitemShut
  {NoStop}%
\bibitem [{\citenamefont {Killoran}\ \emph {et~al.}(2019)\citenamefont
  {Killoran}, \citenamefont {Bromley}, \citenamefont {Arrazola}, \citenamefont
  {Schuld}, \citenamefont {Quesada},\ and\ \citenamefont
  {Lloyd}}]{Killoran2019}%
  \BibitemOpen
  \bibfield  {author} {\bibinfo {author} {\bibfnamefont {N.}~\bibnamefont
  {Killoran}}, \bibinfo {author} {\bibfnamefont {T.~R.}\ \bibnamefont
  {Bromley}}, \bibinfo {author} {\bibfnamefont {J.~M.}\ \bibnamefont
  {Arrazola}}, \bibinfo {author} {\bibfnamefont {M.}~\bibnamefont {Schuld}},
  \bibinfo {author} {\bibfnamefont {N.}~\bibnamefont {Quesada}},\ and\ \bibinfo
  {author} {\bibfnamefont {S.}~\bibnamefont {Lloyd}},\ }\bibfield  {title}
  {\bibinfo {title} {{Continuous-variable quantum neural networks}},\
  }\href@noop {} {\bibfield  {journal} {\bibinfo  {journal} {Phys. Rev.
  Research}\ }\textbf {\bibinfo {volume} {1}},\ \bibinfo {pages} {033063}
  (\bibinfo {year} {2019})}\BibitemShut {NoStop}%
\bibitem [{\citenamefont {Fan}\ and\ \citenamefont {Zubairy}(2018)}]{Fan2018}%
  \BibitemOpen
  \bibfield  {author} {\bibinfo {author} {\bibfnamefont {L.}~\bibnamefont
  {Fan}}\ and\ \bibinfo {author} {\bibfnamefont {M.~S.}\ \bibnamefont
  {Zubairy}},\ }\bibfield  {title} {\bibinfo {title} {{Quantum illumination
  using non-Gaussian states generated by photon subtraction and photon
  addition}},\ }\href@noop {} {\bibfield  {journal} {\bibinfo  {journal} {Phys.
  Rev. A}\ }\textbf {\bibinfo {volume} {98}},\ \bibinfo {pages} {012319}
  (\bibinfo {year} {2018})}\BibitemShut {NoStop}%
\bibitem [{\citenamefont {Gessner}\ \emph {et~al.}(2019)\citenamefont
  {Gessner}, \citenamefont {Smerzi},\ and\ \citenamefont
  {Pezz\'e}}]{Gessner2019}%
  \BibitemOpen
  \bibfield  {author} {\bibinfo {author} {\bibfnamefont {M.}~\bibnamefont
  {Gessner}}, \bibinfo {author} {\bibfnamefont {A.}~\bibnamefont {Smerzi}},\
  and\ \bibinfo {author} {\bibfnamefont {L.}~\bibnamefont {Pezz\'e}},\
  }\bibfield  {title} {\bibinfo {title} {{Metrological Nonlinear Squeezing
  Parameter}},\ }\href@noop {} {\bibfield  {journal} {\bibinfo  {journal}
  {Phys. Rev. Lett.}\ }\textbf {\bibinfo {volume} {122}},\ \bibinfo {pages}
  {090503} (\bibinfo {year} {2019})}\BibitemShut {NoStop}%
\bibitem [{\citenamefont {Tezak}\ \emph {et~al.}(2017)\citenamefont {Tezak},
  \citenamefont {Amini},\ and\ \citenamefont {Mabuchi}}]{Tezak2017}%
  \BibitemOpen
  \bibfield  {author} {\bibinfo {author} {\bibfnamefont {N.}~\bibnamefont
  {Tezak}}, \bibinfo {author} {\bibfnamefont {N.~H.}\ \bibnamefont {Amini}},\
  and\ \bibinfo {author} {\bibfnamefont {H.}~\bibnamefont {Mabuchi}},\
  }\bibfield  {title} {\bibinfo {title} {{Low-dimensional manifolds for exact
  representation of open quantum systems}},\ }\href@noop {} {\bibfield
  {journal} {\bibinfo  {journal} {Phys. Rev. A}\ }\textbf {\bibinfo {volume}
  {96}},\ \bibinfo {pages} {062113} (\bibinfo {year} {2017})}\BibitemShut
  {NoStop}%
\bibitem [{\citenamefont {Lu}\ \emph {et~al.}(2020)\citenamefont {Lu},
  \citenamefont {Li}, \citenamefont {Zou}, \citenamefont {{Al Sayem}},\ and\
  \citenamefont {Tang}}]{Lu2020}%
  \BibitemOpen
  \bibfield  {author} {\bibinfo {author} {\bibfnamefont {J.}~\bibnamefont
  {Lu}}, \bibinfo {author} {\bibfnamefont {M.}~\bibnamefont {Li}}, \bibinfo
  {author} {\bibfnamefont {C.-L.}\ \bibnamefont {Zou}}, \bibinfo {author}
  {\bibfnamefont {A.}~\bibnamefont {{Al Sayem}}},\ and\ \bibinfo {author}
  {\bibfnamefont {H.~X.}\ \bibnamefont {Tang}},\ }\bibfield  {title} {\bibinfo
  {title} {{Towards 1\% single photon nonlinearity with periodically-poled
  lithium niobate microring resonators}},\ }\href@noop {} {\bibfield  {journal}
  {\bibinfo  {journal} {Optica}\ }\textbf {\bibinfo {volume} {7}},\ \bibinfo
  {pages} {1654} (\bibinfo {year} {2020})}\BibitemShut {NoStop}%
\bibitem [{\citenamefont {Heuck}\ \emph {et~al.}(2020)\citenamefont {Heuck},
  \citenamefont {Jacobs},\ and\ \citenamefont {Englund}}]{Heuck2020b}%
  \BibitemOpen
  \bibfield  {author} {\bibinfo {author} {\bibfnamefont {M.}~\bibnamefont
  {Heuck}}, \bibinfo {author} {\bibfnamefont {K.}~\bibnamefont {Jacobs}},\ and\
  \bibinfo {author} {\bibfnamefont {D.~R.}\ \bibnamefont {Englund}},\
  }\bibfield  {title} {\bibinfo {title} {{Controlled-Phase Gate Using
  Dynamically Coupled Cavities and Optical Nonlinearities}},\ }\href@noop {}
  {\bibfield  {journal} {\bibinfo  {journal} {Phys. Rev. Lett.}\ }\textbf
  {\bibinfo {volume} {124}},\ \bibinfo {pages} {160501} (\bibinfo {year}
  {2020})}\BibitemShut {NoStop}%
\bibitem [{\citenamefont {Kinsler}\ and\ \citenamefont
  {Drummond}(1991)}]{Kinsler1991}%
  \BibitemOpen
  \bibfield  {author} {\bibinfo {author} {\bibfnamefont {P.}~\bibnamefont
  {Kinsler}}\ and\ \bibinfo {author} {\bibfnamefont {P.~D.}\ \bibnamefont
  {Drummond}},\ }\bibfield  {title} {\bibinfo {title} {{Quantum dynamics of the
  parametric oscillator}},\ }\href@noop {} {\bibfield  {journal} {\bibinfo
  {journal} {Phys. Rev. A}\ }\textbf {\bibinfo {volume} {43}},\ \bibinfo
  {pages} {6194} (\bibinfo {year} {1991})}\BibitemShut {NoStop}%
\bibitem [{\citenamefont {Quesada}\ and\ \citenamefont
  {Sipe}(2014)}]{Quesada2014}%
  \BibitemOpen
  \bibfield  {author} {\bibinfo {author} {\bibfnamefont {N.}~\bibnamefont
  {Quesada}}\ and\ \bibinfo {author} {\bibfnamefont {J.~E.}\ \bibnamefont
  {Sipe}},\ }\bibfield  {title} {\bibinfo {title} {{Effects of time ordering in
  quantum nonlinear optics}},\ }\href@noop {} {\bibfield  {journal} {\bibinfo
  {journal} {Phys. Rev. A}\ }\textbf {\bibinfo {volume} {90}},\ \bibinfo
  {pages} {063840} (\bibinfo {year} {2014})}\BibitemShut {NoStop}%
\bibitem [{\citenamefont {Kr\"amer}\ \emph {et~al.}(2018)\citenamefont
  {Kr\"amer}, \citenamefont {Plankensteiner}, \citenamefont {Ostermann},\ and\
  \citenamefont {Ritsch}}]{Kraemer2018}%
  \BibitemOpen
  \bibfield  {author} {\bibinfo {author} {\bibfnamefont {S.}~\bibnamefont
  {Kr\"amer}}, \bibinfo {author} {\bibfnamefont {D.}~\bibnamefont
  {Plankensteiner}}, \bibinfo {author} {\bibfnamefont {L.}~\bibnamefont
  {Ostermann}},\ and\ \bibinfo {author} {\bibfnamefont {H.}~\bibnamefont
  {Ritsch}},\ }\bibfield  {title} {\bibinfo {title} {{QuantumOptics.jl: A Julia
  framework for simulating open quantum systems}},\ }\href@noop {} {\bibfield
  {journal} {\bibinfo  {journal} {Comput. Phys. Commun}\ }\textbf {\bibinfo
  {volume} {227}},\ \bibinfo {pages} {109} (\bibinfo {year}
  {2018})}\BibitemShut {NoStop}%
\bibitem [{\citenamefont {Drummond}\ and\ \citenamefont
  {Hillery}(2014)}]{Drummond2014}%
  \BibitemOpen
  \bibfield  {author} {\bibinfo {author} {\bibfnamefont {P.~D.}\ \bibnamefont
  {Drummond}}\ and\ \bibinfo {author} {\bibfnamefont {M.}~\bibnamefont
  {Hillery}},\ }\href@noop {} {\emph {\bibinfo {title} {The Quantum Theory of
  Nonlinear Optics}}}\ (\bibinfo  {publisher} {Cambridge University Press},\
  \bibinfo {year} {2014})\BibitemShut {NoStop}%
\bibitem [{\citenamefont {Werner}\ \emph {et~al.}(1995)\citenamefont {Werner},
  \citenamefont {Raymer}, \citenamefont {Beck},\ and\ \citenamefont
  {Drummond}}]{Werner1995}%
  \BibitemOpen
  \bibfield  {author} {\bibinfo {author} {\bibfnamefont {M.~J.}\ \bibnamefont
  {Werner}}, \bibinfo {author} {\bibfnamefont {M.~G.}\ \bibnamefont {Raymer}},
  \bibinfo {author} {\bibfnamefont {M.}~\bibnamefont {Beck}},\ and\ \bibinfo
  {author} {\bibfnamefont {P.~D.}\ \bibnamefont {Drummond}},\ }\bibfield
  {title} {\bibinfo {title} {{Ultrashort pulsed squeezing by optical parametric
  amplification}},\ }\href@noop {} {\bibfield  {journal} {\bibinfo  {journal}
  {Phys. Rev. A}\ }\textbf {\bibinfo {volume} {52}},\ \bibinfo {pages} {4202}
  (\bibinfo {year} {1995})}\BibitemShut {NoStop}%
\bibitem [{\citenamefont {Helt}\ and\ \citenamefont
  {Quesada}(2020)}]{Helt2020}%
  \BibitemOpen
  \bibfield  {author} {\bibinfo {author} {\bibfnamefont {L.~G.}\ \bibnamefont
  {Helt}}\ and\ \bibinfo {author} {\bibfnamefont {N.}~\bibnamefont {Quesada}},\
  }\bibfield  {title} {\bibinfo {title} {{Degenerate squeezing in waveguides: a
  unified theoretical approach}},\ }\href@noop {} {\bibfield  {journal}
  {\bibinfo  {journal} {J. Phys. Photonics}\ }\textbf {\bibinfo {volume} {2}},\
  \bibinfo {pages} {035001} (\bibinfo {year} {2020})}\BibitemShut {NoStop}%
\bibitem [{\citenamefont {Quesada}\ \emph {et~al.}(2020)\citenamefont
  {Quesada}, \citenamefont {Triginer}, \citenamefont {Vidrighin},\ and\
  \citenamefont {Sipe}}]{Quesada2020-pdc}%
  \BibitemOpen
  \bibfield  {author} {\bibinfo {author} {\bibfnamefont {N.}~\bibnamefont
  {Quesada}}, \bibinfo {author} {\bibfnamefont {G.}~\bibnamefont {Triginer}},
  \bibinfo {author} {\bibfnamefont {M.~D.}\ \bibnamefont {Vidrighin}},\ and\
  \bibinfo {author} {\bibfnamefont {J.~E.}\ \bibnamefont {Sipe}},\ }\bibfield
  {title} {\bibinfo {title} {{Theory of high-gain twin-beam generation in
  waveguides: From Maxwell's equations to efficient simulation}},\ }\href@noop
  {} {\bibfield  {journal} {\bibinfo  {journal} {Phys. Rev. A}\ }\textbf
  {\bibinfo {volume} {102}},\ \bibinfo {pages} {033519} (\bibinfo {year}
  {2020})}\BibitemShut {NoStop}%
\bibitem [{\citenamefont {Agrawal}(2019)}]{Agrawal2019}%
  \BibitemOpen
  \bibfield  {author} {\bibinfo {author} {\bibfnamefont {G.~P.}\ \bibnamefont
  {Agrawal}},\ }\href@noop {} {\emph {\bibinfo {title} {{Nonlinear Fiber
  Optics, 6th edition}}}}\ (\bibinfo  {publisher} {Academic Press},\ \bibinfo
  {year} {2019})\BibitemShut {NoStop}%
\bibitem [{\citenamefont {Vidal}(2003)}]{Vidal2003}%
  \BibitemOpen
  \bibfield  {author} {\bibinfo {author} {\bibfnamefont {G.}~\bibnamefont
  {Vidal}},\ }\bibfield  {title} {\bibinfo {title} {{Efficient Classical
  Simulation of Slightly Entangled Quantum Computers}},\ }\href@noop {}
  {\bibfield  {journal} {\bibinfo  {journal} {Phys. Rev. Lett.}\ }\textbf
  {\bibinfo {volume} {91}},\ \bibinfo {pages} {147902} (\bibinfo {year}
  {2003})}\BibitemShut {NoStop}%
\bibitem [{\citenamefont {Garc{\'i}a-Ripoll}(2006)}]{Garcia-Ripoll2006}%
  \BibitemOpen
  \bibfield  {author} {\bibinfo {author} {\bibfnamefont {J.~J.}\ \bibnamefont
  {Garc{\'i}a-Ripoll}},\ }\bibfield  {title} {\bibinfo {title} {{Time evolution
  of Matrix Product States}},\ }\href@noop {} {\bibfield  {journal} {\bibinfo
  {journal} {New J. Phys.}\ }\textbf {\bibinfo {volume} {8}},\ \bibinfo {pages}
  {305} (\bibinfo {year} {2006})}\BibitemShut {NoStop}%
\bibitem [{\citenamefont {Yanagimoto}\ \emph
  {et~al.}(2020{\natexlab{b}})\citenamefont {Yanagimoto}, \citenamefont {Ng},
  \citenamefont {Jankowski}, \citenamefont {Onodera}, \citenamefont {Fejer},\
  and\ \citenamefont {Mabuchi}}]{Yanagimoto2020_fano}%
  \BibitemOpen
  \bibfield  {author} {\bibinfo {author} {\bibfnamefont {R.}~\bibnamefont
  {Yanagimoto}}, \bibinfo {author} {\bibfnamefont {E.}~\bibnamefont {Ng}},
  \bibinfo {author} {\bibfnamefont {M.~P.}\ \bibnamefont {Jankowski}}, \bibinfo
  {author} {\bibfnamefont {T.}~\bibnamefont {Onodera}}, \bibinfo {author}
  {\bibfnamefont {M.~M.}\ \bibnamefont {Fejer}},\ and\ \bibinfo {author}
  {\bibfnamefont {H.}~\bibnamefont {Mabuchi}},\ }\href@noop {} {\bibinfo
  {title} {{Broadband Parametric Downconversion as a Discrete-Continuum Fano
  Interaction}}} (\bibinfo {year} {2020}{\natexlab{b}}),\ \Eprint
  {https://arxiv.org/abs/2009.01457} {arXiv:2009.01457 [quant-ph]} \BibitemShut
  {NoStop}%
\bibitem [{\citenamefont {Patera}\ \emph {et~al.}(2010)\citenamefont {Patera},
  \citenamefont {Treps}, \citenamefont {Fabre},\ and\ \citenamefont {{de
  Valc\'arcel}}}]{Patera2010}%
  \BibitemOpen
  \bibfield  {author} {\bibinfo {author} {\bibfnamefont {G.}~\bibnamefont
  {Patera}}, \bibinfo {author} {\bibfnamefont {N.}~\bibnamefont {Treps}},
  \bibinfo {author} {\bibfnamefont {C.}~\bibnamefont {Fabre}},\ and\ \bibinfo
  {author} {\bibfnamefont {G.~J.}\ \bibnamefont {{de Valc\'arcel}}},\
  }\bibfield  {title} {\bibinfo {title} {{Quantum theory of synchronously
  pumped type I optical parametric oscillators: characterization of the
  squeezed supermodes}},\ }\href@noop {} {\bibfield  {journal} {\bibinfo
  {journal} {Eur. Phys. J. D}\ }\textbf {\bibinfo {volume} {56}},\ \bibinfo
  {pages} {123} (\bibinfo {year} {2010})}\BibitemShut {NoStop}%
\bibitem [{\citenamefont {Raymer}\ and\ \citenamefont
  {Walmsley}(2020)}]{Raymer2020-review}%
  \BibitemOpen
  \bibfield  {author} {\bibinfo {author} {\bibfnamefont {M.~G.}\ \bibnamefont
  {Raymer}}\ and\ \bibinfo {author} {\bibfnamefont {I.~A.}\ \bibnamefont
  {Walmsley}},\ }\bibfield  {title} {\bibinfo {title} {{Temporal modes in
  quantum optics: then and now}},\ }\href@noop {} {\bibfield  {journal}
  {\bibinfo  {journal} {Phys. Scr.}\ }\textbf {\bibinfo {volume} {95}},\
  \bibinfo {pages} {064002} (\bibinfo {year} {2020})}\BibitemShut {NoStop}%
\bibitem [{\citenamefont {Stanojevic}\ and\ \citenamefont
  {C\^ot\'e}(2009)}]{Stanojevic2009}%
  \BibitemOpen
  \bibfield  {author} {\bibinfo {author} {\bibfnamefont {J.}~\bibnamefont
  {Stanojevic}}\ and\ \bibinfo {author} {\bibfnamefont {R.}~\bibnamefont
  {C\^ot\'e}},\ }\bibfield  {title} {\bibinfo {title} {{Many-body Rabi
  oscillations of Rydberg excitation in small mesoscopic samples}},\
  }\href@noop {} {\bibfield  {journal} {\bibinfo  {journal} {Phys. Rev. A}\
  }\textbf {\bibinfo {volume} {80}},\ \bibinfo {pages} {033418} (\bibinfo
  {year} {2009})}\BibitemShut {NoStop}%
\bibitem [{\citenamefont {Weber}\ \emph {et~al.}(2015)\citenamefont {Weber},
  \citenamefont {H\"oning}, \citenamefont {Niederpr\"um}, \citenamefont
  {Manthey}, \citenamefont {Thomas}, \citenamefont {Guarrera}, \citenamefont
  {Fleischhauer}, \citenamefont {Barontini},\ and\ \citenamefont
  {Ott}}]{Weber2015}%
  \BibitemOpen
  \bibfield  {author} {\bibinfo {author} {\bibfnamefont {T.~M.}\ \bibnamefont
  {Weber}}, \bibinfo {author} {\bibfnamefont {M.}~\bibnamefont {H\"oning}},
  \bibinfo {author} {\bibfnamefont {T.}~\bibnamefont {Niederpr\"um}}, \bibinfo
  {author} {\bibfnamefont {T.}~\bibnamefont {Manthey}}, \bibinfo {author}
  {\bibfnamefont {O.}~\bibnamefont {Thomas}}, \bibinfo {author} {\bibfnamefont
  {V.}~\bibnamefont {Guarrera}}, \bibinfo {author} {\bibfnamefont
  {M.}~\bibnamefont {Fleischhauer}}, \bibinfo {author} {\bibfnamefont
  {G.}~\bibnamefont {Barontini}},\ and\ \bibinfo {author} {\bibfnamefont
  {H.}~\bibnamefont {Ott}},\ }\bibfield  {title} {\bibinfo {title} {{Mesoscopic
  Rydberg-blockaded ensembles in the superatom regime and beyond}},\
  }\href@noop {} {\bibfield  {journal} {\bibinfo  {journal} {Nat. Phys.}\
  }\textbf {\bibinfo {volume} {11}},\ \bibinfo {pages} {157} (\bibinfo {year}
  {2015})}\BibitemShut {NoStop}%
\bibitem [{\citenamefont {Imry}(1997)}]{Imry1997}%
  \BibitemOpen
  \bibfield  {author} {\bibinfo {author} {\bibfnamefont {Y.}~\bibnamefont
  {Imry}},\ }\href@noop {} {\emph {\bibinfo {title} {{Introduction to
  Mesoscopic Physics}}}}\ (\bibinfo  {publisher} {Oxford University Press},\
  \bibinfo {year} {1997})\BibitemShut {NoStop}%
\bibitem [{\citenamefont {Harder}\ \emph {et~al.}(2016)\citenamefont {Harder},
  \citenamefont {Bartley}, \citenamefont {Lita}, \citenamefont {Nam},
  \citenamefont {Gerrits},\ and\ \citenamefont {Silberhorn}}]{Harder2016}%
  \BibitemOpen
  \bibfield  {author} {\bibinfo {author} {\bibfnamefont {G.}~\bibnamefont
  {Harder}}, \bibinfo {author} {\bibfnamefont {T.~J.}\ \bibnamefont {Bartley}},
  \bibinfo {author} {\bibfnamefont {A.~E.}\ \bibnamefont {Lita}}, \bibinfo
  {author} {\bibfnamefont {S.~W.}\ \bibnamefont {Nam}}, \bibinfo {author}
  {\bibfnamefont {T.}~\bibnamefont {Gerrits}},\ and\ \bibinfo {author}
  {\bibfnamefont {C.}~\bibnamefont {Silberhorn}},\ }\bibfield  {title}
  {\bibinfo {title} {{Single-Mode Parametric-Down-Conversion States with 50
  Photons as a Source for Mesoscopic Quantum Optics}},\ }\href@noop {}
  {\bibfield  {journal} {\bibinfo  {journal} {Phys. Rev. Lett.}\ }\textbf
  {\bibinfo {volume} {116}},\ \bibinfo {pages} {143601} (\bibinfo {year}
  {2016})}\BibitemShut {NoStop}%
\bibitem [{\citenamefont {Antonosyan}\ \emph {et~al.}(2014)\citenamefont
  {Antonosyan}, \citenamefont {Solntsev},\ and\ \citenamefont
  {Sukhorukov}}]{Antonosyan2014}%
  \BibitemOpen
  \bibfield  {author} {\bibinfo {author} {\bibfnamefont {D.~A.}\ \bibnamefont
  {Antonosyan}}, \bibinfo {author} {\bibfnamefont {A.~S.}\ \bibnamefont
  {Solntsev}},\ and\ \bibinfo {author} {\bibfnamefont {A.~A.}\ \bibnamefont
  {Sukhorukov}},\ }\bibfield  {title} {\bibinfo {title} {{Single-photon
  spontaneous parametric down-conversion in quadratic nonlinear waveguide
  arrays}},\ }\href@noop {} {\bibfield  {journal} {\bibinfo  {journal} {Opt.
  Commun.}\ }\textbf {\bibinfo {volume} {327}},\ \bibinfo {pages} {22}
  (\bibinfo {year} {2014})}\BibitemShut {NoStop}%
\bibitem [{\citenamefont {Ferrera}\ \emph {et~al.}(2008)\citenamefont
  {Ferrera}, \citenamefont {Razzari}, \citenamefont {Duchesne}, \citenamefont
  {Morandotti}, \citenamefont {Yang}, \citenamefont {Liscidini}, \citenamefont
  {Sipe}, \citenamefont {Chu}, \citenamefont {Little},\ and\ \citenamefont
  {Moss}}]{Ferrera2008}%
  \BibitemOpen
  \bibfield  {author} {\bibinfo {author} {\bibfnamefont {M.}~\bibnamefont
  {Ferrera}}, \bibinfo {author} {\bibfnamefont {L.}~\bibnamefont {Razzari}},
  \bibinfo {author} {\bibfnamefont {D.}~\bibnamefont {Duchesne}}, \bibinfo
  {author} {\bibfnamefont {R.}~\bibnamefont {Morandotti}}, \bibinfo {author}
  {\bibfnamefont {Z.}~\bibnamefont {Yang}}, \bibinfo {author} {\bibfnamefont
  {M.}~\bibnamefont {Liscidini}}, \bibinfo {author} {\bibfnamefont {J.~E.}\
  \bibnamefont {Sipe}}, \bibinfo {author} {\bibfnamefont {S.}~\bibnamefont
  {Chu}}, \bibinfo {author} {\bibfnamefont {B.~E.}\ \bibnamefont {Little}},\
  and\ \bibinfo {author} {\bibfnamefont {D.~J.}\ \bibnamefont {Moss}},\
  }\bibfield  {title} {\bibinfo {title} {{Low-power continuous-wave nonlinear
  optics in doped silica glass integrated waveguide structures}},\ }\href@noop
  {} {\bibfield  {journal} {\bibinfo  {journal} {Nat. Photon.}\ }\textbf
  {\bibinfo {volume} {2}},\ \bibinfo {pages} {737} (\bibinfo {year}
  {2008})}\BibitemShut {NoStop}%
\bibitem [{\citenamefont {Liu}\ \emph {et~al.}(2018)\citenamefont {Liu},
  \citenamefont {Raja}, \citenamefont {Karpov}, \citenamefont {Ghadiani},
  \citenamefont {Pfeiffer}, \citenamefont {Du}, \citenamefont {Engelsen},
  \citenamefont {Guo}, \citenamefont {Zervas},\ and\ \citenamefont
  {Kippenberg}}]{Liu2018}%
  \BibitemOpen
  \bibfield  {author} {\bibinfo {author} {\bibfnamefont {J.}~\bibnamefont
  {Liu}}, \bibinfo {author} {\bibfnamefont {A.~S.}\ \bibnamefont {Raja}},
  \bibinfo {author} {\bibfnamefont {M.}~\bibnamefont {Karpov}}, \bibinfo
  {author} {\bibfnamefont {B.}~\bibnamefont {Ghadiani}}, \bibinfo {author}
  {\bibfnamefont {M.~H.~P.}\ \bibnamefont {Pfeiffer}}, \bibinfo {author}
  {\bibfnamefont {B.}~\bibnamefont {Du}}, \bibinfo {author} {\bibfnamefont
  {N.~J.}\ \bibnamefont {Engelsen}}, \bibinfo {author} {\bibfnamefont
  {H.}~\bibnamefont {Guo}}, \bibinfo {author} {\bibfnamefont {M.}~\bibnamefont
  {Zervas}},\ and\ \bibinfo {author} {\bibfnamefont {T.~J.}\ \bibnamefont
  {Kippenberg}},\ }\bibfield  {title} {\bibinfo {title} {{Ultralow-power
  chip-based soliton microcombs for photonic integration}},\ }\href@noop {}
  {\bibfield  {journal} {\bibinfo  {journal} {Optica}\ }\textbf {\bibinfo
  {volume} {5}},\ \bibinfo {pages} {1347} (\bibinfo {year} {2018})}\BibitemShut
  {NoStop}%
\bibitem [{\citenamefont {McKenna}\ \emph {et~al.}(2021)\citenamefont
  {McKenna}, \citenamefont {Stokowski}, \citenamefont {Ansari}, \citenamefont
  {Mishra}, \citenamefont {Jankowski}, \citenamefont {Sarabalis}, \citenamefont
  {Herrmann}, \citenamefont {Langrock}, \citenamefont {Fejer},\ and\
  \citenamefont {Safavi-Naeini}}]{McKenna2021}%
  \BibitemOpen
  \bibfield  {author} {\bibinfo {author} {\bibfnamefont {T.~P.}\ \bibnamefont
  {McKenna}}, \bibinfo {author} {\bibfnamefont {H.~S.}\ \bibnamefont
  {Stokowski}}, \bibinfo {author} {\bibfnamefont {V.}~\bibnamefont {Ansari}},
  \bibinfo {author} {\bibfnamefont {J.}~\bibnamefont {Mishra}}, \bibinfo
  {author} {\bibfnamefont {M.}~\bibnamefont {Jankowski}}, \bibinfo {author}
  {\bibfnamefont {C.~J.}\ \bibnamefont {Sarabalis}}, \bibinfo {author}
  {\bibfnamefont {J.~F.}\ \bibnamefont {Herrmann}}, \bibinfo {author}
  {\bibfnamefont {C.}~\bibnamefont {Langrock}}, \bibinfo {author}
  {\bibfnamefont {M.~M.}\ \bibnamefont {Fejer}},\ and\ \bibinfo {author}
  {\bibfnamefont {A.~H.}\ \bibnamefont {Safavi-Naeini}},\ }\href@noop {}
  {\bibinfo {title} {{Ultra-low-power second-order nonlinear optics on a
  chip}}} (\bibinfo {year} {2021}),\ \Eprint {https://arxiv.org/abs/2102.05617}
  {arXiv:2102.05617 [physics.optics]} \BibitemShut {NoStop}%
\bibitem [{\citenamefont {Jankowski}\ \emph
  {et~al.}(2021{\natexlab{b}})\citenamefont {Jankowski}, \citenamefont
  {Jornod}, \citenamefont {Langrock}, \citenamefont {Desiatov}, \citenamefont
  {Marandi}, \citenamefont {Lon{\u c}ar},\ and\ \citenamefont
  {Fejer}}]{Jankowski2021-pdc}%
  \BibitemOpen
  \bibfield  {author} {\bibinfo {author} {\bibfnamefont {M.}~\bibnamefont
  {Jankowski}}, \bibinfo {author} {\bibfnamefont {N.}~\bibnamefont {Jornod}},
  \bibinfo {author} {\bibfnamefont {C.}~\bibnamefont {Langrock}}, \bibinfo
  {author} {\bibfnamefont {B.}~\bibnamefont {Desiatov}}, \bibinfo {author}
  {\bibfnamefont {A.}~\bibnamefont {Marandi}}, \bibinfo {author} {\bibfnamefont
  {M.}~\bibnamefont {Lon{\u c}ar}},\ and\ \bibinfo {author} {\bibfnamefont
  {M.~M.}\ \bibnamefont {Fejer}},\ }\href@noop {} {\bibinfo {title} {{Efficient
  Octave-Spanning Parametric Down-Conversion at the Picojoule Level}}}
  (\bibinfo {year} {2021}{\natexlab{b}}),\ \Eprint
  {https://arxiv.org/abs/2104.07928} {arXiv:2104.07928 [physics.optics]}
  \BibitemShut {NoStop}%
\bibitem [{\citenamefont {Leroux}\ \emph {et~al.}(2018)\citenamefont {Leroux},
  \citenamefont {Govia},\ and\ \citenamefont {Clerk}}]{Leroux2018}%
  \BibitemOpen
  \bibfield  {author} {\bibinfo {author} {\bibfnamefont {C.}~\bibnamefont
  {Leroux}}, \bibinfo {author} {\bibfnamefont {L.~C.~G.}\ \bibnamefont
  {Govia}},\ and\ \bibinfo {author} {\bibfnamefont {A.~A.}\ \bibnamefont
  {Clerk}},\ }\bibfield  {title} {\bibinfo {title} {{Enhancing Cavity Quantum
  Electrodynamics via Antisqueezing: Synthetic Ultrastrong Coupling}},\
  }\href@noop {} {\bibfield  {journal} {\bibinfo  {journal} {Phys. Rev. Lett.}\
  }\textbf {\bibinfo {volume} {120}},\ \bibinfo {pages} {093602} (\bibinfo
  {year} {2018})}\BibitemShut {NoStop}%
\bibitem [{\citenamefont {Qin}\ \emph {et~al.}(2018)\citenamefont {Qin},
  \citenamefont {Miranowicz}, \citenamefont {Li}, \citenamefont {L\"u},
  \citenamefont {You},\ and\ \citenamefont {Nori}}]{Qin2018}%
  \BibitemOpen
  \bibfield  {author} {\bibinfo {author} {\bibfnamefont {W.}~\bibnamefont
  {Qin}}, \bibinfo {author} {\bibfnamefont {A.}~\bibnamefont {Miranowicz}},
  \bibinfo {author} {\bibfnamefont {P.-B.}\ \bibnamefont {Li}}, \bibinfo
  {author} {\bibfnamefont {X.-Y.}\ \bibnamefont {L\"u}}, \bibinfo {author}
  {\bibfnamefont {J.~Q.}\ \bibnamefont {You}},\ and\ \bibinfo {author}
  {\bibfnamefont {F.}~\bibnamefont {Nori}},\ }\bibfield  {title} {\bibinfo
  {title} {{Exponentially Enhanced Light-Matter Interaction, Cooperativities,
  and Steady-State Entanglement Using Parametric Amplification}},\ }\href@noop
  {} {\bibfield  {journal} {\bibinfo  {journal} {Phys. Rev. Lett.}\ }\textbf
  {\bibinfo {volume} {120}},\ \bibinfo {pages} {093601} (\bibinfo {year}
  {2018})}\BibitemShut {NoStop}%
\bibitem [{\citenamefont {Ramelow}\ \emph {et~al.}(2019)\citenamefont
  {Ramelow}, \citenamefont {Farsi}, \citenamefont {Vernon}, \citenamefont
  {Clemmen}, \citenamefont {Ji}, \citenamefont {Sipe}, \citenamefont
  {Liscidini}, \citenamefont {Lipson},\ and\ \citenamefont
  {Gaeta}}]{Ramelow2019-ring}%
  \BibitemOpen
  \bibfield  {author} {\bibinfo {author} {\bibfnamefont {S.}~\bibnamefont
  {Ramelow}}, \bibinfo {author} {\bibfnamefont {A.}~\bibnamefont {Farsi}},
  \bibinfo {author} {\bibfnamefont {Z.}~\bibnamefont {Vernon}}, \bibinfo
  {author} {\bibfnamefont {S.}~\bibnamefont {Clemmen}}, \bibinfo {author}
  {\bibfnamefont {X.}~\bibnamefont {Ji}}, \bibinfo {author} {\bibfnamefont
  {J.~E.}\ \bibnamefont {Sipe}}, \bibinfo {author} {\bibfnamefont
  {M.}~\bibnamefont {Liscidini}}, \bibinfo {author} {\bibfnamefont
  {M.}~\bibnamefont {Lipson}},\ and\ \bibinfo {author} {\bibfnamefont {A.~L.}\
  \bibnamefont {Gaeta}},\ }\bibfield  {title} {\bibinfo {title} {{Strong
  Nonlinear Coupling in a $\text{Si}_3\text{N}_4$ Ring Resonator}},\
  }\href@noop {} {\bibfield  {journal} {\bibinfo  {journal} {Phys. Rev. Lett.}\
  }\textbf {\bibinfo {volume} {122}},\ \bibinfo {pages} {153906} (\bibinfo
  {year} {2019})}\BibitemShut {NoStop}%
\bibitem [{\citenamefont {Park}\ \emph {et~al.}(2021)\citenamefont {Park},
  \citenamefont {Stokowski}, \citenamefont {Ansari}, \citenamefont {McKenna},
  \citenamefont {Hwang}, \citenamefont {Fejer},\ and\ \citenamefont
  {Safavi-Naeini}}]{Park2021}%
  \BibitemOpen
  \bibfield  {author} {\bibinfo {author} {\bibfnamefont {T.}~\bibnamefont
  {Park}}, \bibinfo {author} {\bibfnamefont {H.~S.}\ \bibnamefont {Stokowski}},
  \bibinfo {author} {\bibfnamefont {V.}~\bibnamefont {Ansari}}, \bibinfo
  {author} {\bibfnamefont {T.~P.}\ \bibnamefont {McKenna}}, \bibinfo {author}
  {\bibfnamefont {A.~Y.}\ \bibnamefont {Hwang}}, \bibinfo {author}
  {\bibfnamefont {M.~M.}\ \bibnamefont {Fejer}},\ and\ \bibinfo {author}
  {\bibfnamefont {A.~H.}\ \bibnamefont {Safavi-Naeini}},\ }\href@noop {}
  {\bibinfo {title} {{High efficiency second harmonic generation of blue light
  on thin film lithium niobate}}} (\bibinfo {year} {2021}),\ \Eprint
  {https://arxiv.org/abs/2108.06398} {arXiv:2108.06398 [physics.optics]}
  \BibitemShut {NoStop}%
\bibitem [{\citenamefont {Zhang}\ \emph {et~al.}(2017)\citenamefont {Zhang},
  \citenamefont {Wang}, \citenamefont {Cheng}, \citenamefont {Shams-Ansari},\
  and\ \citenamefont {Lon{\v c}ar}}]{Zhang2017}%
  \BibitemOpen
  \bibfield  {author} {\bibinfo {author} {\bibfnamefont {M.}~\bibnamefont
  {Zhang}}, \bibinfo {author} {\bibfnamefont {C.}~\bibnamefont {Wang}},
  \bibinfo {author} {\bibfnamefont {R.}~\bibnamefont {Cheng}}, \bibinfo
  {author} {\bibfnamefont {A.}~\bibnamefont {Shams-Ansari}},\ and\ \bibinfo
  {author} {\bibfnamefont {M.}~\bibnamefont {Lon{\v c}ar}},\ }\bibfield
  {title} {\bibinfo {title} {{Monolithic ultra-high-Q lithium niobate microring
  resonator}},\ }\href@noop {} {\bibfield  {journal} {\bibinfo  {journal}
  {Optica}\ }\textbf {\bibinfo {volume} {4}},\ \bibinfo {pages} {1536}
  (\bibinfo {year} {2017})}\BibitemShut {NoStop}%
\bibitem [{\citenamefont {Ricardo}(2020)}]{Ricardo2020}%
  \BibitemOpen
  \bibfield  {author} {\bibinfo {author} {\bibfnamefont {H.~J.}\ \bibnamefont
  {Ricardo}},\ }\href@noop {} {\emph {\bibinfo {title} {A Modern Introduction
  to Differential Equations}}}\ (\bibinfo  {publisher} {Academic Press},\
  \bibinfo {year} {2020})\BibitemShut {NoStop}%
\bibitem [{\citenamefont {Case}(2008)}]{Case2008}%
  \BibitemOpen
  \bibfield  {author} {\bibinfo {author} {\bibfnamefont {W.~B.}\ \bibnamefont
  {Case}},\ }\bibfield  {title} {\bibinfo {title} {{Wigner functions and Weyl
  transforms for pedestrians}},\ }\href@noop {} {\bibfield  {journal} {\bibinfo
   {journal} {Am. J. Phys.}\ }\textbf {\bibinfo {volume} {76}},\ \bibinfo
  {pages} {937} (\bibinfo {year} {2008})}\BibitemShut {NoStop}%
\end{thebibliography}%
\end{document}